\documentclass[aps,pre,longbibliography,twocolumn,superscriptaddress,showpacs]{revtex4-1}

\usepackage{CJK}

\usepackage{graphicx}

\usepackage{amsmath}
\usepackage{dcolumn}

\usepackage{bm}
\usepackage[normalem]{ulem}
\usepackage{xcolor}
\usepackage{hyperref}

\usepackage{enumitem,kantlipsum}
\numberwithin{equation}{section}

\usepackage{epstopdf}

\newcommand{\gun}{g^{\rm (unc)}}
\newcommand{\gha}{g^{\rm (hard)}}
\newcommand{\gso}{g^{\rm (soft)}}

\newcommand{\bew}{\begin{widetext}}
\newcommand{\ew}{\end{widetext}}

\newcommand{\ii}{{\rm i}}
\newcommand{\bx}{\mathbf{x}}

\newcommand{\bq}{\mathbf{q}}
\newcommand{\bv}{\mathbf{v}}

\newcommand{\br}{\mathbf{r}}

\newcommand{\brh}{\mathbf{r}_h}

\newcommand{\bqh}{\mathbf{q}_h}

\newcommand{\bff}{\mathbf{f}}
\newcommand{\bu}{\mathbf{u}}

\newcommand{\bbr}{\mathbf{r}}

\newcommand{\bM}{\mathbf{M}}
\newcommand{\bk}{\mathbf{k}}

\newcommand{\sep}{ \ \ \ , \ \ \ }

\newcommand{\beq}{\begin{equation}}
\newcommand{\eeq}{\end{equation}}
\newcommand{\beqn}{\begin{eqnarray}}
\newcommand{\eeqn}{\end{eqnarray}}
\newcommand{\pp}{\partial}
\newcommand{\dd}{{\rm d}}
\newcommand{\ee}{{\rm e}}

\newcommand{\fig}{Fig.\ }

\newcommand{\cO}{{\cal O}}

\newcommand{\cG}{{\cal G}}

\newcommand{\la}{\langle}

\newcommand{\tbq}{\tilde{\bq}}
\newcommand{\tbk}{\tilde{\bk}}

\newcommand{\ra}{\rangle}

\newcommand{\vnab}{{\bf \nabla}}

\newcommand{\AM}[1]{\textcolor{black}{#1}}

\begin{document}
\title{Hydrodynamic theory of two-dimensional incompressible polar active fluids with quenched and annealed disorder}
\author{Leiming Chen}
\email{leiming@cumt.edu.cn}
\address{School of Material Science and Physics, China University of Mining and Technology, Xuzhou Jiangsu, 221116, P. R. China}
\author{Chiu Fan Lee}
\email{c.lee@imperial.ac.uk}
\address{Department of Bioengineering, Imperial College London, South Kensington Campus, London SW7 2AZ, U.K.}
\author{Ananyo Maitra}
\email{nyomaitra07@gmail.com}
\address{Laboratoire de Physique Th\'eorique et Mod\'elisation, CNRS UMR 8089,
	CY Cergy Paris Universit\'e, F-95032 Cergy-Pontoise Cedex, France}
\author{John Toner}
\email{jjt@uoregon.edu}
\affiliation{Department of Physics and Institute of Theoretical
Science, University of Oregon, Eugene, OR $97403^1$}
\affiliation{Max Planck Institute for the Physics of Complex Systems, N\"othnitzer Str. 38, 01187 Dresden, Germany}
\date{\today}

	\begin{abstract}
We study the moving phase of two-dimensional (2D) incompressible polar active fluids in the presence of both quenched and annealed disorder. We show that long-range polar order persists \emph{even} in this defect-ridden two-dimensional system. We obtain the large-distance, long-time scaling laws of the velocity fluctuations using three distinct dynamic renormalization group schemes. These are an uncontrolled one-loop calculation in exactly two dimensions, and two $d=(d_c-\epsilon)$-expansions to $O(\epsilon)$, obtained by two different analytic continuations of our 2D model to higher spatial dimensions: a ``hard" continuation which has  $d_c={7\over 3}$, and a ``soft" continuation with  $d_c={5\over 2}$.
Surprisingly, the quenched and annealed parts of the velocity correlation function have the same anisotropy exponent and the relaxational and propagating parts of the dispersion relation have the same dynamic exponent in the nonlinear theory even though they are distinct in the linearized theory. This is due to anomalous hydrodynamics. Furthermore, all three renormalization
schemes yield very similar values for the universal exponents, and, therefore, we expect the numerical values we predict for them to be highly accurate.
 \end{abstract}
	
\maketitle

\section{Introduction}
\label{Intro}

The competition between order and disorder is one of the central themes of statistical mechanics and condensed matter physics. It is well known that in sufficiently low spatial dimensions, disorder always wins. The  ``Mermin-Wagner-Hohenberg" theorem \cite{MW}, proves this for equilibrium systems ``trying" to break a continuous symmetry at finite temperature in two or fewer dimensions.

Once out of equilibrium, however, the Mermin-Wagner theorem no longer applies. And it has been demonstrated for one particular class of non-equilibrium systems, namely active systems, that long-rang order is possible in two dimensions even in the presence of noise
\cite{vicsek_prl95,toner_prl95, toner_pre98}. In particular,  polar self-propelled particles  moving over a frictional substrate (a system often described as a ``dry polar active fluid") can ``flock"; that is, form a state with a non-zero average velocity $\langle\bv\rangle$, even when perturbed by noise.

The aforementioned results all describe systems in which the ``noise" - that is, the random force ``trying" to disorder the system - is ``annealed"; that is,  time-dependent with only short ranged in time temporal correlations. Thermal ``Brownian"  noise is ``annealed" in this sense.

In equilibrium systems, it is known that {\it quenched} disorder \cite{Harris, Geoff, Aharonyrandom,Dfisher} - that is, disorder that is time-{\it independent } - is even more destructive of order than ``annealed" thermal noise. Indeed, even arbitrarily small quenched disorder destroys long-rang ferromagnetic and crystalline order in all spatial dimensions $d\le4$ \cite{Harris, Geoff, Aharonyrandom,Dfisher}.

It is natural, therefore, to wonder what the effect of quenched disorder on active systems is. This question has received  much attention in recent years \cite{toner_prl18, toner_pre18,Chate-quench, Bartolo2017, Bartolo2021, Tailleur2021, Volpe2016,Mishra2020,Mishra2018,Peruani, Ano_disord}. In particular, it has been shown \cite{toner_prl18, toner_pre18} that for three-dimensional dry polar active systems with quenched disorder, long-rang polar order (i.e., a non-zero average velocity $\langle\bv\rangle$) can survive such quenched disorder. But in two dimensions, only quasi-long-rang polar order (i.e.,  $\langle\bv\rangle$ decaying to zero as a power of the system size was found in \cite{toner_prl18, toner_pre18, Peruani}.

 In this paper, we report that it is possible to achieve true {\it long-ranged} order in two dimensions in dry polar active systems with quenched
disorder, if those systems are {\it incompressible}.

In the following, we will first present a hydrodynamic theory of incompressible polar active fluids \cite{chen_natcomm16, chen_njp18, chen_njp15} with {\it both}  annealed disorder  (i.e., time-dependent noise arising from endogenous fluctuations due to, e.g., the errors made by a motile agent  attempting to follow its neighbors \cite{vicsek_prl95}), and quenched disorder (caused by, e.g.,  static random impurities
on the frictional substrate). We then study the system in the linear regime, and subsequently use \emph{three} different dynamic renormalization group (DRG) schemes
to uncover a novel universality class, whose associated scaling exponents fully characterize the scaling behavior of the system in the moving phase. Specifically, in this moving phase, these exponents characterize the fluctuations  $\bu(\bbr,t)$ of the local active fluid velocity about its mean value $\la \bv \ra= v_0 \hat{\bf x}$, where we've defined our coordinate system so that $\hat{\bf x}$ is along the mean velocity spontaneously chosen by the system. That is,   $\bu(\bbr,t) \equiv \bv( \bbr,t)-v_0 \hat{\bf x}$.
 In particular, the overall   real-space velocity auto-correlation $\langle \bu(\br,t)\cdot\bu(0,\mathbf{0})\rangle$ is given by
\bew
\beqn
\langle \bu(\br,t)\cdot\bu(\mathbf{0},0)\rangle&=&|y|^{2\chi}\cG_{_{Q}}\left({|x|\over |y|^\zeta}\right)
+|y|^{2\chi'}\cG_{_{A}}\left({|x|\over |y|^{\zeta}},{|t|\over |y|^{z}}\right)\,,
\label{Correl1}
\eeqn
\ew
where  $\cG_{_{A,Q}}$ are  scaling functions  that are each universal up to an overall multiplicative factor (a {\it different} overall multiplicative factor for each), corresponding to the annealed and quenched parts of the correlations, respectively.

One of the unusual features of our result is that the annealed part $\cG_{_{A}}\left({|x|\over |y|^{\zeta}},{|t|\over |y|^{z}}\right)$ of the correlations
has such a simple scaling form. This is quite different from, e.g., a simple compressible fluid, in which the density fluctuations are associated with  dispersionless propagating sound waves, which corresponds to a dynamic exponent $z=1$ (i.e., distance proportional to time), while the {\it decay} of those modes is diffusive, which corresponds to a dynamic exponent $z=2$. As we'll see, a {\it linear} theory of our system also predicts such a ``double scaling" character-i.e., propagating and diffusive parts with different dynamic exponents-but the full, non-linear theory has the simpler, ``single scaling" form given in (\ref{Correl1}). Thus, this simplicity is an unusual feature of the anomalous hydrodynamics (what is sometimes called ``the breakdown of linearized hydrodynamics") that occurs in our system.

 Another consequence of the anomalous hydrodynamics is that the anisotropy exponent $\zeta$ takes on the same value in both $\cG_{_{A}}\left({|x|\over |y|^{\zeta}},{|t|\over |y|^{z}}\right)$ and $\cG_{_{Q}}\left({|x|\over |y|^\zeta}\right)$  in the full, nonlinear theory. In contrast, the linear theory predicts different values of $\zeta$'s for $\cG_{_{A}}\left({|x|\over |y|^{\zeta_{1}}},{|t|\over |y|^{z}}\right)$ and $\cG_{_{Q}}\left({|x|\over |y|^{\zeta_{2}}}\right)$.

We have obtained the exponents in (\ref{Correl1})  using three different  DRG schemes:  an uncontrolled calculation in exactly two dimensions, and two different  $\epsilon = (d_c -d)$-expansions, {obtained from two different analytic continuations of our 2{D} model to higher spatial dimensions $d$. We call these continuations the ``hard continuation'' and ``soft continuation'', and they lead respectively to $d_c^{\rm hard}=7/3$, and $d_c^{\rm soft}=5/2$.}

In the uncontrolled calculation in exactly two dimensions, we find
\begin{subequations}
\begin{align}
z&={13\over27}\approx 0.48 \,,\\
\zeta&={7\over9}\approx 0.78 \,,\\
\chi&= -{2\over9}\approx-0.22 \,,\\
\chi'&=-{19\over54}\approx-0.35 \,.
\end{align}
\label{numexpunc}
\end{subequations}

The fact that both $\chi$ and $\chi'$  are negative implies that the fluctuations remain finite as the system size goes to infinity. This means that the system can have long-range polar order for sufficiently weak disorder. The same statement is true of the other two schemes.

In the ``hard" continuation, $d_c=7/3$, and a first order in $\epsilon$ expansion gives
\begin{subequations}
\label{eq:exponents}
\begin{align}
z&={2\over 3}-{5\over 9}\epsilon+O(\epsilon^2)\,,\\
\zeta&={2\over 3}+{1\over 3}\epsilon+O(\epsilon^2)\,,\\
\chi &=-{1\over 3}+{1\over 3}\epsilon+O(\epsilon^2)\,,\\
\chi'&=-{1\over 2}+ {4\over 9}\epsilon+O(\epsilon^2)\,,
\end{align}
\end{subequations}
with $\epsilon = 7/3-d$. For $d=2$, $\epsilon =1/3$,  and so the numerical values  to $O(\epsilon)$ are
\begin{subequations}
\begin{align}
z&\approx{13\over27}\approx 0.48 \,,\\
\zeta&\approx{7\over9}\approx 0.78 \,,\\
\chi&\approx -{2\over9}\approx-0.22 \,,\\
\chi'&\approx -{19\over54}\approx-0.35 \,,
\end{align}
\label{numexp7/3}
\end{subequations}
which
are exactly the same as the $(d=2)$-uncontrolled-calculation results (\ref{numexpunc}).
In the ``soft" continuation, $d_c=5/2$, and, defining $\tilde{\epsilon}\equiv5/2-d$, we find
\begin{subequations}
\begin{align}
z&={2\over 3}-{8\over 27}\tilde{\epsilon}+O(\tilde{\epsilon}^2)\,,\\
\zeta&={2\over 3}+{4\over 27}\tilde{\epsilon}+O(\tilde{\epsilon}^2)\,,\\
\chi&=-{1\over 3}+{4\over 27}\tilde{\epsilon}+O(\tilde{\epsilon}^2)\,,\\
\chi'&=-{1\over 2}+{1\over 9}\tilde{\epsilon}+O(\tilde{\epsilon}^2)\,.
\end{align}
\label{soft exp intro}
\end{subequations}
For $d=2$, $\tilde{\epsilon}=1/2$, and the above results give
\begin{subequations}
\begin{align}
z&\approx{14\over27}\approx 0.52 \,,\\
\zeta&\approx{20\over27}\approx 0.74 \,,\\
\chi&\approx -{7\over27}\approx-0.26 \,,\\
\chi'&\approx -{4\over 9}\approx{-0.44} \,,
\end{align}
\end{subequations}
which are very close to the values obtained from both the uncontrolled calculation and the hard continuation.


Our  best estimate of the actual values of these scaling exponents is a suitably weighted average of these three results. In section \ref{best_exponents}, we argue that the best weighting is to assign each of the (equal) ($7/3-\epsilon$)-expansion results and $(d=2)$-uncontrolled-calculation results with $9/4$  times the weight of the ($5/2-\tilde{\epsilon}$)-expansion results. Thus we obtain
\begin{subequations}
\begin{align}
z&\approx{9\left({13\over27}\right)+2\left({14\over27}\right)\over11}={145\over297}\approx 0.49 \,,\\
\zeta&\approx{9\left({7\over9}\right)+2\left({20\over27}\right)\over11}={229\over297}\approx 0.77 \,,\\
\chi&\approx -\left({9\left({2\over9}\right)+2\left({7\over27}\right)\over11}\right)=-{68\over297}\approx -0.23 \,,\\
\chi'&\approx -\left({9\left({19\over54}\right)+2\left({4\over 9}\right)\over11}\right)=-{ 73\over198}\approx{- 0.37}\,.
\end{align}
\label{}
\end{subequations}
We can estimate the likely errors in these numerical values as the difference between these averaged results and the equal hard-continuation and uncontrolled-calculation results. This gives
\begin{subequations}
\label{eq:exponents_num}
\begin{align}
z&= 0.49\pm {0.01},\ \
\zeta=0.77\pm {0.01},\\
\chi&=-0.23\pm {0.01},\ \ \chi'={-0.37\pm 0.02}\,.
\end{align}
\end{subequations}

Readers unconvinced by our arguments for this particular weighting scheme  can be reassured by the fact that {\it any} weighting scheme will give values quite close to these, since all three sets of results being averaged give very  similar numerical values for the exponents. Indeed,  as noted earlier, the uncontrolled calculation and the hard continuation give {\it exactly} the same exponents.

One way to experimentally test these exponents is by measuring the equal-time velocity correlation function. This is dominated by the quenched part, and goes like
\beq
\langle \bu(\br,t)\cdot\bu(\mathbf{0},t)\rangle \sim
\left\{
\begin{array}{ll}
|y|^{2\chi} \ , & {\rm if }\ |y|^\zeta \gg |x|
\\
|x|^{2\chi/\zeta} \ , & {\rm if }\ |y|^\zeta \ll |x|
\end{array}
\right.
\label{corr eq t}
\ ,
\eeq
and for equal-positions,  the change in the correlation function with time is dominated by the annealed part:
\beq
\langle \bu(\br,t+T)\cdot\bu(\br,T)\rangle-\langle \bu(\br,T)\cdot\bu(\br,T)\rangle \propto |t|^{2\chi'/z}\,.
\label{corr eq r}
\eeq
Our best estimate of  the numerical value of $2\chi'/z$, using \eqref{eq:exponents_num},   is
\beq
\frac{2\chi'}{
z}= -1.51\pm.11.
\eeq

\section{Hydrodynamic description}\label{hydro}
We start with a hydrodynamic model of a generic 2D incompressible polar active fluid, moving on a disordered substrate, in the presence of both quenched  (time-independent) and annealed  (time-dependent) noise. As for incompressible passive fluids \cite{forster_prl76,forster_pra77} described by the Navier-Stokes equation, the only hydrodynamic variable in our problem is the velocity field $\bv$. However, in contrast to the Navier-Stokes equation, $\bv$ is hydrodynamic \emph{not} because it is conserved -- it is not, because the substrate is a momentum sink -- but because it is a broken symmetry variable (more precisely, certain components of it are). Furthermore, because our non-equilibrium system breaks detailed balance, and therefore,\ is not constrained by Onsager symmetry, this equation contains additional terms which would have been absent both from the Navier-Stokes equation and from the equation of motion (EOM) of  passive fluid films on substrates. That is, the  EOM of $\bv$ is only constrained by the spatial symmetries of our system, in particular, rotation and translation invariance. This reasoning implies that the EOM for $\bv$ takes the form \cite{toner_prl95,toner_pre98,wensink_pnas12,chen_natcomm16,chen_njp18}
\bew
\beq
\label{eq:maineom}
\pp_t \bv+ \lambda (\bv \cdot \vnab )\bv = -\vnab  \Pi  -(\bv \cdot \vnab \Pi_1) \bv +U(|\bv|) \bv
+\mu_1 \nabla^2 \bv +\mu_2 (\bv \cdot \vnab)^2 \bv
 +\bff_{_Q}(\br)+\bff_{_A}(\br,t)
\ ,
\eeq
\ew
where the ``pressure"   $\Pi$ acts as a Lagrange multiplier to enforce the incompressibility constraint: $\vnab \cdot \bv = 0$.
The $U(|\bv|)$ term in equation \eqref{eq:maineom} makes the local
$\bv$ have a nonzero magnitude $v_0$
in the ordered phase, by the simple expedient of having $U(|\bv|)>0$ for $|\bv|<v_0$,
$U(|\bv|)=0$ for $|\bv|=v_0$, and $U(|\bv|)<0$ for $|\bv|>v_0$. Aside from this, and the assumption that $U(|\bv|)$ is a smooth, analytic function of $|\bv|$, we will make no assumptions about $U(|\bv|)$.
 Similarly, the ``anisotropic pressure" $\Pi_1$ is also a generic analytic function of $|\bv|$. {Finally}, $\bff_{_Q}(\br)$ and $\bff_{_A}(\br,t)$ are respectively the quenched and annealed noise terms, which have  zero means, and correlations:
\begin{subequations}
\begin{align}
\langle f_{_Q}^i(\br)f_{_Q}^j(\br')\rangle&=
2D_{_Q}\delta_{ij}\delta^2(\br-\br')\,,\label{Random_Q}\\
\langle f_{_A}^i(\br,t)f_{_A}^j(\br',t')\rangle&=
2D_{_A}\delta_{ij}\delta^2(\br-\br')\delta(t-t')\, ,\label{Random_A}
\end{align}
\end{subequations}
where the indices $i,j$ enumerate the spatial coordinates.
Note the time-\emph{independence} of the noise $\bff_{_Q}(\br)$; this is what we mean by ``quenched".

In the EOM (\ref{eq:maineom}), we have only included terms that are relevant to the universal behavior based on the DRG analysis that follows, by which we mean terms that can change the long-distance, long-time behavior of the system.   This equation differs from the EOM introduced in \cite{chen_natcomm16} only through the presence of the quenched noise $\bff_{_Q}(\br)$. As we will see, however, this quenched noise radically changes the behavior of the system.

In the moving phase, we focus on the velocity deviation field $\bu$, from the mean flow $v_0 \hat{\bf x}$:   $\bu = \bv- v_0\hat{\bf x}$. We will expand \eqref{eq:maineom} in powers of the fluctuation $\bu$, keeping only ``relevant" terms.  Doing so, we find that the EOM governing  $\bu$  is, in Einstein component notation,
\bew
\beq
\label{eq:u_eom}
\pp_t u_i=-\pp_i\Pi-\gamma\pp_xu_i-b\delta_{ix}\pp_xu_x-\alpha\left(u_x+{u_y^2\over 2v_0}\right)(\delta_{ix}+{u_y\over v_0}\delta_{iy})
+f_{_Q}^i+f_{_A}^i\,,
\eeq
\ew
where $\gamma\equiv\lambda v_0$,
$\alpha\equiv -v_0\left(\dd U\over\dd |\bv|\right)_{|\bv|=v_0}$,
$b\equiv v_0^2\left(\dd \Pi_1\over\dd |\bv|\right)_{|\bv|=v_0}$,
$\mu_{_\perp}\equiv \mu_1$, and $\mu_x \equiv\mu_1+\mu_2v_0^2$.  We have dropped some terms irrelevant to the long-wavelength behavior of the system from \eqref{eq:u_eom}.

We first focus on the linear regime of the above EOM, which we expect to capture the hydrodynamic behavior  over a large range of length scales if the noise is sufficiently small.

\section{Linear regime}\label{lin}

To study the system's behavior in the linear regime, we first spatio-temporally  Fourier transform the linearized version of the EOM \eqref{eq:u_eom}. The convention we use here is the following:
\beq
\bu(\bq,\omega)
\equiv\int_{t, \bbr}
\ee^{\ii(\bq \cdot \bbr-\omega t)}\bu(\bbr,t) \ ,
\eeq
where $\int_{t} \equiv \int\frac{\dd t}{\sqrt{2\pi }}$ and $\int_\bbr \equiv \int\frac{ \dd^2 r }{2\pi }$.
We will continue to use this shorthand notation for integration throughout the paper.   We will also use the same convention in Fourier space; i.e., $\int_{\omega} \equiv \int\frac{\dd \omega}{\sqrt{2\pi }}$ and $\int_\bq\equiv \int\frac{ \dd^2 q }{2\pi}$.

The linearized version of EOM (\ref{eq:u_eom}) in Fourier space is
\bew
\beqn
\Big[-\ii(\omega-\gamma q_x) +\Gamma(\bq) \Big] u_i(\tilde{\bq})+(\alpha+\ii bq_x)\delta_{ix}u_x(\tilde{\bq})&=-\ii q_i\Pi+f_{_Q}^i+f_{_A}^i \,,
\label{Linear2Dy1}
\eeqn
\ew
where we have introduced the composite vector $\tilde{\bq} \equiv  (\bq,\omega)$, and the $\bq$-dependent damping coefficient $\Gamma(\bq)\equiv \mu_{_\perp} q^2_y+\mu_xq_x^2$.

Acting on both sides with the
transverse projection operator
\beq
P_{li}(\bq)=\delta_{li}-\frac{q_lq_i}{q^2} \,,
\label{projdef}
\eeq
 which projects
orthogonal to the spatial wavevector $\bq$,
eliminates   the pressure ($\Pi$) term. Using $P_{li}u_i=u_l$, which follows from  the incompressibility condition $q_x u_x+q_y u_y=0$, setting $l=y$, and using the replacement $u_x=-{q_y\over q_x}u_y$ that follows from the same incompressibility condition, gives us a simple algebraic equation for $u_y(\bq,\omega)$:
\beq
\left[G(\tilde{\bq})\right]^{-1}u_y(\tilde{\bq})=P_{yi}(f_{_Q}^i+f_{_A}^i) \,,
\label{uyeq}
\eeq
where the ``propagator" $G(\tilde{\bq})$ is as follows
\beq
G(\tilde{\bq})\equiv \left[ -\ii\left(\omega-(\gamma+b{q_y^2\over q^2})q_x\right)
+\left(\alpha {q_y^2\over q^2}+\Gamma(\bq)\right)\right]^{-1} \,.
\label{Gdef}
\eeq

The poles in this propagator in the complex $\omega$ plane are the eigenfrequencies of our problem. In the limit $q_y\ll q_x$, which we will show in a moment is the regime of wavevector space that dominates the fluctuations, those eigenfrequencies are given by
\beq
 \omega({\bf q})=\gamma q_x -\ii \bigg(\mu_x q_x^2+\alpha {q_y^2\over q_x^2}\bigg) .
\label{omegalin}
\eeq
These eigenfrequencies have a rather complicated multiple scaling form. That is, they cannot be written in a simple scaling form, but require  a sum of two scaling forms:
\beq
\omega(\bq)= |q_y|^{z_1} f_{\rm real}\left({q_x\over |q_y|^{\zeta_1}}\right) -\ii|q_y|^{z_2}  f_{\rm imaginary}\left({|q_x|\over |q_y|^{\zeta_2}}\right)
\label{omegascalelin}
\eeq
with
\beq
f_{\rm real}(m)=\gamma m \,,
\label{freallinlin}
\eeq
whose form unfortunately makes it impossible to fix $\zeta_1$ and $z_1$ for the linear problem, but does require
\begin{subequations}
\begin{align}
z_1=\zeta_1 \,,
\label{z1}\\
z_2=1,
\label{z2lin}\\
f_{\rm imaginary}(m)=m^2(\mu_x+\alpha m^{-4})\,,
\label{freallinlin}\\
\zeta_2=1/2\ .
\end{align}
\end{subequations}
Note that, although the individual values of $z_1$ and $\zeta_1$ cannot be determined, we cannot possibly have $z_1=z_2=1$ and $\zeta_1=\zeta_2=1/2$, since this violates \eqref{z1}. Hence, we are forced to use the double scaling form \eqref{omegascalelin}.

We will show in sections \eqref{scaling} and \eqref{equal} that this complexity disappears in two dimensions once the effects of nonlinearities are taken into account. These nonlinear effects replace the multiple scaling form \eqref{omegascalelin} with the simple scaling form
\beq
\omega({\bf q})= |q_y|^{z} f_\omega\left({q_x\over |q_y|^{\zeta}}\right) \,,
\label{omegascaletrue}
\eeq
with the unique universal exponents $z$ and $\zeta$ given by equations \eqref{eq:exponents}, and a single, albeit complex, scaling function $f_\omega$.

This simplification of the eigenfrequencies carries through to all of the correlation functions as well. Indeed, {\it every} long-wavelength, long-time property of the system exhibits simple scaling with the single dynamic exponent $z$, and the single anisotropy exponent $\zeta$, given by \eqref{eq:exponents}.

  Solving equation \eqref{uyeq} for $u_y$, autocorrelating the result with itself, and using our expressions (\ref{Random_Q}) and (\ref{Random_A}) for the autocorrelations of the noises, gives the autocorrelation of $u_y(\bq,\omega)$. Using $u_x=-{q_y\over q_x}u_y$, which follows from the incompressibility condition,  then gives the autocorrelations of $u_x(\bq,\omega)$, and the cross-correlations of $u_x(\bq,\omega)$ and $u_y(\bq,\omega)$. We find   
\beqn
\nonumber
\langle u_i(\tilde{\bq})u_j(\tilde{\bq}')\rangle&=&
 C^{ij}_{_{A}}(\tilde{\bq})\delta(\omega+\omega')\delta(\bq+\bq')
 \\
 &&
+C^{ij}_{_{Q}}(\tilde{\bq})\delta(\omega)\delta(\omega')\delta(\bq+\bq')\,,
\label{eq:corre_uu}
\eeqn
where
\begin{subequations}
\label{eq:Ca&Cq}
\begin{align}
\label{2D_linear_Axx}
C_{_{A}}^{xx}(\tilde{\bq})&=
{q_{y}^2\over q^2}C_{_A}(\tbq)
\,,\\
\label{2D_linear_Axy}
C_{_{A}}^{xy}(\tilde{\bq})&=-{q_{x}q_y\over q^2}C_{_A}(\tbq)=C_{_{A}}^{yx}(\tilde{\bq})
\,,\\
\label{2D_linear_Ayy}
C_{_{A}}^{yy}(\tilde{\bq})&={q_x^2\over q^2}C_{_A}(\tbq)
\,,\\
\label{2D_linear_Qxx}
C_{_{Q}}^{xx}(\tilde{\bq})&={q_{y}^2\over q^2}C_{_Q}(\tbq)
\,, \\
\label{2D_linear_Qxy}
C_{_{Q}}^{xy}(\tilde{\bq})&=-{q_{x}q_y\over q^2}C_{_Q}(\tbq)=C_{_{Q}}^{yx}(\tilde{\bq})
\, ,\\
\label{2D_linear_Qyy}
C_{_{Q}}^{yy}(\tilde{\bq})&={q_{x}^2\over q^2}C_{_Q}(\tbq)
\,,
\end{align}
\end{subequations}
with
\begin{subequations}
\begin{align}
C_{_{A}}(\tbq)&={2D_{_A} \over \left[\omega-\left({b q_{y}^2\over q^2}+\gamma\right)q_x\right]^2+\left[{\alpha q_{y}^2\over q^2}+\Gamma(\bq)\right]^2}\,,
\\
C_{_{Q}}(\tbq)&={4\pi D_{_Q}\over \left({b q_{y}^2\over q^2}+\gamma\right)^2q_x^2+\left[{\alpha q_{y}^2\over q^2}+\Gamma(\bq)\right]^2}\,,
\end{align}
\end{subequations}
and the subscripts $A$ and $Q$ denoting the contributions from the annealed and quenched noises, respectively.

The equal-time correlations of $\bu$   can now be obtained by inverse Fourier transformation. We find in the hydrodynamic limit (i.e., $\bq \rightarrow {\bf 0}$), the annealed and quenched parts of these correlations are given  by
\begin{subequations}
\begin{align}
\label{eq:CAlin_realtime}
\langle u_x(\bq,t)u_x(\bq', t)\rangle_{_A}&\approx{ D_{_A}q_y^2\over \alpha q_y^2+\mu_xq^4_x}\delta(\bq+\bq')\ ,
\\
\langle u_y(\bq,t)u_y(\bq', t)\rangle_{_A}&\approx{ D_{_A}q_x^2\over \alpha q_y^2+\mu_xq^4_x}\delta(\bq+\bq')\ ,
\\
\langle u_x(\bq,t)u_x(\bq', t)\rangle_{_Q}&\approx { 2D_{_Q}q_y^2 q^2\over \alpha^2
q_y^4+\gamma^2q^6_x}\delta(\bq+\bq')\ ,
\\
\langle u_y(\bq,t)u_y(\bq', t)\rangle_{_Q}&\approx { 2D_{_Q}q_x^2 q^2\over \alpha^2
q_y^4+\gamma^2q^6_x}\delta(\bq+\bq')\,.
\label{eq:CQlin_realtime}
\end{align}
\end{subequations}
We see that as $q\to 0$,  $\langle u_x(\bq,t)u_x(\bq', t)\rangle_{_A}$ is always  finite.  In contrast, $\langle u_y(\bq,t)u_y(\bq', t)\rangle_{_A}$ diverges in the regime $q_x\gtrsim q_y^{1/2}$  as $1/q^2$.  The  cross-correlations  $\langle u_x(\bq,t)u_x(\bq', t)\rangle_{_Q}$ and $\langle u_y(\bq,t)u_y(\bq', t)\rangle_{_Q}$ diverge
in the regime $q_x\gtrsim q_y^{2/3}$,  where they scale as $1/q$ and $1/q^2$,
respectively. Note that $q_x\gtrsim q_y^{2/3}$ covers a much larger area in $\bq$-space
than $q_x\gtrsim q_y^{1/2}$ does. In spite of these divergences, it is easy to show that the
integrals of  $\langle u_x(\bq,t)u_x(\bq', t)\rangle_{_A,_Q}$ and $\langle u_y(\bq,t)u_y(\bq', t)\rangle_{_A,_Q}$ over $\bq$ and $\bq'$ both
converge in the infrared,  which implies that the corresponding real-space fluctuations
remain finite in the infinite system size limit. This demonstrates that long-range polar
order persists in these systems, at least according to this linear theory.

These observations imply that:
 \noindent i) the $u_y$ fluctuations dominate in the hydrodynamic limit, which is not a surprise since $u_y$ is the ``Goldstone mode", ii) the quenched fluctuations dominate,   again in the hydrodynamic limit,  iii) the long-range  polar order is robust against the quenched disorder,  and iv) the quenched and annealed anisotropy exponents $\zeta_{\rm quenched}=2/3$ and $\zeta_{\rm annealed}=1/2$, respectively.  All of these conclusions except the  last continue to hold even when the nonlinearities are taken into account, as we will show in the next section.

Note that it is the term $\gamma^2q_x^6$ appearing in the denominator in (\ref{eq:CQlin_realtime}) that stabilizes long-rang order in the presence of quenched disorder.  If that term were absent -- as it is in \emph{equilibrium} ``{divergence-free} magnets" (that is, magnets subject to the constraint $\nabla\cdot\bM=0$, where $\bM$ is the magnetization), whose hydrodynamic properties in the presence of annealed noise are equivalent to incompressible flocks \cite{chen_natcomm16} -- its place would be taken by $\mu_x^2q_x^8$. Such a strong divergence of angular fluctuations for $q_y=0$ would  destroy  long-range, and \emph{even} \emph{quasi}-long range polar order, which is exactly what happens in equilibrium {divergence-free} magnets with quenched disorder.

The physical mechanism of this stabilization of order is suggested by the origin of  the $\gamma^2q_x^6$ term in the correlation functions:  the propagation term $\gamma(\partial_x\bu)$ in the EOM \eqref{eq:u_eom}. This term causes fluctuations along $\hat{\bx}$ to propagate with speed $\gamma$. Thus, in a frame of reference comoving with the fluctuations, the quenched disorder looks time-dependent, and, so, more like annealed  disorder. Indeed, for $q_y=0$  both  $\langle u_y(\bq,t)u_y(\bq', t)\rangle_{_Q}$ and $\langle u_y(\bq,t)u_y(\bq', t)\rangle_{_A}$
 scale in the same way as $\bq\to\mathbf{0}$: both are  $\propto 1/q_x^2$. This ``annealization" effect reduces the quenched fluctuations. Nevertheless,
the quenched disorder is not completely ``annealized" as overall the quenched fluctuations are still larger than the annealed.

Now we turn to the general real-space $\bu$-$\bu$ correlation function $\langle \bu(\br,t)\cdot\bu(0,\mathbf 0)\rangle$, which is the inverse Fourier transform of $\langle \bu(\tilde{\bf q})\cdot\bu(\tilde{\bf q}')\rangle$:
\beqn
\langle \bu(\br,t)\cdot\bu(\mathbf 0, 0)\rangle
&=&\int_{\omega,\omega',\bq,\bq'}
\langle \bu(\tilde{\bf q})\cdot\bu(\tilde{\bf q}')\rangle
\,\ee^{\ii(\bq\cdot\br-\omega t)}\nonumber\\
&=&C_{_{A}}(\br,t)+C_{_{Q}}(\br)\,,
\label{}
\eeqn
where
\begin{subequations}
\label{eq:Ca&Cq_real}
\begin{align}
C_{_{A}}(\br,t)
&=\int_{\omega, \bq}\,
{2D_{_A}\ee^{\ii(\bq\cdot\br-\omega t)}\over \left(\omega-\gamma q_x\right)^2+\left({\alpha q_y^2\over q_x^2}+\mu_x q_x^2\right)^2}\,
\,,\\
C_{_{Q}}(\br)
&=\int_\bq\,
{2 D_{_Q}q_x^4\over {\alpha q_y^4+\gamma^2q_x^6}}\,\ee^{\ii\bq\cdot\br}\,.
\label{}
\end{align}
\end{subequations}
In the above expressions, we have simplified the denominators inside the integrals by keeping only the dominant terms (see \eqref{eq:Ca&Cq})  in the limit $q_y\ll q_x$, which we have shown is the regime of wavevector space that dominates the fluctuations.

The scaling behavior of $C_{_{A}}(\br,t)$ and $C_{_{Q}}(\br)$ can now be worked out by changing the variables of integration: We first rewrite $C_{_{A}}(\br,t)$ as
\bew
\beqn
C_{_{A}}(\br,t)
&=&\int_{\delta\omega, \bq}
\left[{2D_{_A}\over \left(\delta\omega\right)^2+\left({\alpha q_y^2\over q_x^2}+\mu_x q_x^2\right)^2}\right]
\ee^{-\ii\left[\delta\omega t-q_x\left(x-\gamma t\right)-q_yy\right]}\,,
\label{2Dlinear_real_CA1}
\eeqn
\ew
where $\delta\omega\equiv\omega-\gamma q_x$, and then introduce the new variables $\Omega$ and $\mathbf Q$ as follows:
\beqn
q_x={Q_x\over |y|^{1\over 2}}\,,~~~q_y={Q_y\over|y|}\,,
~~~\delta\omega={\Omega\over|y|}\,.
\eeqn
Now, rewriting the integral in (\ref{2Dlinear_real_CA1}) in terms of $\Omega$ and $\mathbf Q$, we get
\beqn
C_{_{A}}(\br,t)
&=&|y|^{-{1\over 2}}{\cal C}_{_{A}}\left({|x-\gamma t|\over |y|^{1\over 2}},{|t|\over |y|},\right)
\eeqn
where
\bew
\beqn
{\cal C}_{_{A}}\left({|x-\gamma t|\over |y|^{1\over 2}},{|t|\over |y|}\right)
\equiv\int_{\Omega,{\bf Q}}
\left[{2D_{_A}\over \Omega^2+\left({\alpha Q_y^2\over Q_x^2}+\mu_x Q_x^2\right)^2}\right]
\exp \left[-\ii\left({\Omega t\over|y|}-Q_y-{Q_x (x-\gamma t)\over |y|^{1\over 2}}\right) \right]\,.
\label{}
\eeqn
\ew

Likewise  changing variables of integration on $C_{_{Q}}(\br)$ we get
\beqn
C_{_{Q}}(\br)
&=&|y|^{-{1\over 3}}{\cal C}_{_{Q}}\left(|x|\over |y|^{2\over 3}\right)
\eeqn
where
\beqn
{\cal C}_{_{Q}}\left(|x|\over |y|^{2\over 3}\right)
\equiv\int_{\bf Q}
\left({2 D_{_Q}Q_x^4\over {\alpha^2 Q_y^4+\gamma^2Q_x^6}}\right)
\ee^{\ii\left(Q_y+{|x| Q_x\over |y|^{2\over 3}} \right)}\,.\nonumber\\
\label{}
\eeqn

Finally, the overall correlation of $\langle \bu(\br,t)\cdot\bu(\mathbf 0,0)\rangle$ is given by
\beq
|y|^{-{1\over 3}}{\cal C}_{_{Q}}\left(|x|\over |y|^{2\over 3}\right)
+|y|^{-{1\over 2}}{\cal C}_{_{A}}\left({|x-\gamma t|\over |y|^{1\over 2}},{|t|\over |y|}\right)\,. \label{Real_linear_corr}
\eeq
Therefore, the linear theory recovers  a form similar to \eqref{Correl1}, with
 the  quenched and the annealed anisotropy exponent  given by  $\zeta_{\rm quenched}=2/3$ and $\zeta_{\rm annealed}=1/2$,  respectively, the quenched and annealed roughness exponents given by $\chi=-1/6$  and $\chi'=-1/4$, respectively, and the dynamic exponent  $z=1$. We will now show that these  exponents  are modified by the nonlinearities in the EOM,  and in particular, the two anisotropy exponents become equal.

\section{Nonlinear regime \& DRG analysis}\label{NL}
 We turn now to the full EOM of $\bu$ (\ref{eq:u_eom}). Fourier transforming this, and acting on both sides with the transverse projection operator $P_{li}(\bq)$ \eqref{projdef}, we obtain
\begin{widetext}
\beq
-\ii\omega u_y=P_{yx}\left(\bq\right)\mathcal{F}_{\tbq}
\left[-\alpha\left(u_x+{u_y^2\over 2}\right)\right]+\mathcal{F}_{\tbq}
\left[-\gamma\pp_xu_y+\mu_x\pp_x^2u_y-\lambda u_y\pp_yu_y-\alpha
\left(u_x+{u_y^2\over 2}\right)u_y+f_{_A}^y+f_{_Q}^y\right]\,,
\label{2Dy4}
\eeq
\end{widetext}
where $\mathcal{F}_{\tbq}$ represents the $\tbq$th Fourier component.
In writing this equation, we have rescaled the fields ($\bu \rightarrow \bu v_0$) and the
noise terms (${\mathbf f}_{_{A,Q}} \rightarrow {\mathbf f}_{_{A,Q}} v_0$) to eliminate the $v_0$'s. We have also neglected many terms which are irrelevant due to the fact that the dominant regime of wavevector is $q_y\ll q_x$, as we discovered in our treatment of the linear theory. We will assume, and verify {\it a posteriori}, that this continues to hold true for the non-linear theory.

First, this assumption implies that  the noise terms $P_{yx}f_{A,Q}^x$ are therefore irrelevant in comparison to $P_{yy}f_{A,Q}^y$. Second, due to the incompressibility constraint, the magnitude of $P_{yx}\pp_xu_x$ is as relevant  as $P_{yx}\pp_yu_y$, and the latter, again due to the anisotropic scaling, is less relevant than $P_{yy}\pp_xu_y$. Therefore, $P_{yx}\pp_xu_x$ is also irrelevant in comparison to $P_{yy}\pp_xu_y$. Similarly, $P_{yy}\pp_y^2u_y$ is irrelevant in comparison to $P_{yy}\pp^2_xu_y$.  Naively $P_{yy}\pp^2_xu_y$ is also irrelevant in comparison to $P_{yy}\pp_xu_y$. However, because $P_{yy}\pp_xu_y$ only leads to propagation, not damping, we need to keep the $P_{yy}\pp^2_xu_y$ term. This is very similar to keeping the viscous term in the dynamics of a simple fluid, even though it is formally less relevant than the pressure term, since the pressure only leads to sound {\it propagation}, while it is the viscosity which controls sound {\it damping}.

Finally, in the limit of interest $q_y\ll q_x$, we can approximate $P_{yy}\left(\bq\right)\approx1$.

To evaluate the importance of the nonlinear terms in \eqref{2Dy4}, we first power count (which can be thought of as a zeroth order renormalization group analysis). We rescale time, lengths, and  fields as
\begin{subequations}
\label{Rescaling_2d}
\begin{align}
&t\to te^{z\ell},~~x\to xe^{\zeta\ell},~~y\to ye^{\ell},
\\
&u_y\to u_ye^{\ell\chi},
~~u_x\to u_xe^{\left(\chi+\zeta-1\right)\ell}\,,
\end{align}
\end{subequations}
and keep the form of the resultant EOM unchanged by absorbing the rescaling factors into the coefficients.
Specifically, the coefficients of the linear terms $P_{yx}u_x$, $P_{yy}\pp_xu_y$, and the noise strength are rescaled respectively as
\begin{subequations}
\begin{align}
\alpha&\to\alpha e^{\left(z+2\zeta-2\right)\ell},~~
\gamma\to\gamma e^{\left(z-\zeta\right)\ell},~~
\\
D_{_A}&\to D_{_A}e^{\left(z-2\chi-\zeta-1\right)\ell}\,,~~
D_{_Q}\to D_{_Q}e^{\left(2z-2\chi-\zeta-1\right)\ell}\,,
\end{align}
\end{subequations}
and the coefficients of the nonlinear terms $P_{yx}u_y^2$, $u_y\pp_yu_y$, $u_xu_y$, and $u_y^3$ as
\begin{subequations}
\label{pcnonlinear}
\begin{align}
&{\alpha\over 2}\to{\alpha\over 2}\, \ee^{\left(z+\chi+\zeta-1\right)\ell},~~
\lambda\to\lambda\, \ee^{(z+\chi-1)\ell},~~
\\
&
\alpha\to \alpha\, \ee^{\left(z+\chi+\zeta-1\right)\ell}\,,~~
{\alpha\over 2}\to {\alpha\over 2}\,\ee^{\left(z+2\chi\right)\ell}\,.
\end{align}
\end{subequations}
We choose $z$, $\zeta$, and $\chi$ to fix $\alpha$, $\gamma$ and $D_{_Q}$, which control the size of the dominant fluctuations (i.e., those coming from the quenched noise). This choice leads to the following values of the scaling exponents:
\beq
z={2\over 3}\,,~~~\zeta={2\over 3}\,,~~~\chi=-{1\over 6}\,.
\eeq
\vspace{.2in}

Note that, as expected, these values of $\zeta$ and $\chi$ are identical to  those for the quenched part of the velocity correlations obtained from our linear theory [e.g., see (\ref{Real_linear_corr})].
Technically, the $z$ we get here is the dynamic exponent for the \emph{quenched} correlations, which, however, are purely static.  As a result, its value is different from $z=1$, which is the dynamic exponent for the annealed part of the correlations in (\ref{Real_linear_corr}).

Substituting  these values into (\ref{pcnonlinear}), we find the coefficients of the nonlinear terms $P_{yx}u_y^2$, $u_xu_y$, and $u_y^3$ all diverge as $\ell\to\infty$; specifically
\begin{subequations}
\label{pcnonlinear2}
\begin{align}
{\alpha}\to {\alpha}\,\ee^{\ell/6}\,,~~
{\alpha\over 2}\to {\alpha\over 2}\,\ee^{\ell/3}\,,
\end{align}
\end{subequations}
which implies these nonlinear terms are \emph{relevant} in the hydrodynamic limit, while $\lambda$ vanishes, which implies $u_y\pp_yu_y$ is \emph{irrelevant} and hence can be neglected.

We will deal with the relevant nonlinearities using the DRG approach of  Ref.~\cite{forster_prl76,forster_pra77}. This begins by formally ``solving" the hydrodynamic EOM  (\ref{2Dy4}) for $u_y$ by Fourier transforming in space {\it and time} to get
\begin{widetext}
\beqn
u_y(\tilde{\bq})&=&G(\tilde{\bq})\left[f_{_Q}^y(\tilde{\bq})+f_{_A}^y(\tilde{\bq})-\left(\alpha\over 2\right)
P_{yx}(\bq)\int_{\tilde{\bk}}u_y(\tilde{\bk})u_y(\tilde{\bq}-\tilde{\bk})
-\alpha \int_{\tilde{\bk}}u_x(\tilde{\bq}-\tilde{\bk})u_y(\tilde{\bk})\right.
\nonumber\\
&&\left.-\left(\alpha\over 2\right) \int_{\tilde{\bk},\tilde{\bk'}}
u_y(\tilde{\bq}-\tilde{\bk}-\tilde{\bk}')
u_y(\tilde{\bk})u_y(\tilde{\bk}')\right]\,,\label{2Dy6}
\eeqn
\end{widetext}
where $G(\tilde{\bq})$ is given by (\ref{Gdef}), and we've defined
$\int_{\tilde{\bk}}\equiv
\int  \dd\Omega \dd^2k/ \left(\sqrt{2\pi}\right)^3$.  In the limit{s} of small $\bq$ and $q_y\ll q_x$, $G(\tilde{\bq})$ can be simplified to
\beq
G(\tilde{\bq})\equiv \left[ -\ii\left(\omega-\gamma q_x\right)
+\left(\alpha {q_y^2\over q^2}+\mu_xq_x^2\right)\right]^{-1}
\ .\label{Gdef1}
\eeq

To ``regularize" our theory, we must introduce a short-distance (i.e., large wavevector) cutoff. We do so by restricting the wavevectors in \eqref{2Dy6} to lie within a Brillouin zone whose shape is a strip, infinite in the $q_x$ direction, and of width $2\Lambda$ in the $q_y$-direction. That is, our allowed wavevectors $\bq$ lie in the range $-\infty<q_x<\infty$, $-\Lambda<q_y<\Lambda$, where $\Lambda$ is the ultraviolet cutoff,

Next we
 decompose $u_y(\tbq)$ into ``slow"  components $u_y^<(\tbq)$ and ``fast"  components $u_y^>(\tbq)$,  where $u_y^<(\tbq)$  is
supported in the wave vector space $-\infty<q_x<\infty$, $|q_y|<\Lambda e^{-\dd\ell}$, and $u_y^>$ in the ``momentum shell'' $-\infty<q_x<\infty$, $\Lambda e^{-\dd\ell}<| q_y|<\Lambda$, where $d\ell\ll1$ is an arbitrary rescaling factor.
We likewise decompose the noises $f^y_{_{A,Q}}$ into fast and slow components $f^{y>}_{_{A,Q}}(\tbq)$ and $f^{y<}_{_{A,Q}}(\tbq)$ respectively. Next we solve (\ref{2Dy6})  iteratively for $u_y^>(\tbq)$ in terms of $u_y^<(\tbq)$  and the noises  $f^{y>}_{_{A,Q}}(\tbq)$. We then substitute the solution into the EOM (\ref{2Dy4})
for $u_y^<(\tbq)$, and average over the short wave length  noises $f^{y<}_{_{A,Q}}(\tbq)$.
This  renormalizes the various coefficients in the EOM for $u_y^<(\tbq)$. Following this averaging step, we perform a rescaling step in which we
	rescale time, lengths, and fields
	\begin{subequations}
		\begin{align}
			&t\to te^{z\dd\ell},~~x\to xe^{\zeta\dd\ell},~~y\to ye^{\dd\ell},
			\\
			&~~u_y\to u_ye^{\chi\dd\ell},
			~~u_x\to u_xe^{\left(\chi+\zeta-1\right)\dd\ell}\,,
		\end{align}
	\end{subequations}
to bring the cutoff back to $\Lambda$. Upon repeating this process recursively with the definition $\ell=n\dd\ell$, where $n$ is the number of iterations of this renormalisation process, we obtain a set of recursion relations. The values of the parameters after these $n$ steps are denoted as $\alpha(\ell)$, etc., with $\ell$ treated as a continuous variable. This enables us to write the recursion relations as differential equations.

{\AM In obtaining the recursion relations we also make use of an important symmetry property: due to the rotation invariance of the hydrodynamic EOM, all the ``$\alpha$"s in that equation should remain equal upon renormalization and we choose values of $\chi$ and $\zeta$ that ensure this; i.e.,
\beq
\chi=\zeta-1\,.\label{Chi1}
\eeq
With this, the recursion relations can be written {\it exactly} as
}
\begin{subequations}
\label{eq:RecurRel}
		\begin{align}
{\dd\ln\alpha\over\dd\ell}&=z+2\zeta-2+\eta_{\alpha} \,,\label{fl_Alpha_unc}\\
{\dd\ln\gamma\over\dd\ell}
&=z-\zeta+\eta_{\gamma}\,,\label{fl_gamma_unc}\\
{\dd\ln{\mu_x}\over\dd\ell}&=z-2\zeta+\eta_\mu\,,\label{Fl_mu_x_exact_unc}\\
{\dd \ln D_{_Q}\over\dd\ell}&=2z-3\zeta+1+\eta_{_Q} \,,\label{fl_D_Q_unc}\\
{\dd \ln D_{_A}\over\dd\ell}&=z-3\zeta+1+\eta_{_A}\,.\label{fl_D_A_unc}
	\end{align}
	\end{subequations}
{\AM where $\eta_{\alpha, \gamma,\mu, _Q, _A}$ denote the corrections arising from  averaging the nonlinear terms in the EOM over the short wave length noises. We calculate $\eta_{\alpha, \gamma,\mu, _Q, _A}$ perturbatively using three different schemes. These are: an uncontrolled one-loop calculation in exactly $d=2$, an  $\epsilon=7/3-d$-expansion to $O(\epsilon)$, and an $\tilde{\epsilon}=5/2-d$-expansion to $O(\tilde{\epsilon})$. The details of these calculations are given in Appendix \ref{App:A}. We now describe DRG analysis for each of these three schemes in turn.

\subsection{Uncontrolled calculation in exactly $d=2$\label{Sec:unc}}
We first  present the DRG for the one-loop uncontrolled calculation  in exactly $d=2$. The reason we refer to this calculation as ``uncontrolled" is that, in contrast to the $\epsilon$-expansions  we will describe in the next two sections, which become asymptotically exact
as we approach the critical dimension (i.e., in the limit $\epsilon$ (or $\tilde{\epsilon}$) $\to 0$), this uncontrolled calculation has no such limit in which it becomes exact, since the dimension of our system (i.e., $2$) differs always from the critical dimension by an amount of order $1$.
	
The detailed calculation of the graphical corrections for this scheme  is presented in Appendix \ref{App:A},  and obtains,  to one-loop order:
\begin{subequations}
\label{eta_unc}
		\begin{align}
	\eta_\alpha&=-{1\over27}{\gun}\,,\label{eta_alpha_unc_1}\\
	\eta_\gamma&={8\over 27}{\gun}\,,\label{eta_gamma_unc_1}\\
	\eta_{_Q}&={10\over 27}{\gun}\,,\label{eta_DQ_unc_1}\\
	\eta_{_A}&={16\over27}{\gun}\,,\label{eta_DA_unc_1}
	\\	\eta_\mu&={\gun_\mu}+{2\over 27}{\gun}\,,\label{Anoma_Exp1-2Unc}
	\end{align}
	\end{subequations}
where the two dimensionless couplings are given by
\begin{subequations}
		\begin{align}
		\label{eq:def_gunc}
		{\gun}&={D_{_Q}\over \pi}|\gamma|^{-{7\over 3}}\alpha^{1\over 3}\Lambda^{-{1\over 3}}\,,
\\
	{\gun_\mu}&\propto D_{_Q}
|\gamma|^{-{1}}\mu_x^{-1}\Lambda^{-{1}}\,.
\end{align}
	\end{subequations}

In writing these expressions \eqref{eta_unc} for the graphical corrections, we have ignored all corrections to any parameters from the annealed noise strength $D_{_A}$ other than $D_{_A}$ itself; that is, we have set $D_{_A}=0$ for all corrections except that to $D_{_A}$ itself. And even that correction is evaluated only to linear order in $D_{_A}$. This will be justified {\it a posteriori} by showing that the effective coupling
$\gun_{_A}$ associated with the annealed noise  (which we will calculate below) flows to zero under the DRG transformation.

Because of this, it is possible to show, to {\it all} orders in perturbation theory,  that $\eta_{\alpha}$, $\eta_{\gamma}$, and $\eta_{_Q}$ all depend {\it only} on the coupling $\gun$ defined in (\ref{eq:def_gunc}) at the fixed point. This observation, together with	the fact that  $\gun$   flows to a non-zero stable fixed point, which we will show below, implies an {\it exact} relation, which does not depend on perturbation theory, between  the graphical corrections $\eta_{\gamma,\alpha,{_Q}}$. Since $\eta_\mu$ also depends on $\gun_\mu$, which we will show also flows to a non-zero stable fixed point, we further obtain a \emph{second} exact relation which is again independent of the perturbation theory.

Using the definitions of $\gun$ and $\gun_\mu$ and recursion relations (\ref{eq:RecurRel}), we can construct the following formally exact recursion relations:
	\begin{subequations}
\label{Flow_exactUnc}
\begin{align}
{\dd\,\ln \gun\over \dd\ell}&={1\over 3}-{7\over 3}\eta_{\gamma}+{1\over 3}\eta_{\alpha}+\eta_{_Q}\,\label{Flow_g_exactUnc}\\
{\dd\, \ln \gun_\mu\over \dd\ell}&=1-\eta_{\gamma}-\eta_{\mu}+\eta_{_Q}\,.\label{Flow_g_muexactUnc}
\end{align}
\end{subequations}

Next, inserting (\ref{eta_unc})
into (\ref{Flow_exactUnc}) we get two closed recursion relations for $\gun$ and $\gun_\mu$ :
	\begin{subequations}
\label{rgflow_un}
\begin{align}
	{\dd\gun\over\dd\ell}&={1\over 3}\left(1-\gun\right)\gun\,,\\
   {\dd\gun_\mu\over\dd\ell}&=\left(1-\gun_\mu\right)\gun_\mu\,.
	\end{align}
\end{subequations}
The associated DRG flow diagram is depicted in \fig \ref{fig:flows}, which shows that the flows have one stable fixed point and three unstable fixed points in {the} $\gun$-$\gun_\mu$ plane. The stable fixed point, which generically describes the universal behavior of the system, is at	
	\beqn
	{\gun}^*={\gun_\mu}^*=1\,.\label{g"_1_unc}
	\eeqn

	\begin{figure}
		\begin{center}
			\includegraphics[scale=.45]{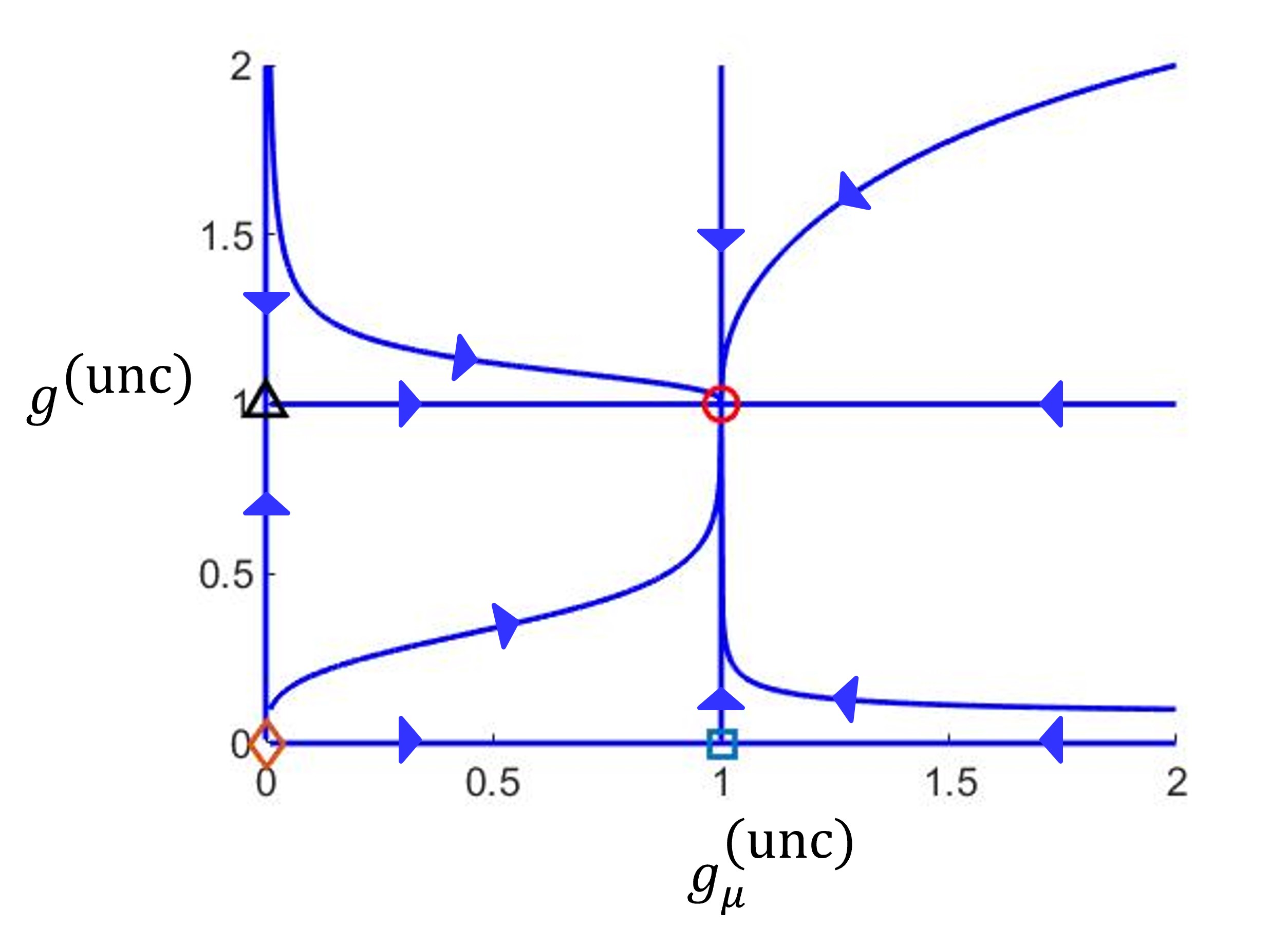}
		\end{center}
		\caption{ Renormalization group flows in $d=2$ in the plane of $g^{\rm (unc)}$  and $g_\mu^{\rm (unc)}$ (\ref{rgflow_un}).
				There are three unstable fixed points (indicated by the triangle, square and diamond symbols), and one  stable fixed point at $(g^{\rm (unc)},g_\mu^{\rm (unc)})=(1,1)$ (indicated by the red circle), which is our focus here.
								}
		\label{fig:flows}
	\end{figure}

At this nonzero stable fixed point,  $\dd \gun/\dd\ell$ and $\dd \gun_\mu/\dd\ell$ are both $0$;  Equations (\ref{Flow_exactUnc})  imply two exact relations between the $\eta$'s:
\begin{subequations}
\label{eq:exact_rel_0}
		\begin{align}
	7\eta_{\gamma}-3\eta_{_Q}-1&=\eta_{\alpha}\,,\label{Scaling_unc_0}
\\
	\label{Scaling_unc}
1-\eta_{\gamma}+\eta_{_Q}&=\eta_{\mu}\,.
	\end{align}
	\end{subequations}

To get quantitative results for $\eta$'s, we insert (\ref{g"_1_unc}) into (\ref{eta_alpha_unc_1}-\ref{eta_DA_unc_1}), which gives
\begin{subequations}
		\begin{align}
	\eta_\alpha&=-{1\over27}\,,\label{eta_alpha_unc_2}\\
	\eta_\gamma&={8\over 27}\,,\label{eta_gamma_unc_2}\\
	\eta_{_Q}&={10\over 27}\,,\label{eta_DQ_unc_2}\\
	\eta_{_A}&={16\over27}\,.\label{eta_DA_unc_2}
	\end{align}
	\end{subequations}
	Inserting (\ref{eta_gamma_unc_2},\ref{eta_DQ_unc_2}) into the {\it exact} relation (\ref{Scaling_unc}), we get
	\beqn
	\eta_\mu={29\over 27}\,.\label{eta_mu_unc_2}
	\eeqn

We will choose  the scaling exponents $\chi$, $\zeta$, and $z$ to keep the coefficients $\gamma$, $\alpha$, and $D_{_Q}$ fixed under renormalization. We will show in section (\ref{scaling}) that, as usual in the DRG \cite{forster_prl76, forster_pra77},  this choice makes these values of  $\chi$, $\zeta$, and $z$ those that appear in the scaling laws (\ref{Correl1}).} Since these coefficients determine the coupling coefficient $\gun$, which goes to a  non-zero constant  at the fixed point, keeping any two of these coefficients fixed automatically fixes the remaining one. For instance, by keeping $\alpha$ and $\gamma$ fixed  (i.e., setting the right-hand side of (\ref{fl_Alpha_unc},\ref{fl_gamma_unc}) to be zero), we get:
	\beq
	\zeta={2+\eta_{\gamma}-\eta_{\alpha}\over 3}\,,~~~  
	z={2-2\eta_{\gamma}-\eta_{\alpha}\over 3}\,. \\ 
	\label{Zeta1_unc}
	\eeq
	Inserting (\ref{Zeta1_unc}) into (\ref{Chi1}), we obtain $\chi$:
	\beq
	\chi={-1+\eta_{\gamma}-\eta_{\alpha}\over 3}\,.\label{Chi1_v2_unc}
	\eeq
The exponent $\chi'$ is also derived in section (\ref{scaling}). Here we only quote the result:
	\beq
	\chi' ={ \eta_{_A}-\eta_{\gamma}-1\over 2}\,.\label{Chi'_unc}
	\eeq
Using the values \eqref{eta_alpha_unc_2}-\eqref{eta_DA_unc_2} of the $\eta$'s, we then obtain the exponents' values explicitly
\begin{subequations}
		\begin{align}
	z&={13\over 27}\,,\\
	\zeta&={7\over 9}\,,\\
	\chi&=-{2\over 9}\,,\\
	\chi'&=-{19\over 54}\,.
	\label{unc exp}
\end{align}
	\end{subequations}
Coincidentally, these scaling exponents will turn out to be exactly equal to those obtained by the $\epsilon=(7/3-d)$-expansion in the ``hard" continuation approach to be discussed next, when $\epsilon$ is taken to be $1/3$.

	The above analysis ignored most of the graphical corrections induced by the annealed noise. The only such correction  we included was in the correction $\eta_{_A}$ to the annealed noise variance $D_{_A}$ itself, and even for that correction, we left out one-loop corrections that involve two annealed noises. In Appendix \ref{app_DA}, we show that those two annealed noise corrections to  $D_{_A}$ itself  are controlled by the ``annealed coupling coefficient"
\beq
\gun_{_A} \propto \frac{\alpha^{1/4} D_A}{ \mu_x^{5/4} \Lambda^{ {1\over2}}}
 \,.
\label{gdadefunc}
\eeq
Corrections to \emph{all} other coefficients, such as $\alpha$, stemming from the annealed noise, are also controlled by the \emph{same} coupling coefficient.

We will now show that  $\gun_{_A} $  flows to zero at the fixed point we've just found, thereby justifying our neglect of the graphical corrections arising from the annealed noise.
From its definition (\ref{gdadefunc}) and the recursion relations (\ref{eq:RecurRel}), we can derive the formally {\it exact} recursion relations for $\gun_{_A}$:
	\beqn
	{\dd \ln \gun_{_A}\over\dd\ell}&=&{1\over4}{\dd\ln\alpha\over\dd\ell}+{\dd \ln D_{_A}\over\dd\ell}-{5\over4}{\dd\ln{\mu_x}\over\dd\ell}
	\nonumber
	\\
	&=&{1\over2}+{(\eta_\alpha+4\eta_{_A}-5\eta_\mu)\over4}\,.\label{Flow_g_A_un}
	\eeqn

Inserting (\ref{eta_alpha_unc_2},\ref{eta_DA_unc_2},\ref{eta_mu_unc_2}) into (\ref{Flow_g_A_un})
gives
\beq
{\dd \ln \gun_{_A}\over\dd\ell}= -{7\over27}\,.\label{Flow_g_A_3}
\eeq
Since this eigenvalue  is less than zero,  we can conclude that the annealed noise is irrelevant at the quenched fixed point. This justifies our neglect of corrections coming from the annealed noise in \eqref{eta_unc}.

	\subsection{``Hard'' continuation}
	\label{Hard_cont_sec}
	In this section, we will obtain DRG recursion relations using an $\epsilon$-expansion method. This presents a problem since the model we described is defined \emph{precisely} in two dimensions. We circumvent this issue by \emph{only} analytically continuing the integrals in Fourier space required for the averaging over the large wavenumber modes in the nonlinear terms, and making the trivial (but important!) changes in the power counting on the rescaling step of the DRG. In this section, we generalize our calculation to dimensions $d>2$ by treating the ``soft" direction ($x$ direction) as one dimensional, while treating the other spatial component $y$ (the ``hard" direction) as $(d-1)$-dimensional. That is, we replace $y\to\brh$, and, in Fourier space, $q_y\to\bqh$.  In particular, the integrals of Fourier variables become $\int_{\tilde{\bq}}\equiv
		\int  \dd\Omega \dd^{d-1}q_h \dd q_x/ \left(\sqrt{2\pi}\right)^{d+1}$. Of course, this extension also changes the recursion relations for $D_{_Q}$ and $D_{_A}$ from \eqref{fl_D_Q_unc} and \eqref{fl_D_A_unc}, since these explicitly depend on the dimensionality $d$. For clarity, we rewrite all the recursion relations again:
	\begin{subequations}
		\label{rr}
		\begin{align}
			{\dd\ln\alpha\over\dd\ell}&=z+2\zeta-2+\eta_\alpha \,,\label{fl_Alpha}\\
			{\dd\ln\gamma\over\dd\ell}
			&=z-\zeta+\eta_{\gamma}\,,\label{fl_gamma}\\
			{\dd\ln{\mu_x}\over\dd\ell}&=z-2\zeta+\eta_\mu\,,\label{Fl_mu_x_exact}\\
			{\dd \ln D_{_Q}\over\dd\ell}&=2z-3\zeta+3-d+\eta_{_Q} \,,\label{fl_D_Q}\\
			{\dd \ln D_{_A}\over\dd\ell}&=z-3\zeta+3-d+\eta_{_A}\,.\label{fl_D_A}
		\end{align}
	\end{subequations}
Note the power counting of $D_{_{Q,A}}$ is changed due to the fact that we're not in $d=2$.
	
The $\eta$'s for this hard-continuation $\epsilon$-expansion are calculated in Appendix \ref{App:A} and, to one-loop order, are identical to our uncontrolled-calculation results (\ref{eta_unc}), i.e.,
\begin{subequations}
\label{etas_hard}
		\begin{align}		
	\eta_{\alpha}&=-{1\over 27}{\gha}\,,~~~ \eta_{\gamma}={8\over 27}{\gha}\,,~~~~~\label{Anoma_Exp1-1}
	\\
	\eta_{_Q}&={10\over 27}{\gha}\,,
	~~~\eta_{_A}={ 16\over 27}{\gha}\,,\label{Anoma_Exp1-2}
	\\
	\eta_\mu&={\gha_\mu}+{2\over 27}{\gha}\,.
	\end{align}
	\end{subequations}
The only difference between these graphical corrections and those of the uncontrolled calculation is the generalization of the dimensionless couplings  $\gun$ and $\gun_\mu$ to higher dimensions. These new  dimensionless couplings $\gha$ and $\gha_\mu$, are given by
	\begin{subequations}
		\begin{align}
			{\gha}&\equiv {S_{d-1}\over (2\pi)^{d-1}}|\gamma|^{-{7\over 3}}\alpha^{1\over 3}\Lambda^{3d-7\over 3}D_{_Q}\,,\label{g_def}\\
			{\gha_\mu}&\propto|\gamma|^{-1}\mu_x^{-1}D_{_Q}\Lambda^{d-3}
\,.\label{g_mu_def}
		\end{align}
	\end{subequations}

		In writing \eqref{etas_hard}, we have, as in the uncontrolled calculation, ignored all corrections to any parameters from the annealed noise strength $D_{_A}$ other than $D_{_A}$ itself; that is, we have set $D_{_A}=0$ for all corrections except that to $D_{_A}$ itself. And even that correction is evaluated only to linear order in $D_{_A}$. This will again be justified {\it a posteriori} by showing that the effective coupling $\gha_{_A}$ associated with the annealed noise actually flow to zero under the DRG transformation.
		
	
As in the last section, we can construct an exact relation between the graphical corrections $\eta_{\gamma,\alpha,{_Q}}$. Using the definition of $\gha$ (\ref{g_def}) and \eqref{rr}
		we construct  a formally {\it exact} recursion relation for $\gha$:
	\beq
	{\dd\,\ln{\gha}\over \dd\ell}=\bigg[{1\over 3}\left(7-3d\right)-{7\over 3}\eta_{\gamma}+{1\over 3}\eta_{\alpha}+\eta_{_Q}\bigg]\,.\label{Flow_g_exact}
	\eeq

	At the fixed point, since ${\dd \gha\over \dd\ell}=0$, Eq.~(\ref{Flow_g_exact}) clearly implies that either $\gha=0$, or
	\beq
	7-3d-7\eta_{\gamma}+\eta_{\alpha}+3\eta_{_Q}=0\,.\label{Exact_Relation1}
	\eeq

		It is easy to see by inspection that the $\gha=0$ fixed point is unstable (with eigenvalue $7/3-d$) for $d<7/3$, and, hence, in the physical case $d=2$. Therefore, in $d=2$, and formally in all spatial dimensions $d<7/3$, the graphical corrections $\eta_{\gamma}$, $\eta_{\alpha}$, and $\eta_{_Q}$ {\it must} obey \eqref{Exact_Relation1} {\it exactly}, at the nonzero stable fixed point, as expected from our previous calculation in the uncontrolled-calculation scheme (\ref{eq:exact_rel_0}a).

Likewise,  we can construct a formally exact recursion relation for  $\gha_\mu$:
		\beq
		{\dd\,\ln\gha_\mu\over \dd\ell}=3-d-\eta_{\gamma}-\eta_{\mu}+\eta_{_Q}\,.\label{Flow_g_muexact}
		\eeq
Reasoning as we just did for $\gha$, this recursion relation implies that the ${\gha_\mu}=0$ fixed point is unstable for any $d<3$. Hence, in all $d<3$ and, in particular, in the physical case $d=2$,
we obtain a second exact relation since ${\dd { \gha_\mu}\over \dd\ell}=0$ at the nonzero stable fixed point:
		\beq
		3-d+\eta_{_Q}-\eta_\gamma-\eta_\mu =  0\,.\label{Exact_Relation2}
		\eeq
Note that this reduces to \eqref{Scaling_unc} for $d=2$.

		Subtracting $1/3$ of equation \eqref{Exact_Relation1} from
			\eqref{Exact_Relation2}, we obtain an exact relation between $\eta_\mu$, $\eta_\gamma$, and $\eta_\alpha$:
			\beq
			\eta_\mu={2+4\eta_\gamma-\eta_\alpha\over3}
			\label{etamuexact}
			\eeq
			which will prove useful later.

To proceed further and obtain quantitative predictions for the exponents, we use the perturbation theory results (\ref{etas_hard})
for the graphical corrections in the recursion relations \eqref{Flow_g_exact}
		and \eqref{Flow_g_muexact}. This gives	
	\begin{subequations}
		\label{rgflow}
		\begin{align}
			{\dd {\gha}\over\dd\ell}&={1\over 3}\left(7-3d-{\gha}\right){\gha}\,,\label{Flow_g}\\
			{\dd {\gha_\mu}\over\dd\ell}&=\left(3-d-{\gha_\mu}\right){\gha_\mu}\,.\label{Flow_g_mu}
		\end{align}
	\end{subequations}
	 The associated DRG flow diagram for $d=2$ is identical to that in the uncontrolled calculation, as shown in \fig \ref{fig:flows}, which shows that the flows have one stable fixed point and three unstable fixed points. The stable fixed point, which generically describes the universal behavior of the system, is at
		\begin{subequations}
			\begin{align}
				{\gha}^* &= 3\epsilon+\cO(\epsilon^2)\,,\label{fix_p_ep}
				\end{align}
				\label{gha*}
		\end{subequations}
where $\epsilon=7/3-d$. Inserting the definition of $\epsilon$ in \eqref{Flow_g_mu}, we can also obtain an expression for the stable fixed point of ${\gha_\mu}$ for $d<7/3$:
	\begin{equation}
	{\gha_\mu}^*= \frac{2}{3} + \epsilon \,.
	\end{equation}
Note that, unlike our result (\ref{gha*}) for ${\gha}^*$, this result is {\it not}, strictly speaking, asymptotically valid in the limit of small $\epsilon$. This is because this value of ${\gha_\mu}^*$ does {\it not} become small for $\epsilon\ll1$; hence, there is no formal justification for dropping terms higher order in $\gha_\mu$ from the recursion relation for $\gha_\mu$. However, this is not a problem, since our results for all of the other $\eta$'s (except $\eta_\mu$) {\it are}  asymptotically correct  in the limit $\epsilon\ll1$. We can therefore obtain quantitatively valid  results for $\epsilon\ll1$ for those other $\eta$'s, and then use the {\it exact} scaling
relation (\ref{etamuexact}) to determine $\eta_\mu$.
		
Inserting this result for the fixed point value of the dimensionless coupling $\gha$  into our earlier expressions (\ref{Anoma_Exp1-1}) and (\ref{Anoma_Exp1-2}) for the graphical corrections $\eta_{\alpha,\gamma,_{Q},_{A}}$ gives their fixed point values explicitly as a functions of $\epsilon$:
\begin{subequations}
\label{etas_hard1}
		\begin{align}		
	\eta_{\alpha}&=-{\epsilon\over 9}\,,~~~ \eta_{\gamma}={8\epsilon\over 9}\,,
	\\
	\eta_{_Q}&={10\epsilon\over 9}\,,
	~~~\eta_{_A}={16\epsilon\over 9}\,.~~~~~
	\end{align}
	\end{subequations}
Inserting these into the exact relation (\ref{etamuexact}) gives our $\epsilon$-expansion result for $\eta_\mu$:
\beq
\eta_\mu={2\over3}+{11\epsilon\over 9}\,.
\label{etamueps}
\eeq

The $O(\epsilon^2)$ corrections can in principle be obtained through higher-loop calculations. We have not attempted this formidable calculation, because we expect our one-loop DRG results to be very quantitatively accurate. This is because,  in $d=2$, $\epsilon=1/3$, which is extremely small for $\epsilon$ expansions.

As in the last section, we have ignored the graphical corrections induced by the annealed noise, which are controlled by the coupling coefficient $\gha_{_A}$. We show in appendix \ref{app_DA} that, for this hard continuation, this coupling is given by
\beq
\gha_{_A} \propto \frac{\alpha^{1/4} D_A}{ \mu_x^{5/4} \Lambda^{ {5\over2}-d}}
\,.
\label{gdadef}
\eeq
 As we did for the uncontrolled approximation, here too we can justify our neglect of the corrections arising from the annealed noise by demonstrating that the coupling $\gha_{_A}$ just defined is irrelevant in the dimension of physical interest $d=2$.

 We do this by using
the definition (\ref{gdadef}) and the recursion relations (\ref{rr}) to derive the formally exact recursion relation for $\gha_{_A}$ :
	\beq
	{\dd \ln \gha_{_A}\over\dd\ell}
	={5\over2}-d+{(\eta_\alpha+4\eta_{_A}-5\eta_\mu)\over4}\,.\label{Flow_g_A_1}
	\eeq

Inserting	
 the fixed point values of $\eta$'s (\ref{etas_hard1}, \ref{etamueps}) into (\ref{Flow_g_A_1}) gives
\beq
{\dd \ln \gha_{_{_A}}\over\dd\ell}=
{59-33d\over27}\,,\label{Flow_g_A_3}
	\eeq
where we have inserted $\epsilon={7\over 3}-d$.

	Since the eigenvalue ${59-33d\over27}$ is less than zero \emph{both} near the critical dimension $d=7/3$, where it is $-{2\over3}$, and in $d=2$, where it is $-{7\over27}$, we can conclude that the annealed noise is irrelevant at the quenched fixed point. This justifies our neglect of  the corrections  to the $\eta$'s coming from the annealed noise in  (\ref{etas_hard}).

Finally, to obtain the scaling exponents, we substitute the fixed point values of   the $\eta$'s (\ref{etas_hard1}) into Eqs.~(\ref{Zeta1_unc},\ref{Chi1_v2_unc},\ref{Chi'_unc}). This gives
	\begin{subequations}
		\label{eq:exponents2nd}
		\begin{align}
			z&={2\over 3}-{5\over 9}\epsilon+O(\epsilon^2)\,,\\
			\zeta&={2\over 3}+{1\over 3}\epsilon+O(\epsilon^2)\,,\\
			\chi &=-{1\over 3}+{1\over 3}\epsilon+O(\epsilon^2)\,,\\
			\chi'&=-{1\over 2}+ {4\over 9}\epsilon+O(\epsilon^2)\,.
		\end{align}
	\end{subequations}

\subsection{``Soft" continuation\label{Sec:Soft}}
In this {section} we obtain the DRG recursion relations using a ``soft" continuation to higher dimensions. In this approach, we treat the ``hard" direction $y$ as one-dimensional, while the ``soft" direction $x$ is extended to $(d-1 )$-dimensions.  In practice, this means we will simply replace $q_x$ in Fourier space with a $(d-1 )$-dimensional vector $\bq_s$ orthogonal to the $y$-direction.

{As in the last section, this modifies the form of the recursion relations for $D_{_Q}$ and $D_{_A}$.  Rewriting the full set of recursion relations for completeness, we have:}
\begin{subequations}
\label{flow_s}
		\begin{align}
{\dd\ln\alpha\over\dd\ell}&=z+2\zeta-2+\eta_{\alpha} \,,\label{fl_Alpha_s}\\
{\dd\ln\gamma\over\dd\ell}
&=z-\zeta+\eta_{\gamma}\,,\label{fl_gamma_s}\\
{\dd\ln{\mu_x}\over\dd\ell}&=z-2\zeta+\eta_\mu\,,\label{Fl_mu_x_exact_s}\\
{\dd \ln D_{_Q}\over\dd\ell}&=2z-(d+1)\zeta+1+\eta_{_Q} \,,\label{fl_D_Q_s}\\
{\dd \ln D_{_A}\over\dd\ell}&=z-(d+1)\zeta+1+\eta_{_A}\,.\label{fl_D_A_s}
\end{align}
	\end{subequations}
{The $\eta$'s for the soft continuation are obtained in  Appendix \ref{App:A} and are}
\begin{subequations}
\label{eta_soft}
		\begin{align}
\eta_\alpha&=0\,,\label{eta_alpha_s_1}\\
\eta_\gamma&={2\over 3}{\gso}\,,\label{eta_gamma_s_1}\\
\eta_{_Q}&={2\over 3}{\gso}\,,\label{eta_DQ_s_1}\\
\eta_{_A}&= \gso \,.\label{eta_DA_s_1}
\\
\eta_\mu
&={\gso_\mu}+{1\over 6}{\gso}\,,\label{eta_mu_s}
\end{align}
	\end{subequations}
where {the two dimensionless couplings are given by}
\begin{subequations}
\label{eq:def_gsoft}
		\begin{align}
{\gso}&\equiv {S_{d-1}D_{_Q}\over \sqrt{2}(2\pi)^{d-1}}|\gamma|^{-{\left({d+5\over3}\right)}}\alpha^{\left({d-1\over 3}\right)}\Lambda^{\left({2d-5\over 3}\right)}\,,\label{g'def}
\\
{\gso_\mu}&\propto
D_{_Q}\mu_x^{-1}\alpha^{\left({d-2\over 3}\right)}|\gamma|^{-\left({d+1\over 3}\right)}\Lambda^{\left({2d-7\over 3}\right)}\,.\label{g'_mu}
\end{align}
	\end{subequations}

Inserting (\ref{eta_alpha_s_1}-\ref{eta_DQ_s_1}) into (\ref{fl_Alpha_s},\ref{fl_gamma_s},\ref{fl_D_Q_s}) and using the definitions of ${\gso}$ and $\gso_\mu$ (\ref{eq:def_gsoft}), we get two closed recursion relations for $\gso$ and $\gso_\mu$:
\begin{subequations}
\label{rgflow_soft}
		\begin{align}
{\dd\ln{{\gso}}\over\dd\ell}&=\left({5-2d\over 3}\right)-\left({2d+4\over 9}\right){\gso}\,,\label{fl_g'_s}
\\
{\dd\ln{{\gso_\mu}}\over\dd\ell}&=\left({7-2d\over 3}\right)+\left({5-4d\over 18}\right){\gso}-{\gso_\mu}\,.\label{fl_g'_smu}
\end{align}
	\end{subequations}
As in the uncontrolled calculation and the hard-continuation scheme, the trivial $\gso=0$ and $\gso_\mu=0$ fixed points are obviously unstable at the physically relevant dimension $d=2$. In fact, here the $\gso=0$ fixed point is unstable for all $d<5/2$ and $\gso_\mu=0$ fixed point is unstable for all $d<7/2$. Below $d=5/2$, the stable fixed point which generically describes hydrodynamics of the system is at
\beqn
\label{FixP_soft}
{\gso}^*&={2\over 3}\tilde{\epsilon}+{O}(\tilde{\epsilon}^2)\,,
	\eeqn
where $\tilde{\epsilon}=5/2-d$. We can now insert the fixed point value of $\gso$ (\ref{FixP_soft}) into \eqref{fl_g'_smu} to obtain the fixed point value of $\gso_\mu$ \emph{for $d<5/2$}:
\begin{equation}
	{\gso_\mu}^*={2\over 3}+{13\over 27}\tilde{\epsilon}\,.
\end{equation}
As for the hard continuation, unlike our result (\ref{FixP_soft}) for ${\gso}^*$, this result is {\it not}, strictly speaking, asymptotically valid in the limit of small $\tilde{\epsilon}$. This is because this value of ${\gso_\mu}^*$ does {\it not} become small for $\tilde{\epsilon}\ll1$; hence, there is no formal justification for dropping terms higher order in $\gso_\mu$ from the recursion relation for $\gso_\mu$. However, again, this is not a problem, since our results for all of the other $\eta$'s (except $\eta_\mu$) {\it are} exact to linear order in $\tilde{\epsilon}\ll1$.
We can therefore obtain quantitatively valid results for $\tilde{\epsilon}\ll1$ for those other $\eta$'s, and then use the {\it exact} scaling
relation (\ref{Scaling_Smu}), which we'll derive below, to determine $\eta_\mu$.

Again, the fact that both ${\gso}$ and ${\gso_\mu}$ flow to a non-zero stable fixed point for $d<5/2$ implies a pair of \emph{exact} relations between $\eta$'s:
	\begin{subequations}
\begin{align}
	\label{Scaling_S}
\left({5-2d\over 3}\right)+\eta_{_Q}-\left({d+5\over 3}\right)\eta_\gamma+\left({d-1\over 3}\right)\eta_\alpha=0\,,\\
\eta_\mu-\left({7-2d\over 3}\right)-\left({d-2\over 3}\right)\eta_\alpha+\left({d+1\over 3}\right)\eta_\gamma-\eta_{_Q}=0
\,.\label{Scaling_Smu}
\end{align}
\end{subequations}

 Inserting (\ref{FixP_soft})   into (\ref{eta_alpha_s_1}-{\ref{eta_DA_s_1}}) yields
\begin{subequations}
		\begin{align}
\eta_\alpha&=0\,,\label{eta_alpha_s_2}\\
\eta_\gamma&={4\over 9}\tilde{\epsilon}+O(\tilde{\epsilon}^2)\,,\label{eta_gamma_s_2}\\
\eta_{_Q}&={4\over 9}\tilde{\epsilon}+O(\tilde{\epsilon}^2)\,,\label{eta_DQ_s_2}\\
\eta_{_A}&={ 2\over 3}\tilde{\epsilon}+O(\tilde{\epsilon}^2)\,.\label{eta_DA_s_2}
\end{align}
	\end{subequations}
{Then inserting (\ref{eta_alpha_s_2},\ref{eta_gamma_s_2},\ref{eta_DQ_s_2})} into the exact relation \eqref{Scaling_Smu}, we get
\beq
\eta_\mu={2\over 3}+{16\over 27}\tilde{\epsilon}+O(\tilde{\epsilon}^2)\,.
\eeq

Using these values in Eqs.~(\ref{Zeta1_unc},\ref{Chi1_v2_unc},\ref{Chi'_unc}) yields the $O(\tilde{\epsilon})$ values for the scaling exponents:
\begin{subequations}
		\begin{align}
z&={2\over 3}-{8\over 27}\tilde{\epsilon}+O(\tilde{\epsilon}^2)\,,\\
\zeta&={2\over 3}+{4\over 27}\tilde{\epsilon}+O(\tilde{\epsilon}^2)\,,\\
\chi&=-{1\over 3}+{4\over 27}\tilde{\epsilon}+O(\tilde{\epsilon}^2)\,,\\
\chi'&=-{1\over 2}+{1\over 9}\tilde{\epsilon}+O(\tilde{\epsilon}^2)\,.
\label{soft exp}
\end{align}
	\end{subequations}
It is easy to check that the numerical values of these exponents in $d=2$ to $\cO$($\tilde{\epsilon}={1\over 2}$) are very close to those obtained in the uncontrolled calculation and the hard-continuation.

As in the other two schemes, we have ignored the graphical corrections induced by the annealed fluctuations.  We'll now show that these graphical corrections are also irrelevant in the soft-continuation scheme. In Appendix \ref{app_DA}, we show that these graphical corrections are controlled by the dimensionless coupling
\beq
\gso_{_A} \propto D_A\alpha^{(d-1)\over 4}\mu_x^{-{(d+3)\over 4}}\Lambda^{ d-3\over 2}\,.
\label{gdadefs}
\eeq
Using (\ref{gdadefs}) and (\ref{flow_s}) we get the recursion relation for $\gso_{_A}$ at the fixed point controlled by the quenched fluctuations:
\beq
{\dd\ln\gso_{_A} \over\dd\ell}= -{2\over 3}+{14\over 27}\tilde{\epsilon}+O(\tilde{\epsilon}^2)\,.
\label{}
\eeq
This eigenvalue is $-{11\over 27}$ in $d=2$ to $O(\tilde{\epsilon}={1\over 2})$,   which is less than $0$, and thus  again  shows the irrelevance of graphical corrections coming from the annealed noises.

\subsection{Best estimate of exponents via weighted average of all three approaches}\label{best_exponents}

The best estimate of the numerical values of the exponents is a suitably weighted average  of  the results obtained from the three schemes. Since the errors are $O(\epsilon^2)$ and $O(\tilde{\epsilon}^2)$ in the hard-continuation and soft-continuation, respectively, and the former is ${4\over 9}$ times the latter in $d=2$, we
will weight the soft continuation with $4/9$ the weight of the hard continuation. Given that the exponents obtained from the uncontrolled calculation are identical to those found by the hard continuation, it seems most sensible to weight those two results equally. We thereby obtain for our best estimates of the exponents:
\begin{subequations}
		\begin{align}
z&={9z(\epsilon={7\over 3}-d)+9z({\rm uncontrolled})+4z(\tilde{\epsilon}={5\over 2}-d)\over 22}\,,\nonumber\\\\
\zeta&={9\zeta(\epsilon={7\over 3}-d)+9\zeta({\rm uncontrolled})+4\zeta(\tilde{\epsilon}={5\over 2}-d)\over 22}\,,\nonumber\\\\
\chi&={9\chi(\epsilon={7\over 3}-d)+9\chi({\rm uncontrolled})+4\chi(\tilde{\epsilon}={5\over 2}-d)\over 22}\,,\nonumber\\\\
\chi'&={9\chi'(\epsilon={7\over 3}-d)+9\chi'({\rm uncontrolled})+4\chi'(\tilde{\epsilon}={5\over 2}-d)\over 22}\ .
\end{align}
	\end{subequations}
We can also estimate the errors as the differences between the hard-continuation result  (or equivalently the uncontrolled-calculation result) and the weighted averages above. Thus, we ultimately obtain:
\begin{subequations}
\label{exp num}
		\begin{align}
z&=0.49\pm 0.01\,,\\
\zeta &=0.77\pm 0.01\,,\\
\chi&=-0.23\pm 0.01\,,\\
\chi'&=-0.37\pm 0.02\,,
\end{align}
	\end{subequations}
which are the numerical values of the exponents quoted in the introduction.

\section{Scaling behavior}\label{scaling}
Now we utilize the DRG to calculate the real time-real space correlation functions $\langle\bu(\br,t)\cdot\bu(\mathbf{0},0)\rangle$. Let's first focus on the quenched part $C_{_Q}\left(\br\right)\equiv\langle\bu(\br,t)\cdot\bu(\mathbf{0},0)\rangle_{_Q}$ (see (\ref{eq:Ca&Cq_real}b)), which is purely static.
Keeping track of the rescalings done on each step of the DRG enables us to relate the correlation functions in the renormalized theory to those in the original (unrenormalised) model. This relation, known as a ``trajectory integral matching" formula, reads, for the real-space correlations \cite{forster_prl76, forster_pra77}:
\beqn
&&C_{_Q}\left(\alpha_0,\gamma_0,D_{_{Q,0}},\br\right)\nonumber\\
&=&e^{2\chi\ell}C_{_Q}\left[\alpha(\ell),\gamma(\ell),D_{_Q}(\ell),|x|e^{-\zeta\ell},|y|e^{-\ell} \right]\,.\label{Traje1}
\eeqn
We have explicitly displayed the renormalized parameters $\alpha$, $\gamma$, and $D_{_Q}$  because they are the parameters that determine $C_{_Q}\left(\br\right)$, as we saw in the linear theory. In \eqref{Traje1} the subscript ``0" denotes the bare values of the parameters. Note that $C_{_Q}\left(\br\right)$ only depends on the absolute values of $x$ and $y$.

Let's now choose the exponents $z$, $\zeta$, and $\chi$,  to be the values
given by \eqref{Zeta1_unc} and (\ref{Chi1_v2_unc}), which keep
 $\alpha$, $\gamma$, and $D_{_Q}$ fixed, and also choose $\ell=\ln\left(\Lambda |y|\right)$. Eq. (\ref{Traje1}) can then be rewritten as
\beqn
C_{_Q}\left(\alpha_0,\gamma_0,D_{_{Q0}},\br\right)
={|y|}^{2\chi}\cG_{_{Q}}\left(|x|\over |y|^{\zeta}\right)\,,\label{}
\eeqn
where
\beqn
\cG_{_{Q}}\left(|x|\over |y|^{\zeta}\right)
\equiv\Lambda^{2\chi}C_{_Q}\left[\alpha_0,\gamma_0,D_{_{Q0}},{|x|\over (|y|\Lambda)^{\zeta}},{1\over\Lambda}\right]\,.\label{}
\eeqn

We have found the quenched part of the correlation function shown in (\ref{Correl1}).
Now we turn to the annealed part of the correlations, for which there is a relation similar to (\ref{Traje1}):
\beqn
&&C_{_A}\left(\alpha_0,\mu_{x0},D_{_{A0}},\br,t\right)\nonumber\\
&=&e^{2\chi\ell}C_{_A}\left[\alpha(\ell),\mu_x(\ell),D_{_A}(\ell),|x|e^{-\zeta\ell},|y|e^{-\ell},|t|e^{-z\ell}\right]\,,
\nonumber\\
\label{Traje2}
\eeqn
where $C_{_A}\left(\br,t\right)\equiv\langle\bu(\br,t)\cdot\bu(\mathbf{0},0)\rangle_{_A}$ (see (\ref{eq:Ca&Cq_real}a)),  whose value is determined by $\alpha$, $\mu$, and $D_{_A}$, as we saw in the linear theory.

Next we make the same choice of $\zeta$, $z$, $\chi$, and $\ell$ as we did in deriving the quenched part of the correlations. In addition to $\alpha$, $\gamma$, and $D_{_Q}$ this choice also fixes $\mu_x$
due to the exact scaling relation \eqref{Scaling_unc},  or equivalently \eqref{Exact_Relation2} or \eqref{Scaling_S}. The three exact scaling relations become identical in $d=2$.

On the other hand, with our choice of the rescaling exponents, $D_{_A}(\ell)$ is {\it not} fixed. Instead, it is straightforward to show that, in all three of our approaches,
\beq
{\dd \ln D_{_A}\over\dd\ell}-{\dd \ln D_{_Q}\over\dd\ell}=\eta_{_A}-\eta_{_Q}-z\,.
\label{D dif}
\eeq
With our choice of exponents, ${\dd \ln D_{_Q}\over\dd\ell}=0$, so we have
\beq
{\dd \ln D_{_A}\over\dd\ell}=\eta_{_A}-\eta_{_Q}-z\,,
\label{D dif}
\eeq
which we can immediately solve for the renormalized annealed noise strength $D_{_A}(\ell)$:
\beq
D_{_A}(\ell)=D_{_{A0}}\exp[(\eta_{_A}-\eta_{_Q}-z)\ell] \,.
\label{DAsol}
\eeq
\vspace{.2in}
We can therefore rewrite (\ref{Traje2}) as
\beqn
&&C_{_A}\left(\alpha_0,\mu_{x0},D_{_{A0}},\br,t\right)\nonumber\\
&=&e^{2\chi\ell}C_{_A}\left[\alpha_0,\mu_{x0}, D_{_A}(\ell),|x|e^{-\zeta\ell},|y|e^{-\ell},|t|e^{-z\ell}\right]\,,
\nonumber\\
\label{Traje3}
\eeqn
with $D_{_A}(\ell)$ given by (\ref{DAsol}).

To proceed, we recall the result from the linear theory that $C_{_A}$ is proportional to $D_{_A}$. The linear theory should be valid on the right-hand side of (\ref{Traje3}) if we choose $\ell$ so that the $y$ argument of $C_{_A}$
on that side
is of order a microscopic length (i.e., $\Lambda^{-1}$, where $\Lambda$ is the ultraviolet cutoff). That is, we'll choose
\beq
\ell=\ell^*(y)=\ln\left(\Lambda |y|\right)
\label{ell*}
\eeq
 again. Doing so, and using the fact that, with this choice of $\ell$, the linear theory works, and making further use of the fact that, in the linear  theory, the annealed correlation function is proportional to $D_{_A}$, we have
\bew
\beqn
C_{_A}\left(\alpha_0,\mu_{x0},D_{_{A0}},\br,t\right)
=\left({|y|\Lambda}\right)^{2\chi}D_{_A}(\ell^*(y))\Theta\left({|x|\over |y|^{\zeta}},{|t|\over |y|^z}\right)\,,
\nonumber\\
~\label{CA_n}
\eeqn
\ew
where we have defined
\beq
\Theta\left({|x|\over |y|^{\zeta}},{|t|\over |y|^z}\right)\equiv {C_{_A}\left(\alpha_0,\mu_{x0},D_{_A}
(\ell^*),{|x|\over (|y|\Lambda)^{\zeta}},{1\over\Lambda},{|t|\over (|y|\Lambda)^z}\right)\over
D_{_A}(\ell_*)} \,.
\label{Thetadef}
\eeq
Note that $\Theta$ is actually independent of $D_{_A}(\ell^*)$, since we have cancelled off its linear dependence on  $D_{_A}(\ell^*)$. Indeed, $\Theta$ depends {\it only} on the ratios
$,{|x|\over |y|^{\zeta}}$ and ${|t|\over |y|^z}$, since $\alpha_0$, $\mu_{x0}$ and $\Lambda$ are constants.

Inserting our result (\ref{DAsol}) for $D_{_A}(\ell)$ into (\ref{CA_n}), and evaluating it at $\ell^*(y)$ gives
\beqn
C_{_A}\left(\alpha_0,\mu_{x0},D_{_{A0}},\br,t\right)
={|y|}^{2\chi'}\cG_{_{A}}\left({|x|\over |y|^{\zeta}},{|t|\over |y|^z}\right)\,,~\label{CA_last}
\eeqn
where
\beq
\cG_{_{A}}\left({|x|\over |y|^{\zeta}},{|t|\over |y|^z}\right)
\equiv\Lambda^{2\chi'}D_{_{A0}}\Theta\left({|x|\over |y|^{\zeta}},{|t|\over |y|^z}\right)\,,
\label{GAdef}
\eeq
and
\beq
\chi' =\chi+{1\over 2}\left( \eta_{_A}-\eta_{_Q}-z\right)\,.\label{Chi'_original_1}
\eeq
Insert (\ref{Zeta1_unc},\ref{Chi1_v2_unc}) into (\ref{Chi'_original_1}) and use the exact scaling relation (\ref{Scaling_unc_0}) to eliminate $\eta_{_Q}$. This leads to
\beq
	\chi' ={ \eta_{_A}-\eta_{\gamma}-1\over 2}\,,\label{}
	\eeq
which is the value of $\chi'$ that we quoted in (\ref{Chi'_unc}).

 Alternatively, the correlations can be derived by performing the inverse Fourier transform of the correlation function obtained in the linear theory (\ref{eq:Ca&Cq}), albeit now with wavenumber-dependent coefficients due to  the renormalization. The Fourier transformed correlations are also of interest in their own right. Again using trajectory matching, we have
\bew
\beqn
\langle u_y(\tbq)u_y(\tbq')\rangle
=\ee^{2\left(d-1+z+\zeta+\chi\right)\ell}\left\langle u'_y(\ee^{\ell}q_x,\ee^{\zeta\ell}q_y,\ee^{z\ell}\omega)
u'_y(\ee^{\ell}q'_x,\ee^{\zeta\ell}q_y',\ee^{z\ell}\omega')\right\rangle\,,
\label{RG_Conne1}
\eeqn
\ew
where $u_y(\tbq)$ and $ u'_y(\ee^{\ell}q_x,\ee^{\zeta\ell}q_y,\ee^{z\ell}\omega)$ represent the velocity field before and after rescaling, respectively.

	Since the non-linear corrections become more important as we go to longer wavelengths {and times} (i.e., smaller {$\tbq$}), it conversely follows that we can best approximate the right-hand side of \eqref{RG_Conne1} using the linear theory if we make $(\ee^{\ell}q_x,\ee^{\zeta\ell}q_y,\ee^{z\ell}\omega)$ on the right-hand side as {\it large} as possible.
We will therefore choose $\ell$ so that this rescaled momentum lies near the Brillouin zone (BZ) boundary. This allows us to evaluate the correlation function on the right-hand side~of Eq. (\ref{RG_Conne1}) using the linear theory (\ref{eq:Ca&Cq}). To determine the value $\ell^*$ of $\ell$ that  is sufficiently near the BZ boundary to allow  this, we use the  criterion
\beqn
\gamma^2|q_x|^6+\alpha^2 |q_y|^4=\alpha^2\Lambda^4\,.
\label{Brillouin1}
\eeqn
Our motivation for this choice is that this makes the correlation function at the renormalized wavevector as small as its largest value on the BZ boundary.
For the rescaled momentum to satisfy this condition, we must have
\beqn
\gamma^2(\ell^*)\left(|q_x|\ee^{\zeta\ell^*}\right)^6+
\alpha^2(\ell^*)\left(|q_y|\ee^{\ell^*}\right)^4
=\alpha^2(\ell^*)\Lambda^4\,,\nonumber\\
\label{Brillouin2}
\eeqn
where the $\ell^*$-dependences of $\alpha(\ell^*)$ and $\gamma(\ell^*)$ is obtained by solving the recursion relations (\ref{fl_Alpha}) and (\ref{fl_gamma}), respectively. These are most easily solved by choosing the rescaling exponents $z$ and $\zeta$ to keep $\gamma$
and $\alpha$ fixed at their bare values. This leads to the values of $z$ and $\zeta$ quoted in \eqref{Zeta1_unc}.

Making this choice and dividing \eqref{Brillouin2} by $\alpha^2$, gives
\beqn
a^2\left(|q_x|\ee^{\zeta\ell^*}\right)^6+
\left(|q_y|\ee^{\ell^*}\right)^4
=\Lambda^4\,,
\label{Brillouin3}
\eeqn
where we've defined the microscopic length $a\equiv {\gamma_0\over\alpha_0}$.

We will seek a scaling solution for $\ee^{\ell^*}$ of \eqref{Brillouin3}
of the form
\beqn
\ee^{\ell^*}={\Lambda\over |q_y|}f_1\left(|q_x|/\Lambda'\over\left(|q_y|/\Lambda\right)^{\zeta}\right)\,,
\label{Ell1}
\eeqn
where $\Lambda'\equiv \Lambda^{2\over 3}a^{-{1\over 3}}$, and $\zeta$ is given by (\ref{Zeta1_unc}).

Inserting this scaling ansatz \eqref{Ell1} into our condition \eqref{Brillouin3} gives
\beqn
\nonumber
\left({a^2\Lambda^{6\zeta}|q_x|^6\over |q_y|^{6\zeta}}\right)\bigg[f_1\left(|q_x|/\Lambda'\over\left(|q_y|/\Lambda\right)^{\zeta}\right)\bigg]^{6\zeta}
&&
\\
+\Lambda^4\bigg[f_1\left(|q_x|/\Lambda'\over\left(|q_y|/\Lambda\right)^{\zeta}\right)\bigg]^4&=&\Lambda^4 \,.
\label{scalingcond1}
\eeqn
Dividing both sides of this expression by $\Lambda^4$ and reorganizing a bit yields
\beqn
\nonumber
\left({a^2\Lambda^{6\zeta-4}|q_x|^6\over |q_y|^{6\zeta}}\right)\bigg[f_1\left(|q_x|/\Lambda'\over\left(|q_y|/\Lambda\right)^{\zeta}\right)\bigg]^{6\zeta}
&&
\\+
\bigg[f_1\left(|q_x|/\Lambda'\over\left(|q_y|/\Lambda\right)^{\zeta}\right)\bigg]^4&=&1 \,.
\label{scalingcond2}
\eeqn
The coefficient of $\bigg[f_1\left(|q_x|/\Lambda'\over\left(|q_y|/\Lambda\right)^{\zeta}\right)\bigg]^{6\zeta}$ in this expression can be re-expressed as
\beqn
\nonumber
\left({a^2\Lambda^{6\zeta-4}|q_x|^6\over |q_y|^{6\zeta}}\right)&=&\left({a^{1/3}\Lambda^{\zeta-2/3}|q_x|\over |q_y|^{\zeta}}\right)^6
\\
\nonumber
&=&\left({a^{1/3}\Lambda^{-2/3}|q_x|\over (|q_y|/\Lambda)^{\zeta}}\right)^6
\\
&=&\left(|q_x|/\Lambda'\over\left(|q_y|/\Lambda\right)^{\zeta}\right)^6=w^6 \,,
\label{coeff}
\eeqn
where we've defined the argument of the scaling function $f_1$ in our ansatz \eqref{Ell1} to be $w$; i.e.,
\beq
w\equiv\left(|q_x|/\Lambda'\over\left(|q_y|/\Lambda\right)^{\zeta}\right) \,.
\label{wdef}
\eeq
Thus, our condition on the scaling function $f_1$ can be rewritten as
\beq
w^6[f_1(w)]^{6\zeta}+[f_1(w)]^4=1 \,.
\label{scalingcondfinal}
\eeq
Since this condition only involves the scaling argument $w$, defined by \eqref{wdef}, and the scaling function $f_1(w)$, which also only depends on $w$, it is clear that our ansatz has worked, and that the scaling  function $f_1(w)$ is determined by the solution of \eqref{scalingcondfinal}. That solution is easily seen to have the following limiting behaviors:
\begin{eqnarray}
f_1(x)=\left\{
\begin{array}{ll}
1\,,&x\ll 1\,,\\
x^{-{1\over\zeta}}\,,&x\gg1\,.
\end{array}
\right.
\end{eqnarray}

Inserting (\ref{Ell1}) into our expression (\ref{RG_Conne1}) for the correlation function in Fourier space, and evaluating the correlation function on the right-hand side~using the linear theory (\ref{eq:corre_uu}), we obtain the velocity correlation function in momentum space:
\bew
\beqn
\langle u_y(\tbq)u_y(\tbq')\rangle
&=&\left[{2D_{A0}\ee^{\eta_{_A}\ell^*}\delta(\omega+\omega')\delta(\bq+\bq')\over \left(\omega-\gamma_0\ee^{\eta_{\gamma}\ell^*} q_x\right)^2+\left({\alpha_0\ee^{\eta_{\alpha}\ell^*} q_y^2\over q_x^2}+\mu_{x0}\ee^{\eta_\mu\ell^*}q_x^2\right)^2}
+{4\pi D_{Q0}\ee^{\eta_{_Q}\ell^*}q_x^4\delta(\omega)\delta(\omega')\delta(\bq+\bq')\over \left(\alpha_0\ee^{\eta_{\alpha}\ell^*}\right)^2 q_y^4+\left(\gamma_0\ee^{\eta_{\gamma}\ell^*}\right)^2q_x^6}\right]\nonumber\\
&=&\left[{2D_{_A}(\bq)\delta(\omega+\omega')\delta(\bq+\bq')\over \left[\omega-\gamma(\bq) q_x\right]^2+\left[{\alpha(\bq) q_y^2\over q_x^2}+\mu_x(\bq) q_x^2\right]^2}
+{4\pi D_{_Q}(\bq) q_x^4\delta(\omega)\delta(\omega')\delta(\bq+\bq')\over \alpha^2(\bq) q_y^4+\gamma^2(\bq) q_x^6}\right]
\label{RG_Conne2}
\eeqn
where
\begin{subequations}
\begin{align}
\gamma(\bq)&=\gamma_0 |q_y|^{-\eta_\gamma}f_\gamma\left(|q_x|\over |q_y|^{\zeta}\right)\,,\\
\alpha(\bq)&=\alpha_0 |q_y|^{-\eta_\alpha}f_\alpha\left(|q_x|\over|q_y|^{\zeta}\right)\,,\\
\mu_x(\bq)&=\mu_{x0} |q_y|^{-\eta_\mu}f_\mu\left(|q_x|\over |q_y|^{\zeta}\right)\,,\\
D_{_{A,Q}}(\bq)&=D_{_{A0,Q0}} |q_y|^{-\eta_{_{A,Q}}}f_{_{A,Q}}\left(|q_x|\over |q_y|^{\zeta}\right)\,,
\end{align}
\label{anomparam}
\end{subequations}
and
\beqn
&&f_{\gamma,\alpha,\mu,_A,_Q}\left(|q_x|\over |q_y|^{\zeta}\right)\nonumber\\
&\equiv& \Lambda^{\eta_{\gamma,\alpha,\mu,_A,_Q}} \left[f_1\left(|q_x|/\Lambda'\over\left(|q_y|/\Lambda\right)^{\zeta}\right)\right]^{\eta_{\gamma,\alpha,\mu,A,Q}}\,.
\eeqn
The subscript ``0" denotes the bare value of the coefficient.

We Fourier transform the momentum-space correlation function to obtain the real-space one. But first it is convenient to write $\langle u_y(\tbq)u_y(\tbq')\rangle$ in a compact form:
\beqn
\langle u_y(\tbq)u_y(\tbq')\rangle
&=&\left[|q_y|^{-\left(2z+\eta_{_A}\right)}F_{_{A}}\left({q_x\over |q_y|^\zeta},{\omega\over |q_y|^z}\right)\delta(\omega+\omega')
+|q_y|^{-(4-4\zeta+\eta_{_Q}-2\eta_\alpha)}F_{_{Q}}\left({q_x\over |q_y|^\zeta}\right)\delta(\omega)\delta(\omega')\right]\delta(\bq+\bq')
\,,\nonumber\\ \label{Correlation_Mom_Scal1}
\eeqn
where
\begin{subequations}
\begin{align}
F_{_{A}}\left({q_x\over |q_y|^\zeta},{\omega\over |q_y|^z}\right)&\equiv
{2D_{_{A0}}f_{_A}\left(|q_x|\over |q_y|^\zeta\right)\over \left[{\omega\over |q_y|^z}-\gamma_0{q_x\over |q_y|^\zeta}f_\gamma\left(|q_x|\over |q_y|^\zeta\right)\right]^2+\left[{\alpha_0 f_\alpha\left(|q_x|\over |q_y|^\zeta\right)\over \left(|q_x|/|q_y|^\zeta\right)^2}+\mu_{x0}\left(|q_x|\over |q_y|^\zeta\right)^2f_\mu\left(|q_x|\over |q_y|^\zeta\right)\right]^2}\,,\\
F_{_{Q}}\left({|q_x|\over |q_y|^\zeta}\right)&\equiv{4\pi D_{_{Q0}}\left(|q_x|\over |q_y|^\zeta\right)^4 f_{_Q}\left(|q_x|\over |q_y|^\zeta\right)\delta(\bq+\bq')\over \left[\gamma_0\left(|q_x|\over |q_y|^\zeta\right)^3f_\gamma\left(|q_x|\over |q_y|^\zeta\right) \right]^2+\left[\alpha_0 f_\alpha\left(|q_x|\over |q_y|^\zeta\right) \right]^2}\,.
\end{align}
\end{subequations}
In deriving (\ref{Correlation_Mom_Scal1}) we have used the exact relations (\ref{eq:exact_rel_0}).

We'll next calculate the annealed part of the real-space correlation functions. Using the scaling form (\ref{Correlation_Mom_Scal1}) we get
\beq
\langle u_y(\br,t)u_y(\mathbf{0},0)\rangle_{_A}
=\int_{\omega,\omega', \bq, \bq'}
\ \ |q_y|^{-\left(2z+\eta_{_A}\right)}F_{_{A}}\left({q_x\over |q_y|^\zeta},{\omega\over |q_y|^z}\right)\delta(\omega+\omega')
\delta(\bq+\bq')e^{-\ii\left(\omega t-\bq\cdot\br\right)} \ .
\label{C_A_real}
\eeq
Changing variables of integration,
\beqn
q_y={{Q}_y\over |y|}\,,~~~q_x={Q_x\over |y|^{\zeta}}\,,~~~\omega={\Omega\over |y|^z}\,,
\eeqn
we write (\ref{C_A_real}) as
\beqn
\langle u_y(\br,t)u_y(\mathbf{0},0)\rangle_{_A}
=|y|^{2\chi'}\cG_{_{A}}\left({|x|\over |y|^\zeta},{|t|\over |y|^z}\right)\,,
\eeqn
where
\beq
\chi'={1\over 2}\left(z-\zeta+\eta_{_A}-1\right)\,, \label{Chi'_original_2}
\eeq
\beq
\cG_{_{A}}\left({|x|\over |y|^\zeta},{|t|\over |y|^z}\right)
\equiv{1\over (2\pi)^{3/2}}\int_{\Omega,{\bf Q}}
\ \ |Q_y|^{-(2z+\eta_{_A})}F_{_{A}}\left({Q_x\over |{Q}_y|^\zeta},{\Omega\over |{Q}_y|^z}\right)
e^{-\ii\left({t\Omega\over |y|^z} -{x Q_x\over|y|^\zeta}-{y Q_y\over |y|}\right)}\,,
\eeq
and the values of $\zeta$, $z$, and $\chi$ are again given by \eqref{Zeta1_unc} and (\ref{Chi1_v2_unc}). Inserting (\ref{Zeta1_unc}) into (\ref{Chi'_original_2}) leads to (\ref{Chi'_unc}) again.

The quenched part of the correlation is obtained in essentially the same way, and the result is
\beqn
\langle u_y(\br,t)u_y(\mathbf{0},0)\rangle_{_Q}
=|y|^{2\chi}\cG_{_{Q}}\left({|x|\over |y|^\zeta}\right)\,,\label{Correl_Q1}
\eeqn
\beq
\cG_{_{Q}}\left({|x|\over |y|^\zeta}\right)
\equiv{1\over 2\pi}\int_{\bf Q}
\ \ |{Q}_y|^{-(4-4\zeta+\eta_{_Q}-2\eta_\alpha)}F_{_{Q}}\left({|Q_x|\over |{Q}_y|^\zeta}\right)
e^{\ii\left({yQ_y\over |y|}+\frac{x Q_x}{|y|^\zeta}\right)}\,,
\eeq
\ew
where the values of $\zeta$ and $\chi$ are again given by \eqref{Zeta1_unc} and (\ref{Chi1_v2_unc}). In deriving (\ref{Correl_Q1}) we have used the exact relation (\ref{Scaling_unc}).

Here, we have focused exclusively on the $u_y$-$u_y$ correlation. This is sufficient to calculate $\langle \bu(\br,t)\cdot\bu(0,\mathbf{0})\rangle$,
since the $u_x$ correlations -- related to the $u_y$-$u_y$ correlation by the incompressibility constraint -- are much smaller than the $u_y$-$u_y$ correlations in the dominant regime of wavevector $q_x\sim |q_y|^\zeta\ll |q_y|$ and can therefore be ignored.
	
\section{Equalization of the  exponents $z$ and $\zeta$.}\label{equal}
As we discovered in section \eqref{lin}, the linear theory leads to two different dynamic exponents $z_{1,2}$, and two different  anisotropy exponents $\zeta_{1,2}$ for the annealed fluctuations, as defined by the expression
\beq
\omega(\vec{q})= |q_y|^{z_1} f_{\rm real}\left({q_x\over |q_y|^{\zeta_1}}\right) -\ii|q_y|^{z_2}  f_{\rm imaginary}\left({|q_x|\over |q_y|^{\zeta_2}}\right)
\label{omegascalelin2}
\eeq
Furthermore, the anisotropy exponent $\zeta_2$  differs from that for the quenched fluctuations. However, the DRG analysis presented in section (\ref{NL}) identified a unique $\zeta$ and $z$. The reason for this is that once nonlinearities are taken into account, all three anisotropy exponents, and both $z$'s, become equal. We will demonstrate this explicitly in this section and discuss how this equalization comes about as a function of dimensionality. This is analogous to the situation in smectics.
	The linear hydrodynamic theory of smectics A predicts a ``second sound'' mode \cite{MPP}, which has the dispersion relation
	\beq
	\omega=\pm c_2(\phi) q - \ii \nu(\phi) q^2\,,
	\label{smecomegalin}
	\eeq
	where $\phi$ is the angle between the wavevector and the layer normal, $c_2(\phi)$ is a direction-dependent sound speed, and $\nu(\phi)$ a direction-dependent viscosity. This has different $z$'s for the real and the imaginary part: $z=1$ for the real part, and $z=2$  for the imaginary part.
	
	However, in three dimensions, once non-linearities are taken into account, one finds \cite{MRT}
	\beq
	\omega=\pm c_2(\phi) q - \ii \nu_{\rm renorm}(q, \phi) q^2\,,
	\label{smecomeganonlin}
	\eeq
	with
	\beq
	\nu_{\rm renorm}(q, \phi)={f(\phi)\over q}
	\label{smecetanonlin}
	\eeq
	which, when inserted into (\ref{smecomeganonlin}) gives
	\beq
	\omega=\pm c_2(\phi) q - \ii f(\phi) q\,,
	\label{smecomeganonlin}
	\eeq
	which has the same value of $z$ (namely, $z=1$) for both the real and the imaginary part.
	
How does the smectic go from having two $z$'s that differ by an amount of $O(1)$ in high dimensions (where linear theory works), to equality in $d=3$? This is because $\nu$ becomes anomalous in a higher dimension than $c_2$; the critical dimension for $\nu$ is \cite{MRT} $d_c^{\nu}=5$, while the critical dimension for $c_2$ (which is proportional to $\sqrt{B}$, where $B$ is the smectic layer compression modulus) is \cite{GP} $d_c^{c_2}=3$. And it's between these two critical dimensions that $z_2$ {\it continuously} evolves from $z_2=2$ to $z_2=1$. In general spatial dimensions between $d=5$ and $d=3$,
	\beq
	\nu_{\rm renorm}(q, \theta)\propto q^{(d-5)/2} \,,
	\label{smecetanonlinarbd}
	\eeq
	which, when inserted into (\ref{smecomeganonlin}), gives
	\beq
	{\rm Im} (\omega)\propto q^{(d-1)/2} \,,
	\label{smecomeganonlinarbd}
	\eeq
	which implies
	\beq
	z_2={d-1\over2} \,.
	\label{z2arbd}
	\eeq
This interpolates continuously between the linear value $z_2=2$ and the $d=3$ nonlinear value $z=1$ as one lowers the dimension from $5$ to $3$.

We now show that an analogous variation of $\mu_x$, which plays the role of viscosity in our model, in dimensions higher than the critical dimension of our model is responsible for the equalization of the exponents. However, unlike the smectic, the incompressible flock model we have described  here is strictly only valid in two dimensions. Nevertheless, we formally analytically continue the model, as in section (\ref{NL}), to higher dimension to understand how the multi $z$ and $\zeta$ linear scaling behavior reduces to one with a unique $z$ and $\zeta$ once nonlinearities are accounted for. Of course, we can use either hard continuation or soft continuation for this examination. Here, we choose to use the former. The critical dimension of the incompressible flock with the hard continuation extension is $7/3$. We first demonstrate that for $d< 7/3$, there is indeed a unique $z$ and $\zeta$ and then show that this comes about because $\mu_x$ starts changing from its linear value for $d<3$. An analogous argument can be constructed for soft continuation as well{, where there is a unique $z$ and $\zeta$ for $d<5/2$ which comes about because $\mu_x$ starts changing from its linear value for $d<7/2$}, but we will not present  it here.

\vspace{.2in}

\subsection{Incompressible flocks in $d<7/3$ (hard continuation)}
We begin by recalling that the result of the DRG analysis is that all linearized expressions become valid for the non-linear problem {\it provided} that we replace all of the constant phenomenological parameters $\gamma$, $\alpha$, $\mu_x$, and $D_{A,Q}$ with the wavevector dependent renormalized quantities $\gamma(\bq)$, $\alpha(\bq)$, $\mu_x(\bq)$, and $D_{_{A,Q}}(\bq)$ given by equations \eqref{anomparam}. Doing this with our expression \eqref{omegascalelin2} for the eigenfrequencies gives
\bew
\beqn
\omega(\bq)&=&\gamma(\bq) q_x -\ii \bigg(\mu_x(\bq) q_x^2+\alpha(\bq) {|\bqh|^2\over q_x^2}\bigg) \nonumber\\
&=& \gamma_0 |\bqh|^{-\eta_\gamma}f_\gamma\left(|q_x|\over |\bqh|^{\zeta}\right)q_x-\ii\bigg\{\mu_{x0} |\bqh|^{-\eta_\mu}f_\mu\left(|q_x|\over |\bqh|^{\zeta}\right) q_x^2+\alpha_0 |\bqh|^{-\eta_\alpha}f_\alpha\left(|q_x|\over |\bqh|^{\zeta}\right) {|\bqh|^2\over |q_x|^2}\bigg\} \,,
\label{omegarenorm}
\eeqn
 where we have extended  the $y$ direction to be $d-1$ dimensional and replaced $q_y$ with $\bq_h$.
The real part of this can be written
\begin{equation}
{\rm Re}(\omega)=\gamma_0 |\bqh|^{-\eta_\gamma}f_\gamma\left(|q_x|\over |\bqh|^{\zeta}\right)q_x
=\gamma_0 |\bqh|^{z_1}{q_x\over |\bqh|^{z_1+\eta_\gamma}}f_\gamma\left(|q_x|\over |\bqh|^{\zeta}\right) \,,
\label{omegareal}
\end{equation}
\ew
where the second equality is true for any value of $z_1$. However, if we choose $z_1$ such that
\beq
z_1+\eta_\gamma=\zeta\label{z1.2}
\eeq
then the factor ${q_x\over |\bqh|^{z_1+\eta_\gamma}}={q_x\over |\bqh|^{\zeta}}$, and the entire factor  ${q_x\over |\bqh|^{z_1+\eta_\gamma}}f_\gamma\left(|q_x|\over |\bqh|^{\zeta}\right)={q_x\over |\bqh|^{\zeta}}f_\gamma\left(|q_x|\over |\bqh|^{\zeta}\right)$, which is obviously a function only of the scaling ratio ${q_x\over |\bqh|^{\zeta}}$. Hence, we can define a new scaling function
\beq
f_{\rm real}(m)\equiv \gamma_0mf_\gamma\left(m\right),
\eeq
and rewrite equation \eqref{omegareal} as
\beq
{\rm Re}(\omega)= |\bqh|^{z_1} f_{\rm real}\left({q_x\over |\bqh|^{\zeta_1}}\right) \,,
\label{omegareal}
\eeq
with $\zeta_1=\zeta$,
{\it provided} that (\ref{z1.2}) is satisfied.
Using our earlier expression \eqref{Zeta1_unc} for $\zeta$ in terms of $\eta_\alpha$ and $\eta_\gamma$, and solving (\ref{z1.2}) for $z_1$, we find
\beq
z_1=\zeta-\eta_\gamma={2+\eta_\gamma-\eta_\alpha\over3}-\eta_\gamma
=z \,,
\label{z_1.2}
\eeq
where in the last equality we have used the result \eqref{Zeta1_unc} for $z$ found earlier.

Thus we have established that the real part of $\omega$ can be written in the form
\beq
{\rm Re} [\omega(\bq)]= |\bqh|^{z} f_{\rm real}\left({q_x\over |\bqh|^{\zeta}}\right) \label{Reomegascalelin2}
\eeq
with $z$ and $\zeta$ the exponents we found in our DRG analysis of section \eqref{NL}, whose values are given by \eqref{Zeta1_unc}.

We will now show that the imaginary part of $\omega$ can also be written in this scaling form, with its dynamic and anisotropy exponents $z_2$ and $\zeta_2$ respectively also being given by $z_2=z$ and $\zeta_2=\zeta$.

The proof is quite straightforward. The imaginary part of $\omega(\bq)$ can, by factoring out $|\bqh|^z$ and reorganizing \eqref{omegarenorm} a bit, be rewritten as
\beqn
\nonumber
{\rm Im}[\omega(\bq)]&=&-|\bqh|^z\bigg\{\mu_{x0}q_x^2 |\bqh|^{-\eta_\mu-z}f_\mu\left(|q_x|\over |\bqh|^{\zeta}\right)
\\
&&+ \alpha_0 \left({|\bqh|^{2-\eta_\alpha-z}\over q_x^2}\right)f_\alpha\left(|q_x|\over |\bqh|^{\zeta}\right) \bigg\}  \,,
\label{Imomega}
\eeqn
Using the relation \eqref{Zeta1_unc} between $z$, $\zeta$,  and $\eta_{\alpha,\gamma}$, and the exact scaling relation
\eqref{etamuexact} between $\eta_\mu$ and $\eta_{\alpha,\gamma}$, we see that
\begin{subequations}
\begin{align}
\nonumber
\eta_\mu+z&=\left({2+4\eta_\gamma-\eta_\alpha\over3}\right)+\left({2-2\eta_{\gamma}-\eta_{\alpha}\over 3}\right)
\\
&={4+2\eta_{\gamma}-2\eta_{\alpha}\over 3}=2\zeta \,, 
\label{mueta+z}
\\
\nonumber
2-z-\eta_\alpha&=2-{2-2\eta_{\gamma}-\eta_{\alpha}\over 3}-\eta_\alpha\\
&={4+2\eta_{\gamma}-2\eta_{\alpha}\over 3}=2\zeta \,.
\label{other exp}
\end{align}
\end{subequations}
With these results in hand, we can rewrite \eqref{Imomega} as
\begin{subequations}
\begin{align}
\nonumber
{\rm Im}[\omega(\bq)]=&-|\bqh|^z\bigg\{\left({q_x\over |\bqh|^{\zeta}}\right)^2\mu_{x0}f_\mu\left({|q_x|\over |\bqh|^{\zeta}}\right)
\\
&+ \left({|\bqh|^{\zeta}\over q_x}\right)^2\alpha_0f_\alpha\left({|q_x|\over |\bqh|^{\zeta}}\right) \bigg\}
\\
=&|\bqh|^zf_{\rm{Im}}\left({|q_x|\over |\bqh|^{\zeta}}\right)  \,,
\label{Imomegascale}
\end{align}
\end{subequations}
where we've defined
\beq
f_{\rm{Im}}(m)\equiv m^2f_\mu(m)\mu_{x0} +m^{-2}f_\alpha(m)\alpha_0 \,.
\label{fimdef}
\eeq

So there {\it is} only one $z$ and $\zeta$ for this problem in the non-linear theory, despite the fact that there are {\it two} $\zeta$'s and $z$'s for the linear theory.

\subsection{Incompressible flocks for $7/3<d<3$ (hard continuation)}
The reader might reasonably wonder how we get from the linear result, with its different values of $z$ and $\zeta$  for the real and imaginary parts of $\omega$, to this result in the nonlinear theory of a single $z$ and a single $\zeta$ for both parts. The answer is that, as we come down in dimension from $d=3$ to $d=7/3$,  the exponents $z_2$ and $\zeta_2$  for the imaginary part of $\omega$ evolve continuously from their linear values  of
$z_{2\rm{lin}}=1$ and $\zeta_{2{\rm lin}}=1/2$ to become equal to the values $z_{1\rm{lin}}=\zeta_{1{\rm lin}}=2/3$ in $d=7/3$. At this point, $z_2$ locks onto $z_1$, and $\zeta_2$ locks onto $\zeta_1$ as the dimension is lowered.

To see this, first note, as is clear from the recursion relation \eqref{Flow_g_mu} for $\gha_\mu$, that $\mu_x$ becomes anomalous {\it not} in $d=7/3$, as $\alpha$ and $\gamma$ do, but, rather, in $d=3$. For $7/3<d<3$, the nonlinear coupling $\gha$ flows to zero, so the scaling relation \eqref{Exact_Relation1}, which was derived by assuming that $\gha$ renormalized to a {\it non-zero} fixed point value, no longer holds. However, because $\mu_x$ grows upon renormalization if its bare value is small (again, as shown by the recursion relation \eqref{Flow_g_mu}, its fixed point value must be {\it non-zero}. Hence, the exact scaling relation \eqref{Exact_Relation2} {\it does} hold, even for $7/3<d<3$. Furthermore, because $\gha\to0$ upon renormalization for $7/3<d<3$, the exponents $\eta_{\alpha, \gamma, _Q}=0$ for $7/3<d<3$. Hence, the scaling relation \eqref{Exact_Relation2} implies that
\beq
\eta_\mu=3-d
\label{etamu intermed}
\eeq
 while $\eta_{\alpha, \gamma, _Q}=0$.

 Hence we can write
\beq
\omega(\bq)=\gamma_0 q_x -\ii \bigg(\mu_x(\bq) q_x^2+\alpha_0{|\bqh|^2\over q_x^2}\bigg)
\label{omegarenorm7/3<d<3}
\eeq
with
\beq
\mu_x(\bq)=\mu_{x0}|\bqh|^{d-3}f_\mu\left({|q_x|\over |\bqh|^{\zeta_2}}\right) \,,
\label{muscale7/3<d<3}
\eeq
where the anisotropy exponent $\zeta_2$ remains to be determined.

\begin{figure}
	\begin{center}
\includegraphics[scale=.14]{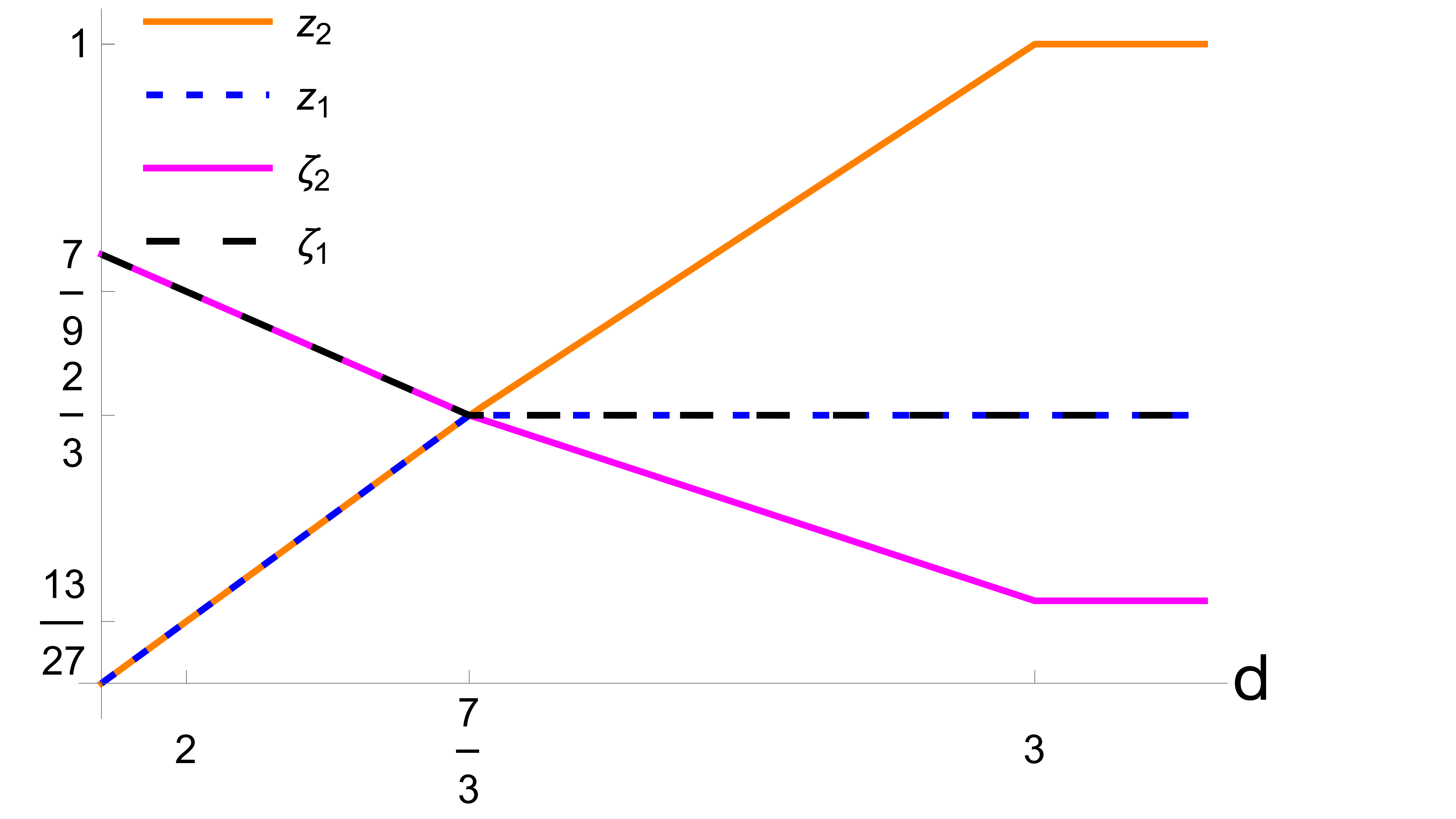}
	\end{center}
\caption{Plot of the exponents $z_{1,2}$ and $\zeta_{1,2}$ versus dimension {in the hard continuation to $O(\epsilon={7\over 3}-d$)}. The exponents $z_2$ and $\zeta_2$ become anomalous (i.e., depart from the values $z_2=1$ and
$\zeta_2=2/3$ below $d=3$. The two $z$'s and the two $\zeta$'s lock onto equality below $d=7/3$. {At $d=3$, $z_{1,2}=\zeta_{1,2}={2\over 3}$; at $d=2$, $z_{1,2}={13\over 27}$, $\zeta_{1,2}={7\over 9}$.} There are small slope discontinuities in $z_2(d)$ and $\zeta_2(d)$ at $d=7/3$; the slope of $z_2$ changes from $1/2$ to $5/9$, while that of $\zeta_2$ changes from $-1/4$ to $-1/3$. Both changes are so small as to be invisible to the naked eye (at least to the aging naked eye of the oldest author!), but are present in this plot.}
	\label{z2}
\end{figure}

We can do so by factoring $|\bq_h|^{z_2}$ out of the imaginary part of $\omega$ in \eqref{omegarenorm7/3<d<3}:
\beqn
\nonumber
{\rm Im}[\omega(\bq)]&=& -|\bqh|^{z_{2}}\bigg[\mu_{x0} q_x^2|\bqh|^{d-3-z_2}f_\mu\left({|q_x|\over |\bqh|^{\zeta_2}}\right)
\\
&&+\alpha_0{|\bqh|^{2-z_2}\over q_x^2}\bigg] \,,
\label{Imomegarenorm7/3<d<3}
\eeqn
where $z_2$ also remains to be determined. We can do so by requiring that \eqref{Imomegarenorm7/3<d<3} take the form
\beqn
{\rm Im}[\omega(\bq)]=|\bqh|^{z_2}f_{\rm{Im}}\left({|q_x|\over |\bqh|^{\zeta_2}}\right)  \,.
\label{Imomegascale7/3<d<3}
\eeqn
Comparing this with
\eqref{Imomegarenorm7/3<d<3}, we see that to make these two forms equal, we must have
\beq
d-3-z_2=-2\zeta_2
\label{cond1}
\eeq
and
\beq
2-z_2=2\zeta_2 \,.
\label{cond2}
\eeq
The simultaneous solution of these two equations is trivially found to be
\beq
\zeta_2= {5-d\over4} \sep z_2={d-1\over2}\,.
\label{zeta 7/3<d<3}
\eeq
Note that, as $d$ is decreased between $3$ and $7/3$, this smoothly interpolates between the values in the linearized  theory for the annealed anisotropy and dynamic exponents $\zeta_2={1\over2}$ and $z_2=1$  in $d=3$, and the values  $\zeta_2=z_2=2/3$ in $d={7/3}$. Once we go below $d=7/3$, the analysis given earlier in this section applies, so the quenched and  annealed anisotropy exponents remain locked together, and both evolve away from the value in the linearized  theory, as described by the $7/3-\epsilon$-expansion. The exponents $z_{1,2}$ and $\zeta_{1,2}$ are plotted as functions of dimension in Fig. \ref{z2}.

\section{Summary \& Outlook}\label{sum}
{In this article, w}e have examined the effects of quenched disorder {on the moving phase of an} incompressible polar active fluid in 2D. {We show that, surprisingly, the polarised phase retains  long-rang order in 2D \emph{even} in the presence of quenched disorder. This is all the more surprising since {
the closest equilibrium analogs of our system, which are}
equilibrium magnets {constrained to have a divergenceless magnetization}, whose long-time, large-scale properties are \emph{exactly} equivalent to incompressible flocks \cite{chen_natcomm16} in the presence of only \emph{annealed} disorder, are \emph{much} more susceptible to quenched disorder and do not form an ordered phase in 2D. }

We {have also} characterise{d} the hydrodynamic properties of the polarised phase, uncovering a novel universality class and calculating the exponents using three distinct DRG analyses. Since the values obtained from all three methods are very close to each other, we expect the{m} to be quantitatively {accurate}. {Our results should be readily testable in agent-based simulations. Further, since quenched disorder is inevitable in all experiments, our results may also be relevant for interpreting experiments on motile cell layers.  }

 {Our work suggests that novel physics may also emerge from the incompressibility constraint in}
 active polar suspensions, which are  a two-component (swimmers and solvent) system that is only incompressible as a whole.

\onecolumngrid
\appendix

\section{Calculating the graphical corrections to the various coefficients \label{App:A}}
In this appendix, we explicitly calculate the corrections to the various coefficients in the EOM for $u_y$ upon averaging over the  short-distance fluctuations.

The hydrodynamic EOM, retaining only relevant terms, is 
\beq
\pp_t u_y(\bq)=P_{yx}\left(\bq\right)\mathcal{F}_{\bq}
\left[-\alpha\left(u_x+{u_y^2\over 2}\right)\right]+\mathcal{F}_{\bq}
\left[-\gamma\pp_xu_y+\mu_x\pp_x^2u_y-{\alpha}
\left(u_x+{u_y^2\over 2}\right)u_y+f_{_A}^y+f_{_Q}^y\right]\,,
\label{EOMA1}
\eeq
which can be re-written as:
\beqn
u_y(\tilde{\bq})&=&
G(\tilde{\bq}) \left[
f_{_A}^y(\tilde{\bq})+f_{_Q}^y(\tilde{\bq})
+\left(\frac{\alpha}{2}\right) P_{yx} (\bq) \int_{\tilde{\bk}}
u_y(\tilde{\bk})u_y(\tilde{\bq}-\tilde{\bk})
-\alpha
\int_{\tilde{\bk}}
u_y(\tilde{\bk})u_x(\tilde{\bq}-\tilde{\bk})\right.\nonumber
\\
&&
\left.-\left(\frac{\alpha}{2}\right)
\int_{\tilde{\bk},\tilde{\bk}'}
u_y(\tilde{\bk})u_y(\tilde{\bk}')u_y(\tilde{\bq}-\tilde{\bk}-\tilde{\bk}')\right]
\label{EOMA2}
\eeqn
with the bare propagator
\beq
G(\tilde{\bq})= \left[ -\ii\left(\omega-\gamma q_x\right)
+\left(\alpha {q_y^2\over q^2}+\mu_xq_x^2\right)\right]^{-1}={\ii\left(\omega-\gamma q_x\right)
+\left(\alpha {q_y^2\over q^2}+\mu_xq_x^2\right)\over\left(\omega-\gamma q_x\right)^2
+\left(\alpha {q_y^2\over q^2}+\mu_xq_x^2\right)^2}
\ .
\label{eq:G0}
\eeq

We remind the reader that the bare correlation functions  (i.e.,  those of the linear theory) are given by
\beqn
\label{eq:corre_uu_A}
\langle u_i(\tilde{\bq})u_j(\tilde{\bq}')\rangle=
 C^{ij}_{_{A}}(\tilde{\bq})\delta(\omega+\omega')\delta(\bq+\bq')
+C^{ij}_{_{Q}}(\tilde{\bq})\delta(\omega)\delta(\omega')\delta(\bq+\bq')\,,~~~
\eeqn
where
\begin{subequations}
\label{eq:Ca&Cq app}
\begin{align}
\label{2D_linear_Axx_A}
C_{_{A}}^{xx}(\tilde{\bq})&=
{q_{y}^2\over q^2}C_{_A}(\tbq)
\,,\\
\label{2D_linear_Axy_A}
C_{_{A}}^{xy}(\tilde{\bq})&=-{q_{x}q_y\over q^2}C_{_A}(\tbq)=C_{_{A}}^{yx}(\tilde{\bq})
\,,\\
\label{2D_linear_Ayy_A}
C_{_{A}}^{yy}(\tilde{\bq})&={q_x^2\over q^2}C_{_A}(\tbq)
\,,\\
\label{2D_linear_Qxx_A}
C_{_{Q}}^{xx}(\tilde{\bq})&={q_{y}^2\over q^2}C_{_Q}(\tbq)
\,, \\
\label{2D_linear_Qxy_A}
C_{_{Q}}^{xy}(\tilde{\bq})&=-{q_{x}q_y\over q^2}C_{_Q}(\tbq)=C_{_{Q}}^{yx}(\tilde{\bq})
\, ,\\
\label{2D_linear_Qyy_A}
C_{_{Q}}^{yy}(\tilde{\bq})&={q_{x}^2\over q^2}C_{_Q}(\tbq)
\,, \\
\end{align}
\end{subequations}
with
\beq
C_{_Q}(\tilde{\bq}) = \frac{ 4 \pi D_Q q_x^4 }{ \gamma^2 q_x^6
+\alpha^2 q_y^4}
\ ,~~~~~
C_{_A}(\tilde{\bq}) = {2 D_{_A}\over (\omega-\gamma q_x)^2+\left(\alpha{q_y^2\over q^2}+\mu_xq_x^2\right)^2} \,.
\eeq
Here the subscripts $A$ and $Q$ denote the contributions from the annealed and quenched noises, respectively.

Note that we can think of (\ref{EOMA2})
as a closed equation for $u_y(\tilde{\bq})$ with $u_x(\tilde{\bq})$ simply  a shorthand notation for $-{q_y\over q_x}u_y(\tilde{\bq})$, which follows from the incompressibility condition.

\begin{figure}
	\begin{center}
\includegraphics[scale=.5]{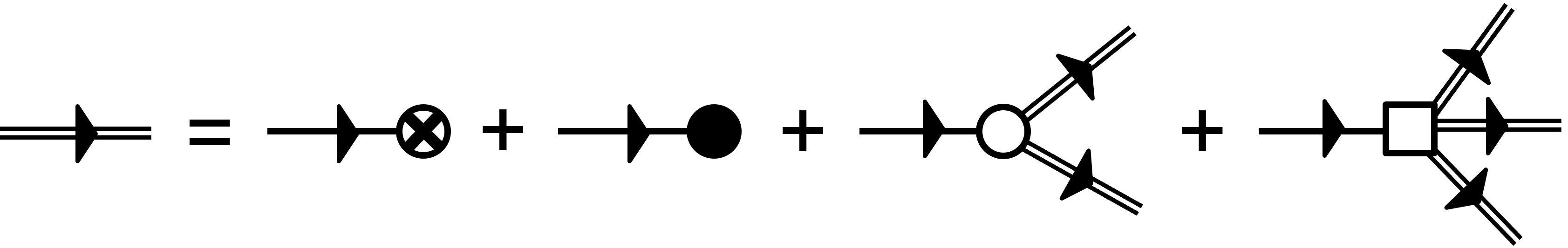}
	\end{center}
\caption{Feynman diagram representation of the formal solution (\ref{EOMA2}) of (\ref{EOMA1}). The circle with an interior cross represents the quenched noise, while the solid circle represents the annealed noise. The meaning of the various
other elements in this figure are given in  figure (\ref{Fig_Rules}).}
	\label{A1}
\end{figure}

\begin{figure}
	\begin{center}
\includegraphics[scale=.5]{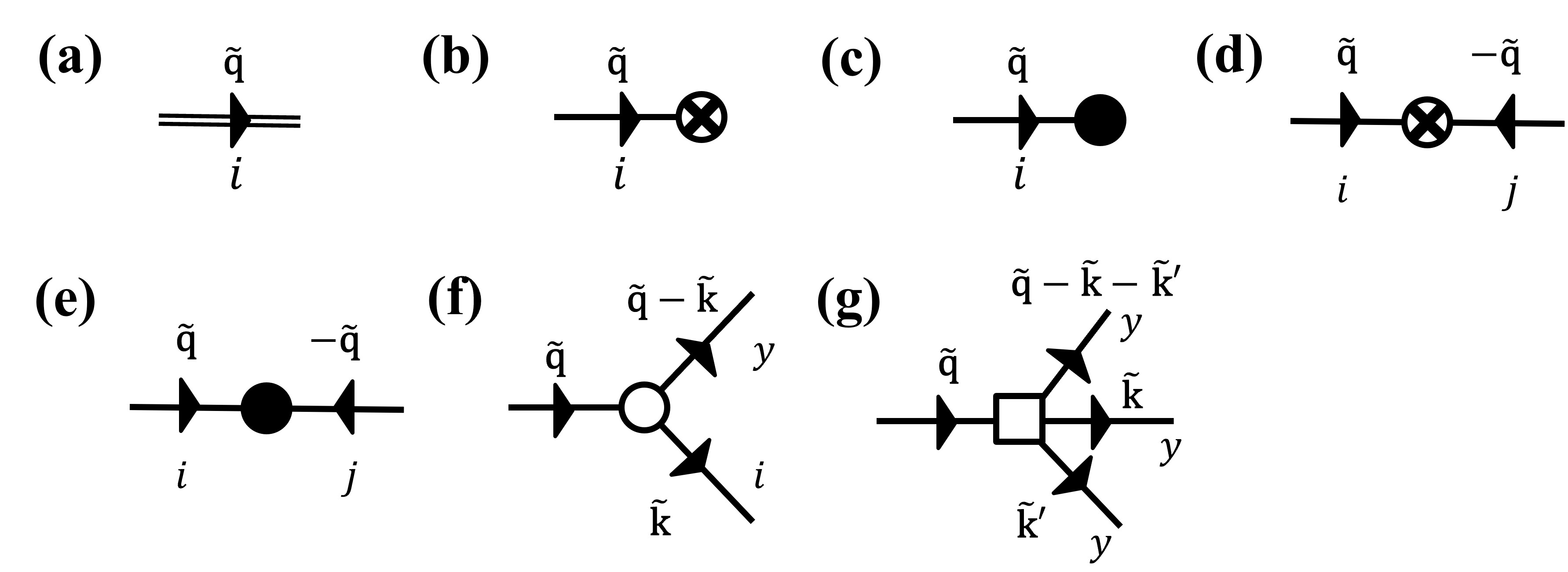}
	\end{center}
\caption{
	{\AM Definitions of the} elements in the Feynman diagrams:
(a) $=u_i(\tilde{\bq})$, (b) $={q_y\left(1-2\delta_{ix}\right)\over q_i}G\left(\tilde{\bq}\right) f_{_Q}^i(\tilde{\bq})$, (c) $={q_y\left(1-2\delta_{ix}\right)\over q_i}G\left(\tilde{\bq}\right) f_A^i(\tilde{\bq})$,
(d) $=
C_{_Q}^{ij}(\tilde{\bq})\delta(\omega)$,
(e) $=
C_{_A}^{ij}(\tilde{\bq})$,
(f) ${\rm{the~circle}}=-\alpha\left(1-{1\over 2}\delta_{yi}\right)\left[\delta_{xi}
+P_{yx}\left(\bq\right)\delta_{yi}\right]$,
(g) the square $= -\alpha/2$.
}
	\label{Fig_Rules}
\end{figure}

As usual (see, e.g., \cite{forster_prl76, forster_pra77}), our formal solution (\ref{EOMA2}) can be represented by Feynman graphs, as illustrated in figures \ref{A1}. The definition of the various pictorial elements in this diagram are given in figure \ref{Fig_Rules}.

\begin{figure}
	\begin{center}
\includegraphics[scale=.5]{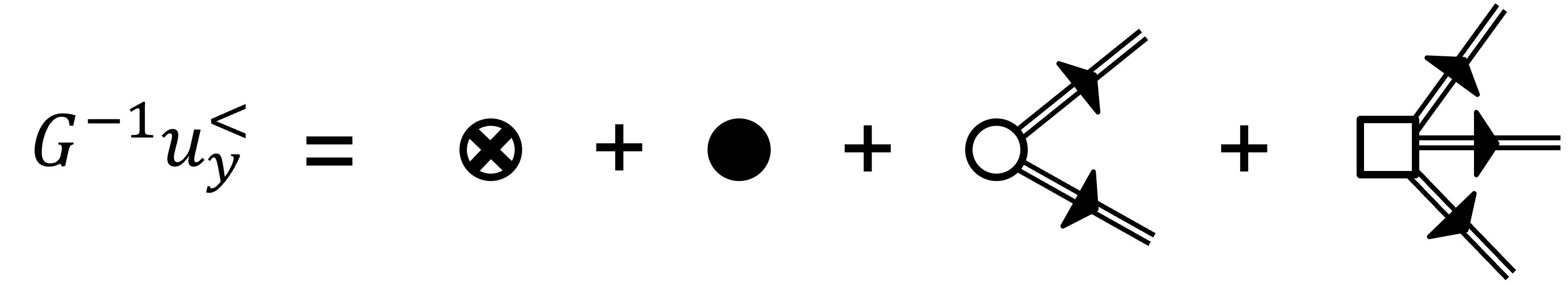}
	\end{center}
\caption{Feynman diagram representation of the  EOM (\ref{EOMA3}). This amounts to ``amputating" the leftmost leg of each of the Feynman diagrams in figure \ref{A1}.}
	\label{A2}
\end{figure}

The most useful way to derive the DRG recursion relations for the various parameters is to divide EOM (\ref{EOMA2}) by $G(\tbq)$, which gives
\beqn
G^{-1}(\tilde{\bq})u_y(\tilde{\bq})&=&
\left[
f_{_A}^y(\tilde{\bq})+f_{_Q}^y(\tilde{\bq})
+\left(\frac{\alpha}{2}\right) P_{yx} (\bq) \int_{\tilde{\bk}}
u_y(\tilde{\bk})u_y(\tilde{\bq}-\tilde{\bk})
-\alpha
\int_{\tilde{\bk}}
u_y(\tilde{\bk})u_x(\tilde{\bq}-\tilde{\bk})\right.\nonumber
\\
&&
\left.-\left(\frac{\alpha}{2}\right)
\int_{\tilde{\bk},\tilde{\bk}'}
u_y(\tilde{\bk})u_y(\tilde{\bk}')u_y(\tilde{\bq}-\tilde{\bk}-\tilde{\bk}')\right]\,.
\label{EOMA3}
\eeqn
Graphically, this amounts to ``amputating" the leftmost leg of each of the Feynman diagrams in Fig. \ref{A1}, which gives  Fig. \ref{A2}.

 Now we decompose $u_y(\tbq)$ into ``slow"  components $u_y^<(\tbq)$ and ``fast"  components $u_y^>(\tbq)$,  where $u_y^<(\tbq)$ is supported in the wave vector space $|q_x|<\infty$, $ |q_y|
<\Lambda e^{-\dd\ell}$, and $u_y^>(\tbq)$ in the ``momentum shell'' $| q_x|<\infty$, $\Lambda e^{-\dd\ell}<|q_y|<\Lambda$, where $\Lambda$ is the ultraviolet cutoff, and $d\ell\ll1$ is an arbitrary rescaling factor. We likewise decompose the noises $f^y_{_{A,Q}}$ into fast and slow components $f^{y>}_{_{A,Q}}$ and $f^{y<}_{_{A,Q}}$ respectively.
	
\begin{figure}
	\begin{center}
\includegraphics[scale=.5]{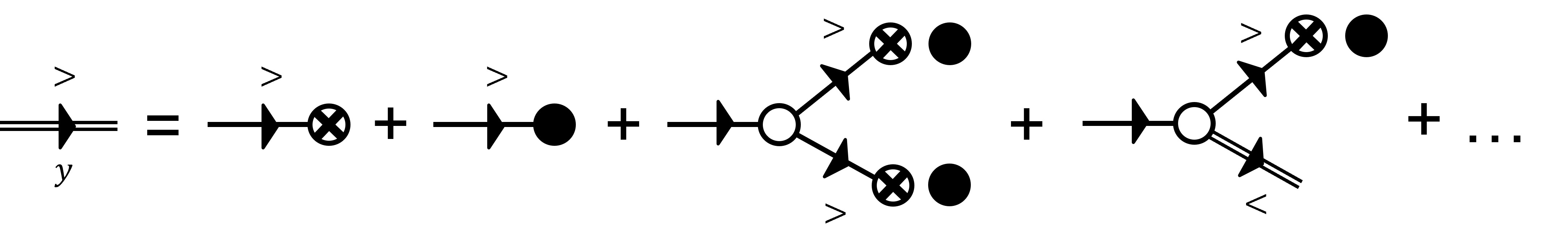}
	\end{center}
\caption{
A diagrammatic expansion of the ``fast" component of $u_y(\tbq)$ after partitioning $u_y(\tbq)$ into ``slow"  components $u_y^<(\tbq)$ and ``fast"  components $u_y^>(\tbq)$.
}
	\label{A3}
\end{figure}

The Feynman graphs are useful for the next step,
which is to solve (\ref{EOMA2}) perturbatively for $u_y^>(\tbq)$ in terms of $u_y^<(\tbq)$  and the noises  $f^{y>}_{_{A,Q}}$. This perturbative solution can be represented by the  graphs
shown in Fig. \ref{A3}.

\begin{figure}
	\begin{center}
\includegraphics[scale=.5]{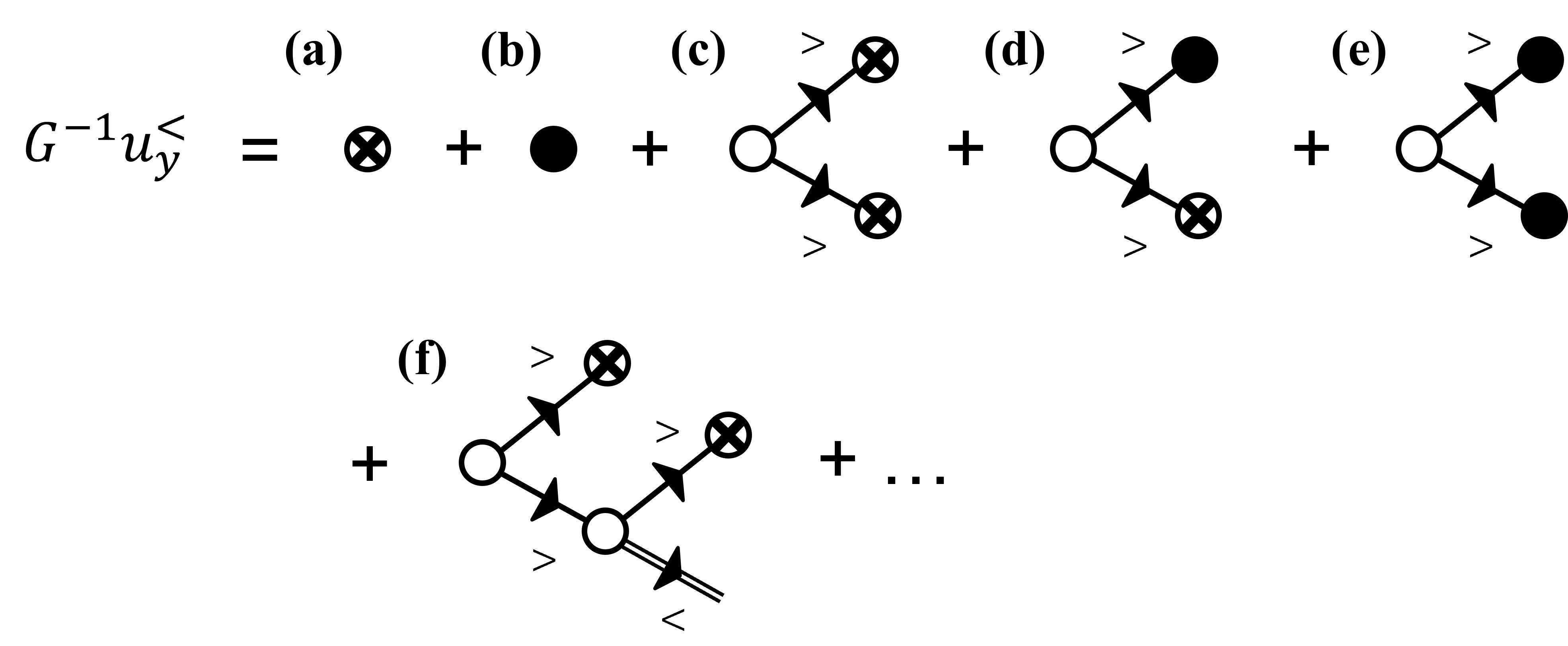}
	\end{center}
\caption{A diagrammatic expansion of $G^{-1}(\tbq)u_y^<(\tbq)$ after partitioning $u_y(\tbq)$ into ``slow"  components $u_y^<(\tbq)$ and ``fast"  components $u_y^>(\tbq)$.
}
	\label{A4}
\end{figure}

We next substitute this solution into the EOM (\ref{EOMA3}) for $u_y^<(\tbq)$; that is, the EOM represented in Fig. \ref{A2}, with the field on the left-hand side taken to be  $u_y^<(\tbq)$. This leads to the series of graphs shown in Fig. \ref{A4}.

Once this is done, one can see at once that the graphs like Fig. \ref{A4}(c), (d), and (e), for which the incoming momentum $\bq$ on the left is  in the long-wavelength (small $\bq$) interior  with every line ending in an  short-wavelength noise $f^{y>}_{_{A,Q}}(\tbq)$, can be thought of as an extra contribution to the long-wavelength noises $f^{y<}_{_{A,Q}}(\tbq)$. Their correlations, therefore,
are renormalizations of the noise correlations, represented by the Feynman graphs shown in Figs.~\ref{fig:noise_DQ} and \ref{Fig_D_A}.

Next, we must average over the short-wavelength  noises $f^{y<}_{_{A,Q}}(\tbq)$. Since the averages of two noises are only
non-zero if their wavevectors and frequencies are equal and opposite,
we represent this averaging by connecting lines that end in noises. The result is components of the graphs shown in figure \ref{Fig_Rules}  (d) and (e), which simply represent the correlation functions.

Performing this connection for the graphs in Fig. \ref{A4} yields the graphs  in Figs. \ref{fig:propagator}.

Note that these graphs are {\it linear} in the slow fields $u^<_y(\tbq)$, as can be seen by the fact that they have precisely one external line coming off to the right. The terms they represent can therefore be pulled over to the left-hand side of the equation schematically represented by figure \ref{A4}, and absorbed into renormalizations of the inverse propagator $G^{-1}(\tbq)$. From the form of $G^{-1}(\tbq)$, we see that the part of such correction  terms proportional to $\ii q_x u_y(\tbq)$  can be absorbed into renormalizations of the linear coefficient $\gamma$. Likewise, terms proportional to  $\left({q_y\over q_x}\right)^2 u_y(\tbq)$ renormalize $\alpha$, and those proportional to $q_x^2 u_y(\tbq)$ renormalize $\mu_x$. This is how we will obtain the graphical corrections to those parameters.

This averaging process also generates graphs which are quadratic and cubic in the slow fields $u^<_y(\tbq)$. This leads to the renormalization of the coefficients of the quadratic and cubic terms. However, as argued in the main text, these coefficients are locked to the coefficient $\alpha$ of the linear term  $\left({q_y\over q_x}\right)^2 u_y(\tbq)$, or equivalently $\left({q_y\over q_x}\right) u_x(\tbq)$. This is because the EOM is rotation invariant, and this symmetry must be preserved under  the DRG transformation. In particular, this means the graphical corrections to these coefficients are all equal to each other. Therefore, in this problem we only need to consider the (one-loop) graphs renormalizing the linear terms and the noise, which doesn't involve the cubic vertex (i.e., the square-shaped knot) at all.

Unusually, in this problem we find that there is one additional correction, which we've not yet described,  that affects the renormalization of {\it all} of the parameters. This arises from the fact that the inverse propagator $G^{-1}(\tbq)$ also gets a contribution from the graphs proportional to $-\ii\omega$. That is, on the right hand side of Fig. \ref{A4}, we also get contributions proportional to $-\ii\omega u_y$. As a result, our final expression for the renormalized  EOM for the slow modes $u^<_y(\tbq)$ in Fourier space takes the form
\beq
 \left(-\ii\omega(1+\eta_\omega d\ell)-\ii\gamma(1+\eta_{\gamma}^{\rm direct}  \dd\ell) q_x
+\alpha(1+\eta_{\alpha}^{\rm direct} \dd\ell) {q_y^2\over q^2}+\mu_x(1+\eta_{\mu}^{\rm direct}  \dd\ell)q_x^2\right)u^<_y(\tbq)=f_{_Q}^{yR}+f_{_A}^{yR}+{\rm NL}\{u_y^<(\tbq)\} \,,
\label{renormedEOM1}
\eeq
where $\gamma\eta_{\gamma}^{\rm direct}\dd\ell$,   $\alpha\eta_{\alpha}^{\rm direct}\dd\ell$,   and $\mu_x\eta_{\mu}^{\rm direct}\dd\ell$ represent the ``direct" corrections to $\gamma$, $\alpha$, and $\mu_x$ calculated as described above,
and  $-\ii\omega\eta_\omega\dd\ell$ likewise represents the correction to the inverse propagator proportional to $-\ii\omega$ described above. The renormalized noises  $f_{_{Q,A}}^{yR}$ are calculated as described above.

The expression ${\rm NL}\{u_y^<(\tbq)\} $ represents the terms non-linear in the slow fields $u_y^<(\tbq)$; as in the original EOM, these will couple $u_y^<(\tbq)$ to the $u_y^<(\tbk)$'s at all $\tbk$ in the (slightly smaller) Brillouin zone.

The coefficient of $-\ii\omega$ in the original EOM was $1$. To make our renormalized EOM look as much like the original as possible, and to avoid having to introduce yet another parameter into our EOM (namely, a coefficient of the $-\ii\omega$ term that is not unity, but a free parameter that can also renormalize), we simply divide the EOM (\ref{renormedEOM1}) by the factor $1+\eta_\omega d\ell$, which gives, in the limit $d\ell\ll1$ in which we're working
\beq
 \left(-\ii\omega-\ii\gamma(1+\eta_{\gamma} \dd\ell) q_x
+\alpha(1+\eta_{\alpha}  \dd\ell) {q_y^2\over q^2}+\mu_x(1+\eta_{\mu}\dd\ell)q_x^2\right)u^<_y(\tbq)=\bigg(f_{_Q}^{yR}+f_{_A}^{yR}+{\rm NL}\{u_y^<(\tbq)\}\bigg)(1-\eta_\omega d\ell) \,,
\label{renormedEOM2}
\eeq
with
\beq
\eta_{\gamma}=\eta_{\gamma}^{\rm direct}-\eta_\omega \sep \eta_{\alpha}=\eta_{\alpha}^{\rm direct}-\eta_\omega \sep \eta_{\mu}=\eta_{\mu}^{\rm direct}-\eta_\omega  \,.
\label{eta_omega_effect_lin}
\eeq
The $\eta_{\gamma}$, $\eta_{\alpha}$, and $\eta_{\mu}$ defined above are those quoted in the main text, which give the graphical corrections to the renormalizations of $\gamma$, $\alpha$, and $\mu_x$. Indeed, defining the renormalized
values $\gamma^R$, $\alpha^R$, and $\mu_x^R$ to be
the coefficients of the appropriate terms in the renormalized EOM (\ref{renormedEOM2}), we have
\beq
\gamma^R=\gamma(1+\eta_{\gamma} \dd\ell)  \sep \alpha^R=\alpha(1+\eta_{\alpha}  \dd\ell)  \sep \mu_x^R=\mu_x(1+\eta_{\mu}\dd\ell) \,.
\label{lin R}
\eeq
Combining these graphical corrections with the effect of the rescalings leads to the recursion relations presented in the main text.

Defining  the effective fully renormalized noises $f_{_{Q,A}}^{yF}$ on the right hand side of this equation via:
\beq
f_{_{Q,A}}^{yF}\equiv f_{_{Q,A}}^{yR}(1-\eta_\omega \dd\ell) \,,
\label{frrdef}
\eeq
we see that these acquire
an extra factor of $(1+\eta_\omega d\ell)$ in addition to their corrections coming from the graphs in Figs \ref{fig:noise_DQ} and \ref{Fig_D_A}. Hence, their {\it correlations} pick up (for small $d\ell$) an extra factor of $(1-2\eta_\omega d\ell)$ (note the ``$2$"), in addition to the direct graphical correction from Figs \ref{fig:noise_DQ} and \ref{Fig_D_A}, whose calculation we described above. That is,
\begin{subequations}
\begin{align}
\langle f^{yF}_{_Q}(\br)f^{yF}_{_Q}(\br')\rangle&=
2D^{R}_{_Q}\delta^2(\br-\br')\,,\label{Random_QRR}\,\\
\langle f^{yF}_{_A}(\br,t)f^{yF}_{{_A}}(\br',t')\rangle&=
2D^{R}_{_A}\delta^2(\br-\br')\delta(t-t')\, ,\label{Random_ARR}
\end{align}
\end{subequations}
where the renormalized noise strengths are given by
\begin{subequations}
\begin{align}
D^{R}_{_{Q}}&=\bigg(1+(\eta_{_{Q}}^{\rm direct}-2\eta_\omega)\dd\ell\bigg)D_{_{Q}}\equiv\bigg(1+\eta_{_{Q}}\dd\ell\bigg)D_{_{Q}}\,,
\\
D^{R}_{_{A}}&=\bigg(1+(\eta_{_{A}}^{\rm direct}-2\eta_\omega)\dd\ell\bigg)D_{_{A}}\equiv\bigg(1+\eta_{_{A}}\dd\ell\bigg)D_{_{A}}
\,,
\label{noiseetaomega}
\end{align}
\end{subequations}
where we've defined
\beq
\eta_{_{Q}}\equiv\eta_{_{Q}}^{\rm direct}-2\eta_\omega \sep \eta_{_{A}}\equiv\eta_{_{A}}^{\rm direct}-2\eta_\omega  \,.
\label{eta_omega_effect_lin_noise}
\eeq
These $\eta_{_{Q,A}}$ are those used in the main text.

Having set up the logic of our approach to the perturbative portion of the DRG, all that remains to
actually do the hard work of evaluating the graphs. We'll do so now.

\subsection{Quenched noise ($D_{_Q}$) renormalization}
\begin{figure}
	\begin{center}
  \includegraphics[scale=.32]{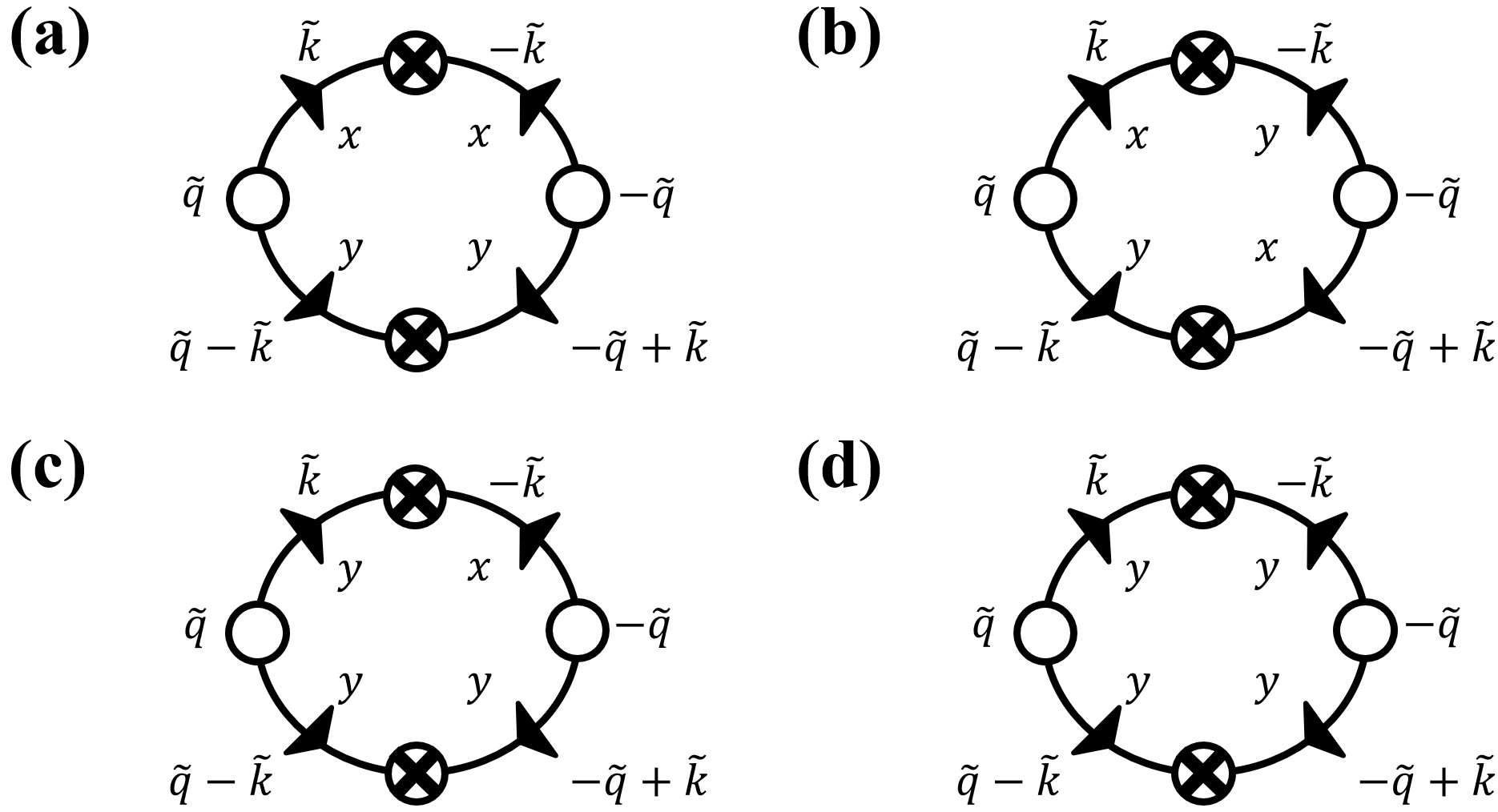}
	\end{center}
\caption{The graphical representations of the correction to $D_Q$.}
	\label{fig:noise_DQ}
\end{figure}

\begin{figure}
	\begin{center}
  \includegraphics[scale=.32]{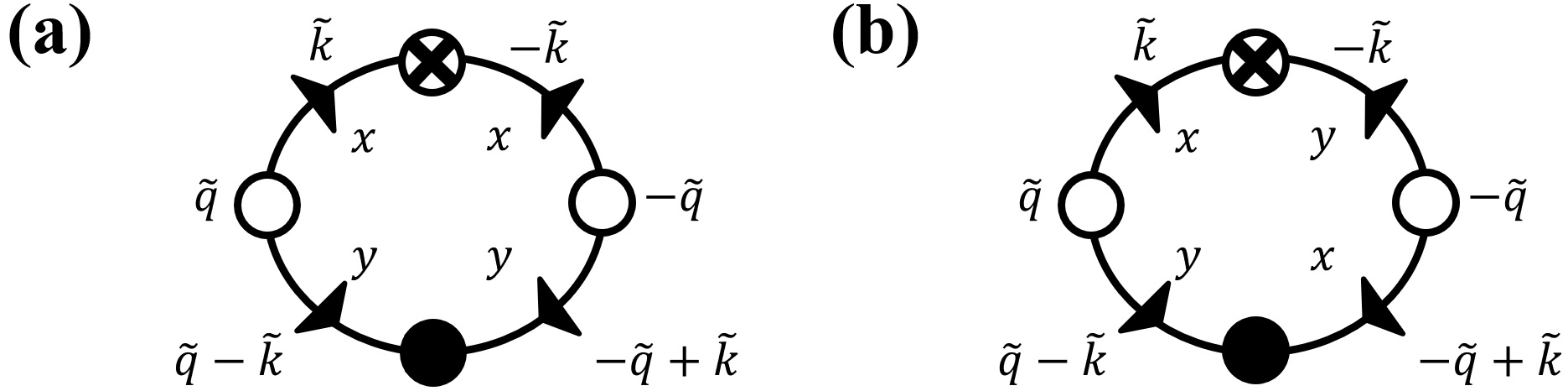}
	\end{center}
\caption{Two particular graphical corrections to $D_{_A}$. They correspond to the graphs (a) and (b) in \fig \ref{fig:noise_DQ} but with one of the quenched noise averages replaced by the annealed noise average.}
	\label{Fig_D_A}
\end{figure}

The graphs in Fig.~\ref{fig:noise_DQ} renormalize the correlations of the quenched noise $f^y_{_{Q}}$. For graphs (a) and (b)
we set the wave vector $\bq$ inside the loop integral to zero to keep only the relevant contributions to the noise correlations. These two graphs thus gives identical results, and the sum of them
gives a contribution to the correlations of the quenched noise:
\beqn
\delta \left[\langle f_{_{Q}}^y(\bq,\omega)f_{_{Q}}^y(-\bq,-\omega)\rangle\right]
&=&2\alpha^2
\int_> \frac{\dd^2 k}{(2 \pi)^2} \frac{\dd \Omega}{2\pi}
\frac{k_y^2}{k_x^2} C_{_Q}(\tilde{\bk}) C_{_Q}(-\tilde{\bk}) \delta(\Omega) \delta(\omega-\Omega)\nonumber
\\
&=&
16 \pi \alpha^2 D_{_Q}^2
\int_> \frac{\dd^2 k}{(2 \pi)^2}
\frac{k_x^6k_y^2}{\left( \gamma^2 k_x^6
+\alpha^2 k_y^4\right)^2}\nonumber
\\
&=&16 \pi \alpha^2 D_{_Q}^2
\delta (\omega)
I_1
\ ,\label{Del_uyy}
\eeqn
where we've defined the integral
\beq
I_1\equiv \int_{>} {\dd^2k\over (2\pi)^2}\,
{k_y^2k_x^6\over \left(\gamma^2 k_x^6+\alpha^2 k_y^4\right)^2}
\eeq
which is calculated in appendix (\ref{Sec:I1}).

Since $\langle f_{_Q}^y(\bq,\omega)f_{_Q}^y(-\bq,-\omega)\rangle=4\pi D_{_Q}
\delta (\omega)$, (\ref{Del_uyy}) implies a correction to the quenched noise strength $D_{_Q}$:
\beq
\label{EQAM:DQ_corr}
\delta D_{_Q}=4\alpha^2D_{_Q}^2I_1
\,.
\eeq

Graph (c) has a prefactor of $q_x^3q_y/q^4$ and graph (d) has a prefactor of $q_x^2q_y^2/q^4$. Since in the dominant wave vector regime $q_y\ll q_x$, these prefactors $\ll1$ and therefore, these graphs contribute corrections that are subdominant to the one in \eqref{EQAM:DQ_corr}.

Inserting the values of $I_1$ in \eqref{EQAM:DQ_corr} for the three different perturbation schemes, we get
\begin{subequations}
\begin{align}
\text{uncontrolled $d=2$:}~~~\delta D_{_Q}
&={2\over 9}\gun D_{_Q}\dd\ell\,,\label{A_DQ_U}\\
\text{hard continuation:}~~~\delta D_{_Q}
&={2\over 9}\gha D_{_Q}\dd\ell\,,\label{A_DQ_H}\\
\text{soft continuation:}~~~\delta D_{_Q}
&={1\over 3}\gso D_{_Q}\dd\ell\,,\label{A_DQ_S}
\end{align}
\end{subequations}
where $\gun= {D_{_Q}\over \pi}|\gamma|^{-{7\over 3}}\alpha^{1\over 3}\Lambda^{-{1\over 3}}$, $\gha={S_{d-1}\over (2\pi)^{d-1}}|\gamma|^{-{7\over 3}}\alpha^{1\over 3}\Lambda^{3d-7\over 3}D_{_Q}$, and $\gso= {S_{d-1}D_{_Q}\over \sqrt{2}(2\pi)^{d-1}}|\gamma|^{-{\left({d+5\over3}\right)}}\alpha^{\left({d-1\over 3}\right)}\Lambda^{\left({2d-5\over 3}\right)}$ as defined in the main text.

The results (\ref{A_DQ_U},\ref{A_DQ_H},\ref{A_DQ_S} ) imply a contribution to the exponent $\eta_{_Q}$,
which we call $\eta_{_Q}^{\rm direct}$ (see (\ref{noiseetaomega}) for its definition):
\begin{subequations}
\label{A_Eta_Q_Direct}
\begin{align}
\text{uncontrolled $d=2$:}~~~\eta_{_Q}^{\rm direct}
&={2\over 9}\gun \,,\label{A_Eta_Q_Direct_U}\\
\text{hard continuation:}~~~\eta_{_Q}^{\rm direct}
&={2\over 9}\gha \,,\label{A_Eta_Q_Direct_H}\\
\text{soft continuation:}~~~\eta_{_Q}^{\rm direct}
&={1\over 3}\gso \,.\label{A_Eta_Q_Direct_S}
\end{align}
\end{subequations}

%

\subsection{Annealed noise ($D_{_A}$) renormalization}
Like the correction to $D_{_Q}$, the graphical corrections to $D_{_A}$ consist of the first two diagrams in \fig \ref{fig:noise_DQ}, albeit with one of the quenched noise averages in each (indicated by a crossed circle) replaced by the annealed noise average (indicated by a black circle).
Hence there are four diagrams in total, two of which
are shown in \fig \ref{Fig_D_A}; the other two are simple permutations of these.

 Again setting  $\bq$ to zero inside the loop integrals, the contributions of all four diagrams are identical. The sum of them gives a contribution to the correlations of the annealed noise:
\beqn
\delta\left[\langle f_{_A}^y(\bq,\omega)f_{_A}^y(-\bq,-\omega)\rangle \right]
&=&4\alpha^2
\int_{>} {\dd^2k\over (2\pi)^2}\,
\left(k_y\over k_x\right)^2
\left[2D_{_A}\over \gamma^2 k_x^2+\left(\alpha{k_y^2\over k^2}+\mu_xk_x^2\right)^2\right]
\left(2 D_{_Q}k_x^4\over \gamma^2 k_x^6+\alpha^2 k_y^4\right)\nonumber\\
&\approx&16\alpha^2D_{_Q}D_{_A}
\int_{>} {\dd^2k\over (2\pi)^2}\,
{k_x^6k_y^2\over \left(\gamma^2 k_x^6+\alpha^2 k_y^4\right)^2}\nonumber\\
&=&16\alpha^2D_{_Q}D_{_A}
I_1
\,.\label{ia}
\eeqn

Since $\langle f_{_A}^y(\bq,\omega)f_{_A}^y(-\bq,-\omega)\rangle_{_A}=2D_{_A}
$, (\ref{ia}) implies a correction to the annealed noise strength $D_{_A}$:
\beq
\delta D_{_A}=8\alpha^2D_{_Q}D_{_A}I_1
\eeq
Again, using the values of $I_1$ for the various schemes yields
\begin{subequations}
\begin{align}
\text{uncontrolled $d=2$:}~~~\delta D_{_A}
&={4\over 9}\gun D_{_A}\dd\ell\,,
\\
\text{hard continuation:}~~~\delta D_{_A}
&={4\over 9}\gha D_{_A}\dd\ell\,,
\\
\text{soft continuation:}~~~\delta D_{_A}
&={2\over 3}\gso D_{_A}\dd\ell\,.
\end{align}
\end{subequations}
These imply
\begin{subequations}
\label{A_Eta_A_Direct}
\begin{align}
\text{[uncontrolled $d=2$:]}~~~\eta_{_A}^{\rm direct}
&={4\over 9}\gun \,,\label{A_Eta_A_Direct_U}\\
\text{hard continuation:}~~~\eta_{_A}^{\rm direct}
&={4\over 9}\gha \,,\label{A_Eta_A_Direct_H}\\
\text{soft continuation:}~~~\eta_{_A}^{\rm direct}
&={2\over 3}\gso \,.\label{A_Eta_A_Direct_S}
\end{align}
\end{subequations}

\begin{figure}
	\begin{center}
  \includegraphics[scale=.3]{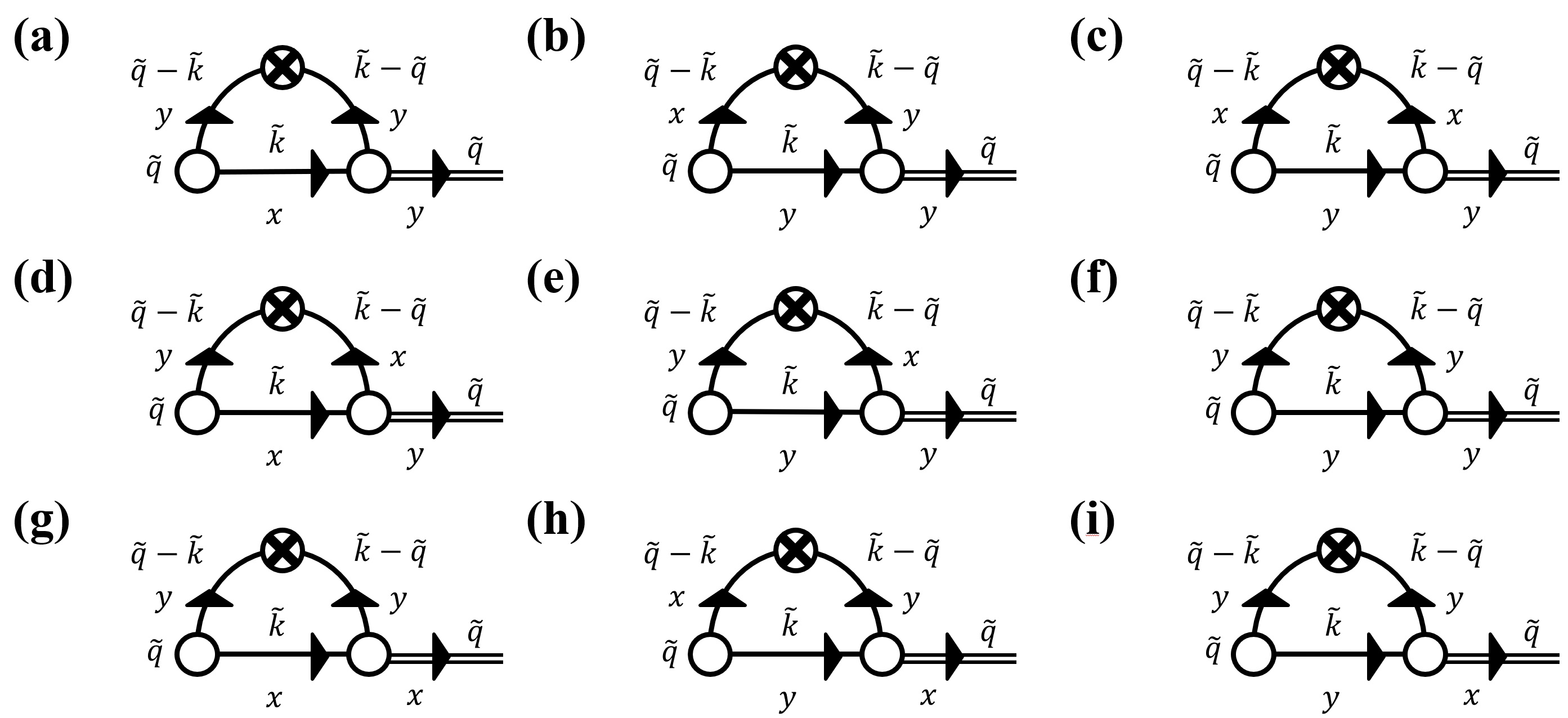}
	\end{center}
\caption{The graphical representations of potential corrections to $G^{-1}(\tilde{\bq})u_y(\tilde{\bq})$.}
	\label{fig:propagator}
\end{figure}

\subsection{Renormalization of $\gamma$}
\label{Sec:gamma}
To look for renormalizations of $\gamma$, we seek  graphs in Fig.~\ref{fig:propagator} which
have the prefactor $\ii q_x$ and  one outgoing  $u_y(\tbq)$  leg. Since
graphs (e) and (f) both have $q_y$ in the  prefactor, and (g) to (i) all have outgoing leg
$u_x(\tbq)$, none of them can contribute to the renormalization of $\gamma$ . This leaves us
with graphs (a) to (d) in Fig.~\ref{fig:propagator} to analyze.

Graph (a) represents a contribution to the right-hand side of the EOM (\ref{EOMA3}):
\beqn
&&2\times\left(-\alpha\right)\times\left(-{\alpha\over 2}\right)
u_y(\tilde{\bq})
\int_{>} {\dd^2k\over (2\pi)^2}{\dd\Omega\over 2\pi}\,
\left(-{k_y\over k_x}\right)G(\tilde{\bk}) P_{yx}(\bk)C_{_Q}(\bq-\bk)\delta(\omega-\Omega)\nonumber\\
&=&\alpha^2
u_y(\tilde{\bq})
\int_{>} {\dd^2k\over (2\pi)^2}\,\left[1\over -\ii\left(\omega-\gamma k_x
\right)
+\alpha{k_y^2\over k^2}\right]
\left[2 D_{_Q}\left(q_x-k_x\right)^4\over \gamma^2 \left(q_x-k_x\right)^6+\alpha^2 \left(q_y-k_y\right)^4\right]\left(k_y\over k_x\right)^2
\,.\label{gamma1}
\eeqn

Graph (b) represents a contribution to the right-hand side of EOM (\ref{EOMA3}):
\beqn
&&2\times\left(-\alpha\right)\times\left(-{\alpha\over 2}\right)
u_y(\tilde{\bq})
\int_{>} {\dd^2k\over (2\pi)^2}{\dd\Omega\over 2\pi}\,
\left(-{q_y-k_y\over q_x-k_x}\right)
G(\tilde{\bk})
P_{yx}(\bk)C_{_Q}(\bq-\bk)\delta(\omega-\Omega)\nonumber\\
&=&\alpha^2
u_y(\tilde{\bq})
\int_{>} {\dd^2k\over (2\pi)^2}\,\left[1\over -\ii\left(\omega-\gamma k_x
\right)
+\alpha{k_y^2\over k^2}\right]
\left[2 D_{_Q}\left(q_x-k_x\right)^4\over \gamma^2 \left(q_x-k_x\right)^6+\alpha^2 \left(q_y-k_y\right)^4\right]\left(k_y\over k_x\right)
\left(q_y-k_y\over q_x-k_x\right)
\,.
\nonumber\\
\label{gamma2}
\eeqn

Graph (c) represents a contribution to the right-hand side of the EOM (\ref{EOMA3}):
\beqn
&&2\times\left(-\alpha\right)\times\left(-{\alpha\over 2}\right)
u_y(\tilde{\bq})
\int_{>} {\dd^2k\over (2\pi)^2}{\dd\Omega\over 2\pi}\,
\left(q_y-k_y\over q_x-k_x\right)^2G(\tilde{\bk}) C_{_Q}(\bq-\bk)\delta(\omega-\Omega)\nonumber\\
&=&\alpha^2
u_y(\tilde{\bq})
\int_{>} {\dd^2k\over (2\pi)^2}\,\left[1\over -\ii\left(\omega-\gamma k_x
\right)
+\alpha{k_y^2\over k^2}\right]
\left[2 D_{_Q}\left(q_x-k_x\right)^4\over \gamma^2 \left(q_x-k_x\right)^6+\alpha^2 \left(q_y-k_y\right)^4\right]\left(q_y-k_y\over q_x-k_x\right)^2
\,.
\nonumber\\
\label{gamma3}
\eeqn

Graph (d) represents a contribution to the right-hand side of the EOM (\ref{EOMA3}):
\beqn
&&2\times\left(-\alpha\right)\times\left(-{\alpha\over 2}\right)
u_y(\tilde{\bq})
\int_{>} {\dd^2k\over (2\pi)^2}{\dd\Omega\over 2\pi}\,
\left(-{k_y-q_y\over k_x-q_x}\right)G(\tilde{\bk})
\left(-{k_y\over k_x}\right)C_{_Q}(\bq-\bk)\delta(\omega-\Omega)\nonumber\\
&=&\alpha^2
u_y(\tilde{\bq})
\int_{>} {\dd^2k\over (2\pi)^2}\,\left[1\over -\ii\left(\omega-\gamma k_x
\right)
+\alpha{k_y^2\over k^2}\right]
\left[2 D_{_Q}\left(q_x-k_x\right)^4\over \gamma^2 \left(q_x-k_x\right)^6+\alpha^2 \left(q_y-k_y\right)^4\right]\left(k_y\over k_x\right)
\left(q_y-k_y\over q_x-k_x\right)
\,,
\nonumber\\
\label{gamma4}
\eeqn
which is the same as that of graph (b).


The integrands in the above formulae are functions of $\omega$ and $q$ and thus can be expanded in powers of $\omega$ and $\bq$. It is easy to check that the zeroth order terms do not vanish. In fact, they lead to a contribution proportional to $u_y$ without any derivatives, which appears to be more relevant than all other existing linear terms in (\ref{EOMA3}). However, the generation of this new term comes from the renormalization of the mean velocity. It can be canceled off if $u_x$ is boosted by a  constant, which corresponds to the correction to the mean velocity. In what follows, we will ignore this trivial correction, and focus on the contributions linear in $\omega$ and $q$. Contributions higher order in $\bq$ and $\omega$ lead only to irrelevant contributions, and will therefore also be dropped.

We will deal with the linear order piece in $\omega$ in the next section, which in fact leads a general renormalization of the temporal derivative of $u_y$ in (\ref{EOMA3}). Here, we will focus on  the linear order piece in $q$ by expanding the integrands to linear order in $q$ while setting $\omega$ to $0$.

Specifically, the linear order piece in $q_x$ of (\ref{gamma1}) is
\beqn
&&\ii q_x
u_y(\tilde{\bq})
\left[8\gamma\alpha^2D_{_Q}\int_{>} {\dd^2k\over (2\pi)^2}\,
{k_y^2k_x^6\over \left(\gamma^2 k_x^6+\alpha^2 k_y^4\right)^2}
-12\gamma^3\alpha^2D_{_Q}\int_{>} {\dd^2k\over (2\pi)^2}\,
{k_y^2k_x^{12}\over \left(\gamma^2 k_x^6+\alpha^2 k_y^4\right)^3}\right]\nonumber\\
&=&\ii q_x
u_y(\tilde{\bq})
\left(8\gamma\alpha^2D_{_Q}I_1-12\gamma^3\alpha^2D_{_Q}I_2\right)
\eeqn
where $I_2$ is calculated in appendix (\ref{Sec:I2}).

The linear order piece in $q_x$ of (\ref{gamma2}) is
\beqn
&&\ii q_x
u_y(\tilde{\bq})
\left[6\gamma\alpha^2D_{_Q}\int_{>} {\dd^2k\over (2\pi)^2}\,
{k_y^2k_x^6\over \left(\gamma^2 k_x^6+\alpha^2 k_y^4\right)^2}
-12\gamma^3\alpha^2D_{_Q}\int_{>} {\dd^2k\over (2\pi)^2}\,
{k_y^2k_x^{12}\over \left(\gamma^2 k_x^6+\alpha^2 k_y^4\right)^3}\right]\nonumber\\
&=&\ii q_x
u_y(\tilde{\bq})
\left(6\gamma\alpha^2D_{_Q}I_1-12\gamma^3\alpha^2D_{_Q}I_2\right)\,.
\eeqn

The linear order piece in $q_x$ of (\ref{gamma3}) is
\beqn
&&\ii q_x
u_y(\tilde{\bq})
\left[4\gamma\alpha^2D_{_Q}\int_{>} {\dd^2k\over (2\pi)^2}\,
{k_y^2k_x^6\over \left(\gamma^2 k_x^6+\alpha^2 k_y^4\right)^2}
-12\gamma^3\alpha^2D_{_Q}\int_{>} {\dd^2k\over (2\pi)^2}\,
{k_y^2k_x^{12}\over \left(\gamma^2 k_x^6+\alpha^2 k_y^4\right)^3}\right]\nonumber\\
&=&\ii q_x
u_y(\tilde{\bq})
\left(4\gamma\alpha^2D_{_Q}I_1-12\gamma^3\alpha^2D_{_Q}I_2\right)\,.
\eeqn

The linear order piece in $q_x$ of (\ref{gamma4}) is the same as that of (\ref{gamma2}).

In summary, the sum of contributions linear both in $q_x$ and $u_y(\tbq)$ to the right-hand side of (\ref{EOMA3}) is
\beq
\ii q_x
u_y(\tilde{\bq})
\left(24\gamma\alpha^2D_{_Q}I_1-48\gamma^3\alpha^2D_{_Q}I_2\right)\,.
\eeq
Pulling this piece to the left-hand side of (\ref{EOMA3}), we find the following correction to $\gamma$:
\beq
\delta\gamma=-24\gamma\alpha^2D_{_Q}I_1+48\gamma^3\alpha^2D_{_Q}I_2\,.
\eeq
Inserting the values of  $I_1$ and $I_2$ for the various schemes, we obtain
\begin{subequations}
\begin{align}
\text{uncontrolled $d=2$:}~~~\delta \gamma
&={2\over 9}\gun\gamma\dd\ell\,,
\\
\text{hard continuation:}~~~\delta \gamma
&={2\over 9}\gha\gamma\dd\ell\,,
\\
\text{soft continuation:}~~~\delta \gamma
&={1\over 2}\gso\gamma\dd\ell\,.
\end{align}
\end{subequations}
Again, this implies
\begin{subequations}
\label{A_Eta_gamma_Direct}
\begin{align}
\text{uncontrolled $d=2$:}~~~\eta_{\gamma}^{\rm direct}
&={2\over 9}\gun \,,\label{A_Eta_gamma_Direct_U}\\
\text{hard continuation:}~~~\eta_{\gamma}^{\rm direct}
&={2\over 9}\gha \,,\label{A_Eta_gamma_Direct_H}\\
\text{soft continuation:}~~~\eta_{\gamma}^{\rm direct}
&={1\over 2}\gso \,.\label{A_Eta_gamma_Direct_S}
\end{align}
\end{subequations}

\subsection{Renormalization of the temporal derivative}
Here we seek one-loop contributions to $-\ii\omega u_y(\tbq)$, which amounts to finding graphs in Fig.~\ref{fig:propagator} which have a prefactor $\ii \omega$ and  one outgoing $u_y(\tbq)$  leg. Since graphs (e) to (i) either have $q_y$ in the prefactor or have an outgoing  $u_x(\bq)$  leg, none of them can generate contributions to $-\ii\omega u_y(\tbq)$. Therefore, only graphs (a) to (d) can. The contribution of these graphs to the right-hand side of the (\ref{EOMA3}) has been given in appendix (\ref{Sec:gamma}).
To find the linear order piece in $\omega$, we expand the integrands in (\ref{gamma1}-\ref{gamma4}) to linear order in $\omega$ and set $q=0$. Once this is done, all  four graphs become equal.The sum of them gives the following contribution to the right-hand side of (\ref{EOMA3}):
\beqn
&&4\alpha^2\ii\omega
u_y(\tilde{\bq})
\int_{>} {\dd^2k\over (2\pi)^2}\,
\left[k^4\over \left(\ii\gamma k_x^3+\alpha k_y^2\right)^2\right]
\left(2 D_{_Q}k_x^4\over \gamma^2 k_x^6+\alpha^2 k_y^4\right)\left(k_y\over k_x\right)^2
\nonumber\\
&=&4\alpha^2\ii\omega
u_y(\tilde{\bq})
\int_{>} {\dd^2k\over (2\pi)^2}\,
{2D_{_Q}k_x^2k_y^2k^4\left(\alpha k_y^2-\ii\gamma k_x^3\right)^2\over
\left(\gamma^2 k_x^6+\alpha^2 k_y^4\right)^3}\nonumber\\
&=&4\alpha^2\ii\omega
P_{yy}(\bq)u_y(\tilde{\bq})
\int_{>} {\dd^2k\over (2\pi)^2}\,
{2D_{_Q}k_x^2k_y^2k^4\left(\alpha^2 k_y^4-\gamma^2 k_x^6\right)\over
\left(\gamma^2 k_x^6+\alpha^2 k_y^4\right)^3}\nonumber\\
&\approx&8D_{_Q}\alpha^2\ii\omega
u_y(\tilde{\bq})
\left[\alpha^2\int_{>} {\dd^2k\over (2\pi)^2}\,
{k_x^6k_y^6\over
\left(\gamma^2 k_x^6+\alpha^2 k_y^4\right)^3}
-\gamma^2\int_{>} {\dd^2k\over (2\pi)^2}\,
{k_x^{12}k_y^2\over
\left(\gamma^2 k_x^6+\alpha^2 k_y^4\right)^3}\right]\nonumber\\
&=&8D_{_Q}\alpha^2\ii\omega
u_y(\tilde{\bq})
\left(\alpha^2I_3-\gamma^2I_2\right)
\,,
\eeqn
where $I_{2,3}$ are calculated in appendices (\ref{Sec:I2}) and (\ref{Sec:I3}), respectively.

Pulling this piece to the left-hand side of (\ref{EOMA3}) and inserting the values of $I_{2,3}$ for  the various schemes, we get the following contribution to $-\ii\omega u_y(\tbq)$:
\begin{subequations}
\begin{align}
\text{uncontrolled $d=2$:}~~~
\delta\left[-\ii\omega u_y(\tbq)\right]
&=-{2\over 27}\gun\dd\ell\left[-\ii\omega u_y(\tbq)\right]\,.
\\
\text{hard continuation:}~~~
\delta\left[-\ii\omega u_y(\tbq)\right]
&=-{2\over 27}\gha\dd\ell\left[-\ii\omega u_y(\tbq)\right]\,.
\\
\text{soft continuation:}~~~
\delta\left[-\ii\omega u_y(\tbq)\right]
&=-{1\over 6}\gso\dd\ell\left[-\ii\omega u_y(\tbq)\right]\,.
\end{align}
\end{subequations}
This implies
\begin{subequations}
\label{A_Eta_Time}
\begin{align}
\text{uncontrolled $d=2$:}~~~
\eta_\omega
&=-{2\over 27}\gun\,,\label{A_Eta_Time_U}
\\
\text{hard continuation:}~~~
\eta_\omega
&=-{2\over 27}\gha\,,\label{A_Eta_Time_H}
\\
\text{soft continuation:}~~~
\eta_\omega
&=-{1\over 6}\gso\,.\label{A_Eta_Time_S}
\end{align}
\end{subequations}

\subsection{Renormalization of $\alpha$}

Here we seek one-loop contributions to ${q_y^2\over q^2}u_y(\tbq)$ or ${q_xq_y\over q^2}u_x(\tbq)$, where the latter is equivalent to the former by the incompressibility condition $u_x(\tbq)=-{q_y\over q_x}u_y(\tbq)$.
Therefore, we look for graphs in Fig.~\ref{fig:propagator} which either have the prefactor $q_y^2\over q^2$ with an outgoing leg $u_y(\tbq)$ or have the prefactor $q_xq_y\over q^2$ with the outgoing leg $u_x(\tbq)$.

Graphs (a) to (d), (g), and (h) have neither the prefactor  $q_xq_y\over q^2$ nor $q_y^2/q^2$, and graphs (e) and (f) do have the prefactor  $q_x q_y/q^2$ but with the outgoing leg $u_y(\tbq)$. Hence these graphs  cannot  renormalize $\alpha$. The only contributing graph is therefore graph (i), which has exactly the prefactor $q_xq_y/q^2$ and the outgoing leg $u_x(\tbq)$. This graph gives the following contribution to the right-hand side of (\ref{EOMA3}):
\beqn
&&2\times\left(-{\alpha\over 2}\right)\times\left(-\alpha\right)
P_{yx}(\tilde{\bq})u_x(\tilde{\bq})
\int_{>} {\dd^2k\over (2\pi)^2}{\dd\Omega\over 2\pi}\,G(\tilde{\bk}) C_{_Q}(\bq-\bk)\delta(\omega-\Omega)\nonumber\\
&=&\alpha^2
{q_y^2\over q^2}u_y(\tilde{\bq})
\int_{>} {\dd^2k\over (2\pi)^2}{\dd\Omega\over 2\pi}\,\left[1\over -\ii\left(\omega-\gamma k_x
\right)
+\alpha{k_y^2\over k^2}\right]
\left[4\pi D_{_Q}\left(q_x-k_x\right)^4\delta(\omega-\Omega)\over \gamma^2 \left(q_x-k_x\right)^6+\alpha^2 \left(q_y-k_y\right)^4\right]\nonumber\\
&=&\alpha^2
{q_y^2\over q^2}u_y(\tilde{\bq})
\int_{>} {\dd^2k\over (2\pi)^2}\,\left[1\over -\ii\left(\omega-\gamma k_x
\right)+\alpha{k_y^2\over k^2}\right]
\left[2D_{_Q}\left(q_x-k_x\right)^4\over \gamma^2 \left(q_x-k_x\right)^6+\alpha^2 \left(q_y-k_y\right)^4\right]\,,
\eeqn
where in the first equality we have used the incompressibility condition $u_x(\tbq)=-{q_y\over q_x}u_y(\tbq)$. To keep only the relevant contributions we set $\omega=0$, $q=0$ in the integrand, which leads to
\beqn
&&\alpha^2{q_y^2\over q^2}u_y(\tilde{\bq})\int_{>} {\dd^2k\over (2\pi)^2}\,
{2D_{_Q}k^2k_x^4\over \left(\ii\gamma k_x^3+\alpha k_y^2\right)\left(\gamma^2 k_x^6+\alpha^2 k_y^4\right)}\nonumber\\
&=&2\alpha^2D_{_Q}{q_y^2\over q^2}u_y(\tilde{\bq})\int_{>} {\dd^2k\over (2\pi)^2}\,
{\left(-\ii\gamma k_x^3+\alpha k_y^2\right)k^2k_x^4\over \left(\gamma^2 k_x^6+\alpha^2 k_y^4\right)^2}\nonumber\\
&=&2\alpha^3D_{_Q}{q_y^2\over q^2}u_y(\tilde{\bq})\int_{>} {\dd^2k\over (2\pi)^2}\,
{k^2k_y^2k_x^4\over \left(\gamma^2 k_x^6+\alpha^2 k_y^4\right)^2}
\nonumber\\
&\approx&2\alpha^3D_{_Q}{q_y^2\over q^2}u_y(\tilde{\bq})\int_{>} {\dd^2k\over (2\pi)^2}\,
{k_y^2k_x^6\over \left(\gamma^2 k_x^6+\alpha^2 k_y^4\right)^2}
\nonumber\\
&=&2\alpha^3D_{_Q}I_1{q_y^2\over q^2}u_y(\tilde{\bq})
\eeqn
where in the second equality we have neglected the imaginary part since it vanishes after the integration, and in the ``$\approx$" we have only kept the dominant part
(that is, the ``$k_x^2$" component of ``$k^2$").

Pulling this to the left-hand side of (\ref{EOMA3}) and using the explicit value for $I_1$ for the various schemes, we get the following correction to $\alpha$:
\begin{subequations}
\begin{align}
\text{uncontrolled $d=2$:}~~~\delta \alpha
&=-{1\over 9}\gun\alpha\dd\ell\,,
\\
\text{hard continuation:}~~~\delta \alpha
&=-{1\over 9}\gha\alpha\dd\ell\,,
\\
\text{soft continuation:}~~~\delta \alpha
&=-{1\over 6}\gso\alpha\dd\ell\,.
\end{align}
\end{subequations}
This implies
\begin{subequations}
\label{A_Eta_alpha_Direct}
\begin{align}
\text{uncontrolled $d=2$:}~~~\eta_\alpha^{\rm direct}
&=-{1\over 9}\gun\,,\label{A_Eta_alpha_Direct_U}
\\
\text{hard continuation:}~~~\eta_\alpha^{\rm direct}
&=-{1\over 9}\gha\,,\label{A_Eta_alpha_Direct_H}
\\
\text{soft continuation:}~~~\eta_\alpha^{\rm direct}
&=-{1\over 6}\gso\,.\label{A_Eta_alpha_Direct_S}
\end{align}
\end{subequations}

\subsection{Renormalization of $\mu_x$\label{Sec:mu_x}}
The one-loop graphical correction to $\mu_x$ can also be derived from the graphs in Fig. \ref{fig:propagator}. However, our purpose here is NOT to get the exact result. We only need to know the dependence of $\eta_\mu^{\rm direct}$  on $\gamma$, $\alpha$, and $D_{_Q}$, because that is sufficient for us to derive an exact relation between $\eta_{\mu,\gamma,\alpha,_Q}$ at the DRG fixed point, which
allows us to determine $\eta_\mu$ through $\eta_{\gamma,\alpha,_Q}$. Therefore, for this purpose we only need to analyze one of the first four graphs in Fig.~\ref{fig:propagator}, for instance, graph (a). This graph represents a contribution to the right-hand side of (\ref{EOMA3})  given by (\ref{gamma1}). Now we seek a term proportional to $q_x^2u_y(\tbq)$. It is the easiest that we expand the denominator of the integrand to $O(q_x^2)$ and set $\omega=0$ and $\bq=\mathbf {0}$ in the numerator. Doing this and focusing on the $q_x^2$ piece we get
\beqn
\alpha^2q_x^2
u_y(\tilde{\bq})
\int_{>} {\dd^2k\over (2\pi)^2}\,\left(k^2\over \ii\gamma{k_x^3}
+\alpha{k_y^2}\right)
\left(12 D_{_Q}k_x^2\over \gamma^2 k_x^6+\alpha^2 k_y^4\right)\left(k_y\over k_x\right)^2
\,.\label{mu_x1}
\eeqn

By rationalizing the numerator,  the imaginary part of (\ref{mu_x1}) will vanish during the integration since it is odd in $k_x$. So we are left with
\beqn
\alpha^2q_x^2
u_y(\tilde{\bq})
\int_{>} {\dd^2k\over (2\pi)^2}\,
{12 \alpha D_{_Q}k^2 k_y^4\over (\gamma^2 k_x^6+\alpha^2 k_y^4)^2}
\approx 12\alpha^3D_{_Q}q_x^2
u_y(\tilde{\bq})I_4
\,,\label{mu_x2}
\eeqn
where
\beqn
I_4\equiv\int_{>} {\dd^2k\over (2\pi)^2}\,
{k_x^2 k_y^4\over (\gamma^2 k_x^6+\alpha^2 k_y^4)^2}\,,
\eeqn
 and in the ``$\approx$'' we have kept only the most diverging piece by approximating $k^2$ as $k_x^2$. The quantity $I_4$ is calculated in appendix (\ref{Sec:I4}). Inserting the value of $I_4$ for the various schemes into (\ref{mu_x2}),
we get
 \begin{subequations}
\begin{align}
 \text{uncontrolled $d=2$:}~~~
 &\left({D_{_Q}\over {\pi|\gamma|}}\Lambda^{-1}\dd\ell\right)
 q_x^2u_y(\tilde{\bq})\,,
 \label{}\\
 \text{hard continuation:}~~~
 &\left[{S_{d-1}\over (2\pi)^{d-1}}{D_{_Q}\over |\gamma|}\Lambda^{d-3}\dd\ell\right]
 q_x^2u_y(\tilde{\bq})\,,
 \label{}\\
 \text{soft continuation:}~~~
 &	\left[{5\pi\over 3\sin\left(5\pi\over 12\right)}{S_{d-1}\over (2\pi)^d}D_{_Q}\alpha^{d-2\over 3}|\gamma|^{-{d+1\over 3}}\dd\ell\right]
 q_x^2u_y(\tilde{\bq})\,.
\label{}
\end{align}
\end{subequations}

Based on the result from graph (a), we conclude that the {\it total} one-loop graphical contributions to $\mu_x$ can be written as
 \begin{subequations}
\begin{align}
 \text{uncontrolled $d=2$:}~~~
 &\delta\mu_x=\left(\gun_\mu\dd\ell\right) \mu_x\,,
 \label{}\\
 \text{hard continuation:}~~~
 &\delta\mu_x=\left(\gha_\mu\dd\ell\right) \mu_x\,,
 \label{}\\
 \text{soft continuation:}~~~
 &\delta\mu_x=	\left(\gso_\mu\dd\ell\right)\mu_x\,.
\label{}
\end{align}
\end{subequations}
where
\begin{subequations}
\begin{align}
&\gun_\mu\propto D_{_Q}|\gamma|^{-{1}}\mu_x^{-1}\Lambda^{-{1}}\,,\\
&\gha_\mu\propto|\gamma|^{-1}\mu_x^{-1}D_{_Q}\Lambda^{d-3}\,,\\
&\gso_\mu\propto D_{_Q}\mu_x^{-1}\alpha^{d-2\over 3}|\gamma|^{-{d+1\over 3}}\Lambda^{2d-7\over 3}\,.
\end{align}
\end{subequations}
This implies
 \begin{subequations}
 \label{A_Eta_mu_Direct}
\begin{align}
 \text{uncontrolled $d=2$:}~~~
  &\eta_\mu^{\rm direct}=\gun_\mu\,,\label{A_Eta_mu_Direct_U}
 \\
 \text{hard continuation:}~~~
 &\eta_\mu^{\rm direct}=\gha_\mu\,,\label{A_Eta_mu_Direct_H}
 \\
 \text{soft continuation:}~~~
 &\eta_\mu^{\rm direct}=\gso_\mu\,.\label{A_Eta_mu_Direct_S}
\end{align}
\end{subequations}

\subsection{Putting it all together}

Inserting \eqref{A_Eta_Q_Direct}, \eqref{A_Eta_A_Direct},  \eqref{A_Eta_gamma_Direct}, \eqref{A_Eta_Time}, \eqref{A_Eta_alpha_Direct}, \eqref{A_Eta_mu_Direct} into \eqref{eta_omega_effect_lin} and \eqref{eta_omega_effect_lin_noise},
we obtain $\eta_{\gamma,\alpha,\mu,_Q,_A}$ to one-loop order for the three schemes, as quoted in the main text. Specifically, for the uncontrolled calculation in $d=2$
\beqn
 \eta_\gamma={8\over 27}\gun\,, ~
 \eta_\alpha=-{1\over 27}\gun\,, ~
 \eta_\mu=\gun_\mu+{2\over 27}\gun\,,~
 \eta_{_Q}={10\over 27}\gun\,, ~
 \eta_{_A}={16\over 27}\gun\,;
\eeqn
for the hard continuation
\beqn
 \eta_\gamma={8\over 27}\gha\,,~
 \eta_\alpha=-{1\over 27}\gha\,,~
 \eta_\mu=\gha_\mu+{2\over 27}\gha\,,~
 \eta_{_Q}={10\over 27}\gha\,,~
 \eta_{_A}={16\over 27}\gha\,;
\eeqn
for the soft continuation
\beqn
 \eta_\gamma={2\over 3}\gso\,,~
 \eta_\alpha=0\,,~
 \eta_\mu=\gso_\mu+{1\over 6}\gso\,,~
 \eta_{_Q}={2\over 3}\gso\,,~
 \eta_{_A}=\gso\,.
\eeqn

\begin{figure}
	\begin{center}
  \includegraphics[scale=.32]{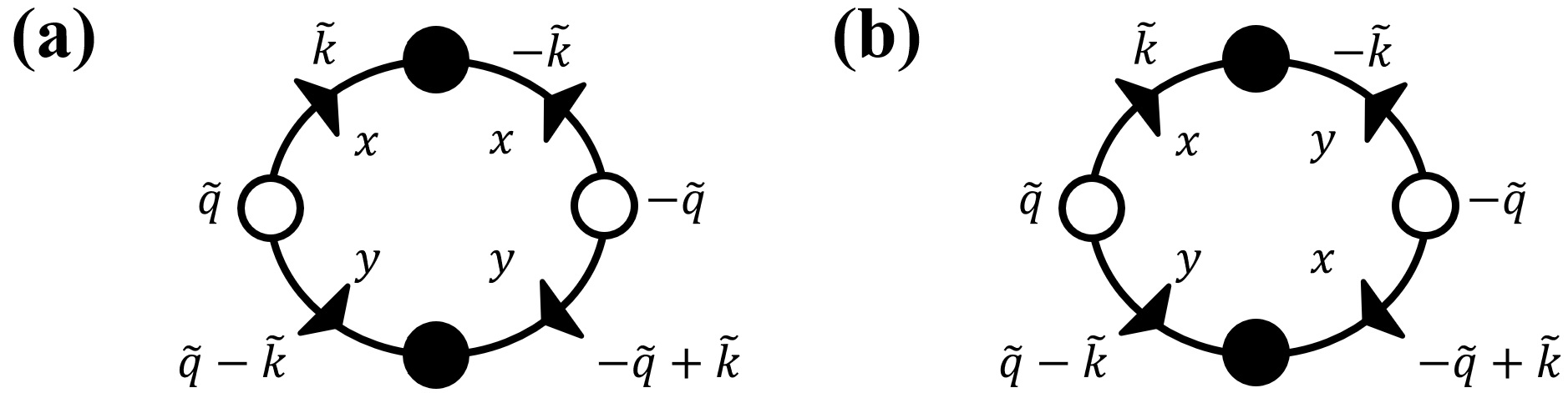}
	\end{center}
\caption{Two particular graphical corrections to $D_{_A}$ purely from the annealed fluctuations. They correspond to the graphs (a) and (b) in \fig \ref{fig:noise_DQ} but with both the quenched noise averages replaced by the annealed noise averages.}
	\label{Fig_D_A_Anneal}
\end{figure}

\section{Derivation of the annealed couplings}
\label{app_DA}
In this section we derive the dimensionless coupling $g_{_A}$ associated with the renormalization coming purely from the annealed fluctuations.
 We do this by calculating  the purely $D_{_A}$-dependent contributions to any of the one-loop graphical corrections to $D_{_A}$, $\alpha$, and $\mu_x$  (All of these give the same expression for $g_{_A}$). Here we choose to calculate the one-loop graphical correction to $D_{_A}$. The Feynman diagrams for this calculation are given in Fig. \ref{Fig_D_A_Anneal},  which are essentially the same as those in Fig. \ref{fig:noise_DQ},  albeit with both the quenched noise averages (indicated by crossed circles) replaced by the annealed noise averages (indicated by black circles). Setting $\tbq=\tilde{0}$ inside the loops, the sum of two graphs represents the following contribution to the correlations of the annealed noise:
\beqn
\delta\left[\langle f_{_A}^y(\bq,\omega)f_{_A}^y(-\bq,\omega)\rangle \right]
=2\alpha^2\int_{>} {\dd^2k\over (2\pi)^2}\int {\dd\Omega\over 2\pi}C_{_{A}}^{xx}(\tbk)C_{_{A}}^{yy}(\tbk)\,,
\eeqn
which implies a correction to the noise strength $D_{_A}$:
\beqn
\delta D_{_A}
=\alpha^2\int_{>} {\dd^2k\over (2\pi)^2}\int {\dd\Omega\over 2\pi}C_{_{A}}^{xx}(\tbk)C_{_{A}}^{yy}(\tbk)\,,
\eeqn
Inserting equations \eqref{eq:Ca&Cq app} for the annealed correlation functions
we obtain
\beqn
\delta D_{_A}=4\alpha^2D_{_A}^2\int_{>} {\dd^2k\over (2\pi)^2}\int {\dd\Omega\over 2\pi} {k_y^2\over k_x^2}\left({1\over (\Omega-\gamma k_x)^2+\left[{\alpha k_{y}^2\over k^2}+\mu_xk_x^2\right]^2}\right)^2\,.
\eeqn
Shifting frequency from $\Omega$ to $\Omega'$ defined by
\beq
\Omega'\equiv\Omega-\gamma k_x
\label{omegashift}
\eeq
shows that $\gamma$ drops out, and leaves a simple integral over the shifted $\Omega'$. Doing that integral gives, up to $O(1)$ multiplicative factors,
\beqn
\delta D_{_A}
&=&\alpha^2D_{_A}^2\int_{>} {\dd^2k\over (2\pi)^2} {k_y^2\over k_x^2}\left({1\over \alpha{ k_{y}^2\over k^2}+\mu_xk_x^2}\right)^3\times O(1)
\nonumber\\
&=&\alpha^2D_{_A}^2\int_{>} {\dd k_y\over 2\pi} k_y^2\int_{-\infty}^\infty {dk_x\over2\pi} {k_x^4\over \bigg(\alpha k_{y}^2+\mu_xk_x^4\bigg)^3}\times O(1)\,.
\label{dadaexpunc}
\eeqn
We can pull the $\alpha$, $\mu_x$, and $k_y$ dependence of the integral over $k_x$ out by the simple rescaling of the variable of integration from $k_x$ to $Q_x$ via
\beq
k_x\equiv \left({\alpha\over\mu_x}\right)^{1\over4}\sqrt{|k_y|} Q_x \,,
\label{rescalekxunc}
\eeq
which gives
\beqn
\delta D_{_A}={\alpha^{1/4}D_{_A}^2\over\mu_x^{5\over4}}\int_{>} {\dd k_y\over 2\pi} k_y^{-{3\over2}}\times O(1)
=\left[\alpha^{1/4} D_A/ (\mu_x^{5/4} \Lambda^{ {1\over2}})\right]D_{_A}\dd\ell\times O(1)
\equiv \gun_{_A}D_{_A}d\ell
\label{dadafinalunc}
\eeqn
where we've defined, up to $O(1)$ multiplicative factors,
\beq
\gun_{_A} \propto\alpha^{1/4} D_A/ (\mu_x^{5/4} \Lambda^{ {1\over2}}) \,,
\label{gdadefuncapp}
\eeq
and the $O(1)$ factor also includes
\beq
\int_{-\infty}^\infty {dQ_x\over2\pi} {Q_x^4\over \bigg(1+Q_x^4\bigg)^3}={5\sqrt{2}\over128} \,.
\eeq

This confirms our claim in the main text that the corrections to $D_{_A}$ coming purely from
$D_{_A}$ itself are proportional
to $\gun_{_A} $ as defined in \eqref{gdadefuncapp}. Hence, our demonstration in the main text that $\gun_{_A} $ flows to zero upon renormalization at our fixed point shows that the {\it purely} annealed noise-generated corrections to the annealed noise can be safely ignored.

Generalizing (\ref{dadaexpunc}) to $d$-dimensions via the ``hard" continuation gives
\beqn
\delta D_{_A}=\alpha^2D_{_A}^2\int_{>} {\dd^{d-1}k_h\over (2\pi)^{d-1}} k_h^2\int_{-\infty}^\infty {\dd k_x\over 2\pi}{k_x^4\over \bigg(\alpha k_{h}^2+\mu_xk_x^4\bigg)^3}\times O(1)\,.
\label{dadaexph}
\eeqn

As in $d=2$, we can pull the $\alpha$, $\mu_x$, and $k_y$ dependence of the integral over $k_x$ out by the simple rescaling of the variable of integration from $k_x$ to $Q_x$ via
\beq
k_x\equiv \left({\alpha\over\mu_x}\right)^{1\over4}\sqrt{k_h} Q_x \,,
\label{rescalekh}
\eeq
which gives
\beqn
\delta D_{_A}={\alpha^{1/4}D_{_A}^2\over\mu_x^{5\over4}}\int_{>} {\dd^{d-1} k_h\over (2\pi)^{d-1}} k_h^{-{3\over2}}\times O(1)
=\left[\alpha^{1/4} D_A/ (\mu_x^{5/4} \Lambda^{ {5\over2}-d})\right]D_{_A}\dd\ell\times O(1)
\equiv \gha_{_A}D_{_A}d\ell
\label{dadafinal7/3}
\eeqn
where we've defined, up to $O(1)$ multiplicative factors,
\beq
\gha_{_A} \propto\frac{\alpha^{1/4} D_A}{ \mu_x^{5/4} \Lambda^{ {5\over2}-d}}
 \,,
\label{gdadef7/3eps}
\eeq
which again is the expression for $\gha_{_A}$ quoted in the main text.

Generalizing (\ref{dadaexpunc}) to higher dimensions via the ``soft" continuation gives
\beq
\delta D_{_A}=\alpha^2D_{_A}^2\int_{>} {\dd k_y\over 2\pi} k_y^2\int_{-\infty}^\infty
{\dd^{d-1}k_s\over (2\pi)^{d-1}}{k_s^4\over \bigg(\alpha k_{y}^2+\mu_xk_s^4\bigg)^3} \times O(1)\,.
\label{dadaexps}
\eeq

As in $d=2$, we can pull the $\alpha$, $\mu_x$, and $k_y$ dependence of the integral over $\bk_s$ out by the simple rescaling of the variables of integration from $\bk_s$ to ${\bf Q}_s$ via
\beq
\bk_s\equiv \left({\alpha\over\mu_x}\right)^{1\over4}\sqrt{k_y} {\bf Q}_s \,,
\label{rescaleks}
\eeq
which gives
\beqn
\delta D_{_A}={\alpha^{d-1\over4}D_{_A}^2\over\mu_x^{{d+3\over4}}}\int_{>} {\dd k_y\over 2\pi} k_y^{{d-5\over2}}\times O(1)
=\left[D_A\alpha^{(d-1)\over 4}\mu_x^{-{(d+3)\over 4}}\Lambda^{ d-3\over 2}\right]D_{_A}d\ell\times O(1)
\equiv \gso_{_A}D_{_A}d\ell
\label{dadafinal5/2}
\eeqn
where we've defined, up to $O(1)$ multiplicative factors,
\beq
\gso_{_A} \propto D_A\alpha^{(d-1)\over 4}\mu_x^{-{(d+3)\over 4}}\Lambda^{ d-3\over 2}
 \,,
\label{gdadef5/2eps}
\eeq
which again is the expression for $\gso_{_A}$ quoted in the main text.

\section{Evaluation of the integrals $I_1$, $I_2$, $I_3$, $I_4$}
In this section, we will calculate the integrals $I_1$, $I_2$, $I_3$ and $I_4$ for the uncontrolled calculation in $d=2$ , the hard continuation, and the soft continuation.
\subsection{The Integral $I_1$
}
\label{Sec:I1}
\subsubsection{Uncontrolled calculation in exactly $d=2$}
We now calculate the integral $I_1$  in exactly $d=2$,  integrating over $\Lambda \ee^{-\dd\ell}<|k_y|<\Lambda $, $-\infty<k_x<\infty$:
\beqn
I_1\equiv\int_{>} {\dd^2k\over (2\pi)^2}\,
{k_y^2k_x^6\over \left(\gamma^2 k_x^6+\alpha^2 k_y^4\right)^2}
={1\over \pi^2}\int_{\Lambda \ee^{-\dd\ell}}^{\Lambda}
\dd k_y\int_0^\infty \dd k_x\,
{k_y^2k_x^{6}\over \left(\gamma^2 k_x^6+\alpha^2 k_y^4\right)^2}\,.\label{I1_unc}
\eeqn
Making the change of variable of integration
\beqn
k_x\equiv\left(\alpha\over|\gamma|\right)^{1\over 3}k_y^{2\over 3}(\tan\theta)^{1\over 3}
\label{k_x_unc}
\eeqn
in the $k_x$ integral in \eqref{I1_unc} gives
\beqn
I_1&=&{1\over 3\pi^2}\left(\alpha\over|\gamma|\right)^{11\over3}{1\over\alpha^4}
\int_{\Lambda \ee^{-\dd\ell}}^{\Lambda}
\dd k_y k_y^{-{4\over3}}\int_0^{\pi\over 2} \dd \theta\,
\cos^2\theta\left(\tan\theta\right)^{4\over 3}\nonumber\\
&=&{1\over 3\pi^2}\left(\alpha\over|\gamma|\right)^{7\over 3}{\Lambda^{-{1\over 3}}\over\alpha^4}
\dd\ell\int_0^{\pi\over 2} \dd \theta\,
\cos^2\theta\left(\tan\theta\right)^{4\over 3}\nonumber\\
&=&{1\over 3\pi^2}|\gamma|^{-{7\over 3}}\alpha^{-{5\over 3}}\Lambda^{-{1\over 3}}
\dd\ell\int_0^{\pi\over 2} \dd \theta\,
\left(\cos\theta\right)^{2\over 3}\left(\sin\theta\right)^{4\over 3}\nonumber\\
&=&{1\over 18\pi}|\gamma|^{-{7\over 3}}\alpha^{-{5\over 3}}\Lambda^{-{1\over 3}}
\dd\ell\nonumber\\
&=&{1\over 18}{\gun\over\alpha^2D_{_Q}}\dd\ell
\,,\label{unc_I1}
\eeqn
where  we have defined
\beq
\gun\equiv {D_{_Q}\over \pi}|\gamma|^{-{7\over 3}}\alpha^{1\over 3}\Lambda^{-{1\over 3}}\,.\label{g"}
\eeq

Note that, unsurprisingly, both our expression for $\gun$ and our result for $I_1$ agree with the results \eqref{g_def2} and \eqref{I1hard}  of the hard continuation, to be discussed next, if we set $d=2$ in those expressions.

\subsubsection{Hard continuation}
In the ``hard" continuation, which we will now present, we treat the ``soft" direction $x$ as one-dimensional, while the ``hard" direction $y$ is  extended to $d-1$-dimensions.  In practice, this means we will simply replace $k_y$ in Fourier space with a $d-1$-dimensional vector $\bk_h$ orthogonal to the $x$-direction. This gives:
\beqn
I_1\equiv\int_{>} {\dd^dk\over (2\pi)^d}\,
{k_h^2k_x^6\over \left(\gamma^2 k_x^6+\alpha^2 k_h^4\right)^2}
={2\over (2\pi)^d}\int_{\Lambda \ee^{-\dd\ell}|\bk_h|<\Lambda}
\dd^{d-1} k_h\int_0^\infty \dd k_x\,
{k_h^2k_x^6\over \left(\gamma^2 k_x^6+\alpha^2 k_h^4\right)^2}\,.\label{i1_v2}
\eeqn
Making the change of variable of integration
\beqn
k_x\equiv\left(\alpha\over|\gamma|\right)^{1\over 3}k_h^{2\over 3}(\tan\theta)^{1\over 3}
\label{k_x}
\eeqn
in the $k_x$ integral in (\ref{i1_v2}) gives
\beqn
I_1&=&{2S_{d-1}\over 3(2\pi)^d}\left(\alpha\over|\gamma|\right)^{7\over 3}{1\over\alpha^4}
\int_{\Lambda \ee^{-\dd\ell}}^{\Lambda}\dd k_h k_h^{d-{10\over 3}}
\int_0^{\pi\over 2} \dd \theta\, \sin^{4\over 3}\theta\cos^{2\over 3}\theta
\nonumber\\
&=&{2S_{d-1}\over 3(2\pi)^d}
\left(1\over|\gamma|^7\alpha^5\right)^{1\over 3}
\Lambda^{3d-7\over 3}\dd\ell
\int_0^{\pi\over 2} \dd \theta\, \sin^{4\over 3}\theta\cos^{2\over 3}\theta
\nonumber\\
&=&{S_{d-1}\over 18(2\pi)^{d-1}}
\left(1\over|\gamma|^7\alpha^5\right)^{1\over 3}
\Lambda^{3d-7\over 3}\dd\ell
\nonumber\\
&=&{\gha\over 18\alpha^2D_{_Q}}\dd\ell
\,,
\label{I1hard}
\eeqn
where
\beqn
\label{g_def2}
\gha\equiv {S_{d-1}\over (2\pi)^{d-1}}
|\gamma|^{-{7\over 3}}\alpha^{1\over 3}D_{_Q}
\Lambda^{3d-7\over 3}\,.
\eeqn
\subsubsection{Soft continuation}
{\AM In the ``soft" continuation, we treat the ``soft" direction $x$ as $d-1$ dimensional, while the ``hard" direction $y$ is  taken to be one-dimensional.  In practice, this means we will simply replace $k_x$ in Fourier space with a $d-1$-dimensional vector ${\bf k}_s$ orthogonal to the $y$-direction. This gives:}
\beqn
I_1\equiv\int_{>} {\dd^dk\over (2\pi)^d}\,
{k_y^2k_s^6\over \left(\gamma^2 k_s^6+\alpha^2 k_y^4\right)^2}
={2S_{d-1}\over (2\pi)^d}\int_{\Lambda \ee^{-\dd\ell}}^{\Lambda}
\dd k_y\int_0^\infty \dd k_s\,
{k_y^2k_s^{d+4}\over \left(\gamma^2 k_s^6+\alpha^2 k_y^4\right)^2}\,.\label{I1_s}
\eeqn
Making the change of variable of integration
\beqn
k_s\equiv\left(\alpha\over|\gamma|\right)^{1\over 3}k_y^{2\over 3}(\tan\theta)^{1\over 3}
\label{k_s}
\eeqn
in the $k_s$ integral in \eqref{I1_s} gives
\beqn
I_1&=&{2S_{d-1}\over 3(2\pi)^d}\left(\alpha\over|\gamma|\right)^{\left({d+5\over3}\right)}{1\over\alpha^4}
\int_{\Lambda \ee^{-\dd\ell}}^{\Lambda}
\dd k_y k_y^{\left({2d-8\over3}\right)}\int_0^{\pi\over 2} \dd \theta\,
\cos^2\theta\left(\tan\theta\right)^{\left({d+2\over 3}\right)}\nonumber\\
&=&{2S_{d-1}\over 3(2\pi)^d}\left(\alpha\over|\gamma|\right)^{\left({d+5\over3}\right)}{\Lambda^{\left({2d-5\over 3}\right)}\over\alpha^4}
\dd\ell\int_0^{\pi\over 2} \dd \theta\,
\cos^2\theta\left(\tan\theta\right)^{\left({d+2\over 3}\right)}\nonumber\\
&=&{2S_{d-1}\over 3(2\pi)^d}|\gamma|^{-{\left({d+5\over3}\right)}}\alpha^{\left({d-7\over 3}\right)}\Lambda^{\left({2d-5\over 3}\right)}
\dd\ell\int_0^{\pi\over 2} \dd \theta\,
\left(\cos\theta\right)^{4-d\over 3}\left(\sin\theta\right)^{d+2\over 3}\nonumber\\
&=&{S_{d-1}\over 12\sqrt{2}(2\pi)^{d-1}}|\gamma|^{-{\left({d+5\over3}\right)}}\alpha^{\left({d-7\over 3}\right)}\Lambda^{\left({2d-5\over 3}\right)}
\dd\ell\nonumber\\
&=&{1\over 12}{\gso\over\alpha^2D_{_Q}}\dd\ell
\,,\label{Soft_I1}
\eeqn
where in the penultimate equality we have set $d={5\over 2}$ in the integrand of the integral over $\theta$, and in the final equality we have defined
\beq
\gso\equiv {S_{d-1}D_{_Q}\over \sqrt{2}(2\pi)^{d-1}}|\gamma|^{-{\left({d+5\over3}\right)}}\alpha^{\left({d-1\over 3}\right)}\Lambda^{\left({2d-5\over 3}\right)}\,.\label{g'}
\eeq


\subsection{integral $I_2$
\label{Sec:I2}}
\subsubsection{Uncontrolled calculation in exactly $d=2$}
In exactly $d=2$
\beqn
I_2\equiv\int_{>} {\dd^2k\over (2\pi)^2}\,
{k_y^2k_x^{12}\over \left(\gamma^2 k_x^6+\alpha^2 k_y^4\right)^3}
={1\over \pi^2}\int_{\Lambda \ee^{-\dd\ell}}^{\Lambda} \dd k_y
\int_0^\infty \dd k_x\,{k_y^2k_x^{12}\over \left(\gamma^2 k_x^6+\alpha^2 k_y^4\right)^3}\,.
\label{unc_I2}
\eeqn
Inserting (\ref{k_x_unc}) into (\ref{unc_I2}) we get
\beqn
I_2&=&{1\over 3\pi^2}\left(\alpha\over|\gamma|\right)^{13\over 3}{1\over\alpha^6}
\int_{\Lambda \ee^{-\dd\ell}}^{\Lambda}
\dd k_y k_y^{-{4\over3}}\int_0^{\pi\over 2} \dd \theta\,
\cos^4\theta\left(\tan\theta\right)^{10\over 3}\nonumber\\
&=&{1\over 3\pi^2}|\gamma|^{-{13\over 3}}\alpha^{-{5\over 3}}
\Lambda^{-{1\over 3}}
\dd\ell\int_0^{\pi\over 2} \dd \theta\,
\cos^{2\over 3}\theta\sin^{10\over 3}\theta\nonumber\\
&=&{7\over 216}{\gun\over\gamma^2\alpha^2D_{_Q}}
\dd\ell
\,.
\eeqn
\subsubsection{Hard continuation}
Continuing to higher dimensions using the hard continuation described above gives:
\beqn
I_2\equiv\int_{>} {\dd^dk\over (2\pi)^d}\,
{k_h^2k_x^{12}\over \left(\gamma^2 k_x^6+\alpha^2 k_h^4\right)^3}
={2\over (2\pi)^d}\int_{\Lambda \ee^{-\dd\ell}}^{\Lambda} \dd^{d-1} k_h
\int_0^\infty \dd k_x\,
{k_h^2k_x^{12}\over \left(\gamma^2 k_x^6+\alpha^2 k_h^4\right)^3}
\,.\label{i2}
\eeqn
Changing variables and inserting (\ref{k_x}) into (\ref{i2}) we get
\beqn
I_2&=&{2S_{d-1}\over 3(2\pi)^d}\left(\alpha\over|\gamma|\right)^{13\over 3}{1\over\alpha^6}
\int_{\Lambda \ee^{-\dd\ell}}^{\Lambda}\dd k_h k_h^{d-{10\over 3}}
\int_0^{\pi\over 2} \dd \theta\, \sin^{10\over 3}\theta\cos^{2\over 3}\theta
\nonumber\\
&=&{7S_{d-1}\over 216(2\pi)^{d-1}}
|\gamma|^{-{13\over 3}}\alpha^{-{5\over 3}}
\Lambda^{3d-7\over 3}\dd\ell
\nonumber\\
&=&{7\gha\over 216\gamma^2\alpha^2D_{_Q}}\dd\ell\,.
\eeqn
\subsubsection{Soft continuation}
{Continuing to higher dimensions using the soft continuation defined above yields:}
\beqn
I_2\equiv\int_{>} {\dd^dk\over (2\pi)^d}\,
{k_y^2k_s^{12}\over \left(\gamma^2 k_s^6+\alpha^2 k_y^4\right)^3}
={2S_{d-1}\over (2\pi)^d}\int_{\Lambda \ee^{-\dd\ell}}^{\Lambda} \dd k_y
\int_0^\infty \dd k_s\,
{k_y^2k_s^{d+10}\over \left(\gamma^2 k_s^6+\alpha^2 k_y^4\right)^3}\,.\label{Soft_I2}
\eeqn
Inserting (\ref{k_s}) into (\ref{Soft_I2}) we obtain
\beqn
I_2&=&{2S_{d-1}\over 3(2\pi)^d}\left(\alpha\over|\gamma|\right)^{d+11\over 3}{1\over\alpha^6}
\int_{\Lambda \ee^{-\dd\ell}}^{\Lambda}
\dd k_y k_y^{2d-8\over3}\int_0^{\pi\over 2} \dd \theta\,
\cos^4\theta\left(\tan\theta\right)^{d+8\over 3}\nonumber\\
&=&{2S_{d-1}\over 3(2\pi)^d}|\gamma|^{-\left({d+11\over 3}\right)}\alpha^{\left({d-7\over 3}\right)}
\Lambda^{\left({2d-5\over 3}\right)}
\dd\ell\int_0^{\pi\over 2} \dd \theta\,
\cos^{1\over 2}\theta\sin^{7\over 2}\theta\nonumber\\
&=&{5\over 96}{\gso\over\gamma^2\alpha^2D_{_Q}}
\dd\ell
\,,
\eeqn
where in the integrals over $\theta$ we have set $d=5/2$.


\subsection{integral $I_3$
\label{Sec:I3}}
\subsubsection{Uncontrolled calculation in exactly $d=2$}
In exactly $d=2$,
\beqn
I_3\equiv \int_{>} {\dd^2k\over (2\pi)^2}\,
{k_x^6k_y^6\over
	\left(\gamma^2 k_x^6+\alpha^2 k_y^4\right)^3}
={1\over \pi^2}\int_{\Lambda \ee^{-\dd\ell}}^{\Lambda} \dd k_y
\int_0^\infty \dd k_x\,
{k_y^6k_x^{6}\over \left(\gamma^2 k_x^6+\alpha^2 k_y^4\right)^3}\,.\label{unc_I3}
\eeqn
Inserting (\ref{k_x_unc}) into (\ref{unc_I3}) we get
\beqn
I_3&=&{1\over 3\pi^2}\left(\alpha\over|\gamma|\right)^{7\over 3}{1\over\alpha^6}
\int_{\Lambda \ee^{-\dd\ell}}^{\Lambda}
\dd k_y k_y^{-{4\over3}}\int_0^{\pi\over 2} \dd \theta\,
\cos^4\theta\left(\tan\theta\right)^{4\over 3}\nonumber\\
&=&{1\over 3\pi^2}|\gamma|^{-{7\over 3}}\alpha^{-{11\over 3}}
\Lambda^{-{1\over 3}}
\dd\ell\int_0^{\pi\over 2} \dd \theta\,
\cos^{8\over 3}\theta\sin^{4\over 3}\theta\nonumber\\
&=&{5\over 216}{\gun\over\alpha^4D_{_Q}}
\dd\ell
\,.
\eeqn
\subsubsection{Hard continuation}
Continuing to higher dimensions using the hard continuation gives
\beqn
I_3\equiv \int_{>} {\dd^dk\over (2\pi)^d}\,
{k_x^6k_h^6\over
\left(\gamma^2 k_x^6+\alpha^2 k_h^4\right)^3}
={2\over (2\pi)^d}\int_{\Lambda \ee^{-\dd\ell}}^{\Lambda} \dd^{d-1} k_h
\int_0^\infty \dd k_x\,{k_x^6k_h^6\over
\left(\gamma^2 k_x^6+\alpha^2 k_h^4\right)^3}\,.\label{i3}
\eeqn
Changing variables and inserting (\ref{k_x}) into (\ref{i3}) we get
\beqn
I_3&=&{2S_{d-1}\over 3(2\pi)^d}\left(\alpha\over|\gamma|\right)^{7\over 3}{1\over\alpha^6}
\int_{\Lambda \ee^{-\dd\ell}}^{\Lambda}\dd k_h k_h^{d-{10\over 3}}
\int_0^{\pi\over 2} \dd \theta\, \sin^{4\over 3}\theta\cos^{8\over 3}\theta
\nonumber\\
&=&{5S_{d-1}\over 216(2\pi)^{d-1}}
|\gamma|^{-{7\over 3}}\alpha^{-{11\over 3}}
\Lambda^{3d-7\over 3}\dd\ell
\nonumber\\
&=&{5\gha\over 216\alpha^4D_{_Q}}\dd\ell\,.
\eeqn
\subsubsection{Soft continuation}
Continuing to higher dimensions using the soft continuation gives
\beqn
I_3\equiv\int_{>} {\dd^dk\over (2\pi)^d}\,
{k_s^6k_y^6\over
	\left(\gamma^2 k_s^6+\alpha^2 k_y^4\right)^3}
={2S_{d-1}\over (2\pi)^d}\int_{\Lambda \ee^{-\dd\ell}}^{\Lambda} \dd k_y
\int_0^\infty \dd k_s\,
{k_y^6k_s^{d+4}\over \left(\gamma^2 k_s^6+\alpha^2 k_y^4\right)^3}\,.\label{Soft_I3}
\eeqn
Inserting (\ref{k_s}) into (\ref{Soft_I3}) gives
\beqn
&=&{2S_{d-1}\over 3(2\pi)^d}\left(\alpha\over|\gamma|\right)^{\left({d+5\over3}\right)}{1\over\alpha^6}
\int_{\Lambda \ee^{-\dd\ell}}^{\Lambda}
\dd k_y k_y^{2d-8\over3}\int_0^{\pi\over 2} \dd \theta\,
\cos^4\theta\left(\tan\theta\right)^{d+2\over 3}\nonumber\\
&=&{2S_{d-1}\over 3(2\pi)^d}|\gamma|^{-{\left({d+5\over3}\right)}}\alpha^{d-13\over 3}
\Lambda^{2d-5\over 3}
\dd\ell\int_0^{\pi\over 2} \dd \theta\,
\cos^{5\over 2}\theta\sin^{3\over 2}\theta\nonumber\\
&=&{1\over 32}{\gso\over\alpha^4D_{_Q}}\dd\ell\,,
\eeqn
where in the integrals over $\theta$ we have set $d=5/2$.

\subsection{integral $I_4$}
\label{Sec:I4}
\subsubsection{Uncontrolled calculation in exactly $d=2$}
In exactly $d=2$
\beqn
I_4\equiv\int_{>} {\dd^2k\over (2\pi)^2}\,
{k_x^2 k_y^4\over (\gamma^2 k_x^6+\alpha^2 k_y^4)^2}
={1\over \pi^2}\int_{\Lambda \ee^{-\dd\ell}}^{\Lambda} \dd k_y
\int_0^\infty \dd k_x\,{k_x^2 k_y^4\over (\gamma^2 k_x^6+\alpha^2 k_y^4)^2}
\,.\label{i4_un}
\eeqn
Inserting (\ref{k_x_unc}) into (\ref{i4_un}) we get
\beqn
I_4&= &{1\over 3\pi^2}{1\over\alpha^3|\gamma|}\int_{\Lambda\ee^{-\dd\ell}}^\Lambda \dd k_y\,k_y^{-2}
\int_0^{\pi\over 2} \dd \theta\,\cos^2\theta\nonumber\\
&=&{1\over 6\pi}{1\over\alpha^3|\gamma|}\int_{\Lambda\ee^{-\dd\ell}}^\Lambda \dd k_y\,k_y^{-2}\nonumber\\
&=&{1\over 6\pi}{1\over \alpha^3|\gamma|}\Lambda^{-1}\dd\ell
\,.\label{}
\eeqn
\subsubsection{Hard continuation}
Continuing to higher dimensions using the hard continuation gives
\beqn
I_4\equiv\int_{>} {\dd^dk\over (2\pi)^d}\,
{k_x^2 k_h^4\over (\gamma^2 k_x^6+\alpha^2 k_h^4)^2}
={2\over (2\pi)^d}\int_{\Lambda \ee^{-\dd\ell}}^{\Lambda} \dd^{d-1} k_h
\int_0^\infty \dd k_x\,{k_x^2 k_h^4\over (\gamma^2 k_x^6+\alpha^2 k_h^4)^2}
\,.\label{i4}
\eeqn
Changing variables and inserting (\ref{k_x}) into (\ref{i4}) we get
\beqn
I_4&= &{2S_{d-1}\over 3(2\pi)^d}{1\over\alpha^3|\gamma|}\int_{\Lambda\ee^{-\dd\ell}}^\Lambda \dd k_h\,k_h^{d-4}
\int_0^{\pi\over 2} \dd \theta\,\cos^2\theta\nonumber\\
&=&{S_{d-1}\over 12(2\pi)^{d-1}}{1\over\alpha^3|\gamma|}\int_{\Lambda\ee^{-\dd\ell}}^\Lambda \dd k_h\,k_h^{d-4}\nonumber\\
&=&{S_{d-1}\over 12(2\pi)^{d-1}}{1\over \alpha^3|\gamma|}\Lambda^{d-3}\dd\ell
\,.\label{}
\eeqn

\subsubsection{Soft continuation}
Continuing to higher dimensions using the soft continuation gives
\beqn
I_4\equiv\int_{>} {\dd^dk\over (2\pi)^d}\,
{k_s^2 k_y^4\over (\gamma^2 k_s^6+\alpha^2 k_y^4)^2}
={2S_{d-1}\over (2\pi)^d}\int_{\Lambda \ee^{-\dd\ell}}^{\Lambda} \dd k_y
\int_0^\infty \dd k_s\,
{k_s^d k_y^4\over (\gamma^2 k_s^6+\alpha^2 k_y^4)^2}\,.\label{Soft_I_4}
\eeqn
Inserting (\ref{k_s}) into (\ref{Soft_I_4}) we get
\beqn
I_4&=&{2S_{d-1}\over 3(2\pi)^d}\alpha^{d-11\over 3}|\gamma|^{-{d+1\over 3}}\dd\ell
\int_0^{\pi\over 2} \dd \theta\,
(\cos\theta)^{8-d\over 3}(\sin\theta)^{d-2\over 3}\nonumber\\
&=&{2S_{d-1}\over 3(2\pi)^d}\alpha^{d-11\over 3}|\gamma|^{-{d+1\over 3}}\dd\ell
\int_0^{\pi\over 2} \dd \theta\,
\cos^{11\over 6}\theta \sin^{1\over 6}\theta\nonumber\\
&=&{5\pi\over 36\sin\left(5\pi\over 12\right)}{S_{d-1}\over (2\pi)^d}\alpha^{d-11\over 3}|\gamma|^{-{d+1\over 3}}\dd\ell\,,
\eeqn
where in the integrals over $\theta$ we have set $d=5/2$.

\newpage
\twocolumngrid

\begin{acknowledgments}
L.C. acknowledges support by the National Science
Foundation of China (under Grant No. 11874420).
J.T.
 thanks the Max Planck Institute for the Physics of Complex Systems,
Dresden, Germany, for their support through a Martin
Gutzwiller Fellowship during this period. L.C. also thanks the MPI-PKS, where the early stage of this work was performed, for their support. AM was supported by a TALENT fellowship awarded by the CY Cergy Paris universit\'e.
\end{acknowledgments}


%

\end{document}